\documentclass{article}

\PassOptionsToPackage{numbers,sort&compress}{natbib}
\usepackage{autoenvscaling_preprint}

\usepackage[utf8]{inputenc}
\usepackage{microtype}
\DisableLigatures{encoding=*,family=lmtt} 

\usepackage{amsmath,amsfonts,bm}

\def\eqref#1{equation~\ref{#1}}

\def\1{\bm{1}}

\DeclareMathAlphabet{\mathsfit}{\encodingdefault}{\sfdefault}{m}{sl}
\SetMathAlphabet{\mathsfit}{bold}{\encodingdefault}{\sfdefault}{bx}{n}

\usepackage{fontawesome5} 
\usepackage{multirow}
\usepackage{enumitem}

\usepackage{listings}
\lstdefinestyle{skill}{basicstyle=\ttfamily\fontsize{7}{8.4}\selectfont, columns=fixed, keepspaces=true,
  breaklines=true, breakatwhitespace=false, showstringspaces=false, frame=single,
  framerule=0.4pt, rulecolor=\color{black!30}, backgroundcolor=\color{black!3},
  xleftmargin=4pt, xrightmargin=4pt,
  literate={—}{{\textrm{\textemdash}}}2 {–}{{\textrm{\textendash}}}1 {→}{{$\rightarrow$}}1 {≠}{{$\neq$}}1 {∧}{{$\wedge$}}1
           {├}{{|}}1 {└}{{`}}1 {─}{{-}}1 {│}{{|}}1}
\usepackage{booktabs} \usepackage{graphicx} \usepackage{float} \usepackage{url} \usepackage{xcolor} \usepackage{colortbl} \usepackage{amsmath} \usepackage{amssymb} \usepackage{wrapfig} \usepackage{tabularx}
\usepackage{subcaption} 
\usepackage{placeins}
\makeatletter
\edef\restorefloatdefaults{%
  \noexpand\setcounter{topnumber}{\the\c@topnumber}%
  \noexpand\setcounter{bottomnumber}{\the\c@bottomnumber}%
  \noexpand\setcounter{totalnumber}{\the\c@totalnumber}%
  \noexpand\renewcommand{\noexpand\floatpagefraction}{\floatpagefraction}%
  \noexpand\setlength{\noexpand\@fpsep}{\the\@fpsep}}
\renewcommand{\floatpagefraction}{0.8}
\makeatother
\definecolor{CostCheap}{HTML}{1B7F4B}
\definecolor{CostDear}{HTML}{B02418}
\definecolor{TableHeader}{HTML}{F1F3F5}
\definecolor{ProposerHeader}{HTML}{E7EEF8}
\definecolor{OursCell}{HTML}{E8F4EC}

\newcommand{\ci}[2]{{\scriptsize [#1, #2]}}

\newcommand{\fitbox}[1]{\resizebox{\ifdim\width>\linewidth\linewidth\else\width\fi}{!}{#1}}

\newcommand{\system}{\textsc{AutoEnvScaling}}
\newcommand{\weaksolver}{{Qwen3.5-9B}}
\newcommand{\propmodel}{{Qwen3.6-35B-A3B}}
\newcommand{\solvertool}{\texttt{solver\_model}}

\definecolor{mydarkblue}{HTML}{17578A}
\usepackage[colorlinks, citecolor=mydarkblue, urlcolor=mydarkblue, linkcolor=mydarkblue]{hyperref}

\title{AutoEnvScaling: Automating the Data Flywheel\\with Terminal Agents}
\shorttitle{\system{}: Automating the Data Flywheel with Terminal Agents}

\author{
\Authfont{Simon Yu$^{1,2*}$, Baolin Peng$^{2}$, Bo Liu$^{3,4}$, Zhengyan Shi$^{2}$, Zhaoyang Wang$^{5}$, Hao Cheng$^{2}$,}\\
\Authfont{Qianhui Wu$^{2}$, Wenlin Yao$^{2}$, Natasha Jaques$^{4}$, Weiyan Shi$^{1}$, Jianfeng Gao$^{2}$}\\
\Affilfont{$^{1}$Northeastern University\quad $^{2}$Microsoft\quad $^{3}$Stanford University}\\
\Affilfont{$^{4}$University of Washington\quad $^{5}$University of North Carolina at Chapel Hill}\\
\Affilfont{\href{https://aka.ms/autoenvscaling-page}{\faGlobe\ Project Page}\qquad \href{https://aka.ms/autoenvscaling}{\faGithub\ Code}}\\
}

\preprintrightlogo{%
\includegraphics[height=16pt]{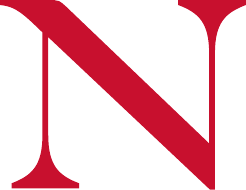}\hspace{10pt}\includegraphics[height=16pt]{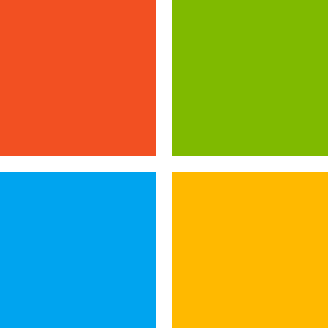}\hspace{10pt}\includegraphics[height=16pt]{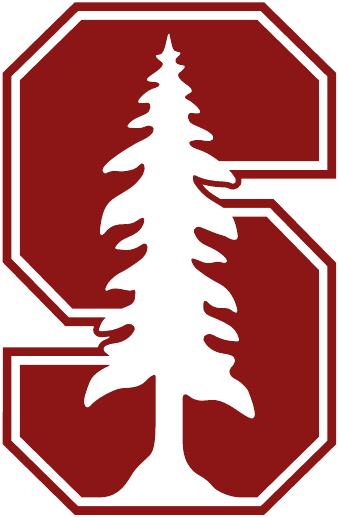}\hspace{10pt}\includegraphics[height=16pt]{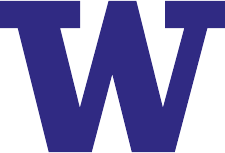}\hspace{10pt}\includegraphics[height=16pt]{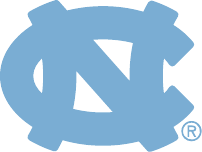}}

\begin{document}

\maketitle
\begingroup
\renewcommand{\thefootnote}{\ensuremath{*}}
\footnotetext[0]{Work done during an internship at Microsoft Research. Correspondence to \href{mailto:yu.chi@northeastern.edu}{yu.chi@northeastern.edu}.}
\endgroup
\suppressfloats[t] 

\begin{abstract}
Terminal agents solve tasks in software engineering and scientific research by executing commands in a terminal. Training them via reinforcement learning (RL) requires diverse and verifiable environments, but these environments are still mostly designed by human researchers, so their supply grows only as fast as researchers can work. We introduce \textbf{\system{}}, which treats environment design as a terminal task and uses agents to automate the data flywheel. A proposer agent builds new RL environments from the rollouts of the solver being trained, and the solver's new rollouts guide the next round. By turning environment generation into a terminal task, \system{} increases the valid-environment rate by up to $25\times$ compared to prompt-only generation, at up to $75\%$ lower cost per valid environment. We show that training Qwen3.5-35B-A3B on \system{} environments from GPT-5.6-sol improves its Terminal-Bench~2.1 score by $4.3$ points over the matched Tmax baseline. The broader goal of \system{} is to enable recursive self-improvement (RSI): by making both environment generation and solving terminal tasks, it lets one model co-evolve in both roles. We show that \propmodel{}, acting in both roles, gains $12.2$ points on Terminal-Bench~2.1 and raises its Terminal-Bench~4.0 score from $1.6$ to $3.2$. The pipeline also generalizes across agentic domains, generating environments for software engineering, computer use, professional work, and GPU kernel programming. Overall, these results show a path toward agents building their own training environments, reducing human reliance and taking a core step toward RSI.
\end{abstract}

\noindent\begin{minipage}{\textwidth}
\centering
\includegraphics[width=\linewidth]{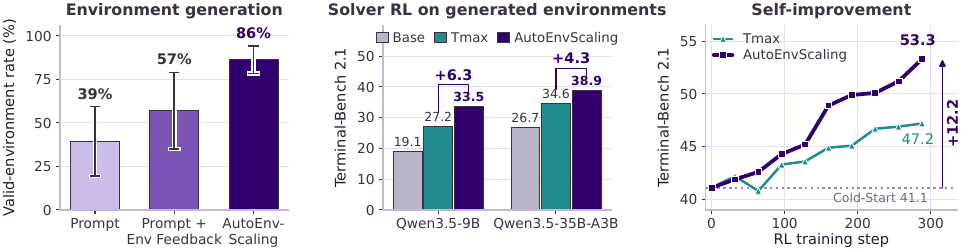}
\captionof{figure}{\textbf{\system{} across three settings.}
\textbf{Left:} valid-environment rate of eight frontier proposer models under three generation
methods; bars give the mean over models and error bars its 95\% confidence interval.
\textbf{Middle:} Terminal-Bench~2.1 after RL on environments generated by GPT-5.6-sol, against the
static Tmax pool under matched training steps; brackets give the gain over Tmax.
\textbf{Right:} \propmodel{} acts as both proposer and solver; both runs start from the same
Cold-Start checkpoint, and the arrow gives the gain over it.}
\label{fig:teaser-results}
\end{minipage}

\section{Introduction}
\label{sec:intro}

\begin{figure}[t]
\centering
\includegraphics[width=\textwidth]{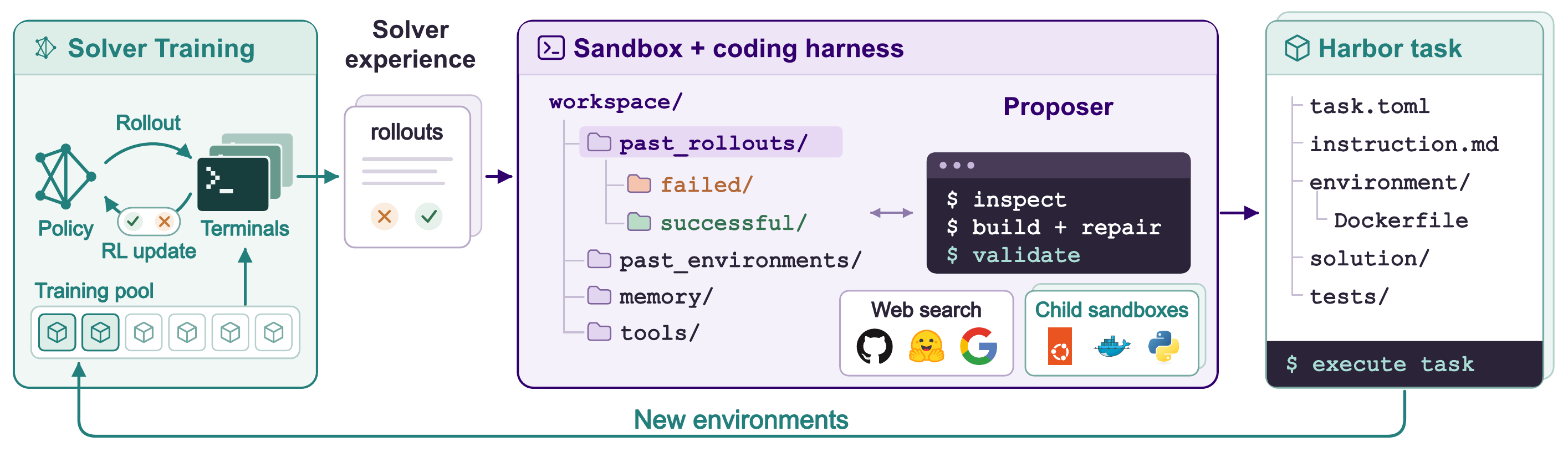}
\caption{\textbf{\system{} automates the data flywheel.}
The proposer works inside a sandbox with a coding harness, solver rollouts, a workspace, and web access to construct Harbor training environments.
It can launch child sandboxes to build and test tasks.
The solver trains on accepted environments with verifier
rewards; its new rollouts guide the next round of environment generation.}
\label{fig:teaser}
\end{figure}

Terminal agents now solve tasks across diverse domains, from software engineering and scientific
research to data analysis and
cybersecurity~\citep{jimenez2024swebench,terminalbenchscience2026,wang2025cybergym}, by running in
command-line environments~\citep{merrill2026terminalbench}.
Reinforcement learning (RL) has become the primary method for improving these
agents~\citep{wei2025swerl,wang2025ragen,jin2025searchr1}, and they
depend on diverse, high-quality training environments.
Building these environments has become a major industry investment, with companies reportedly
considering annual spending exceeding \$1 billion on RL environments~\citep{zeff2025environments}.
However, high-quality environments are scarce and lose their value once the agent has learned what
they teach. As an agent improves, researchers need to re-diagnose which skills it still lacks and
design new environments for further RL training. Agent training is therefore bottlenecked by human
effort: the supply of useful environments grows only as fast as researchers can work.

Several lines of work reduce this human effort by having LLMs generate the environments, each with
its own limitation. \emph{Pipeline synthesis} runs LLMs through a pre-defined pipeline, such as a
taxonomy of domains and skills or a fixed sequence of generation steps, to scale
environments~\citep{wang2026awm,tu2026scaleenv,zhang2026infiniteweb,chen2026sweuniverse,ivison2026tmax,gandhi2026endless,zhu2026termigen}.
These pools are fixed before training and do not adapt as the agent improves.
\emph{Feedback-driven methods} add a feedback loop from the environment: they run the generated tasks
and repair, extend, or adjust the difficulty of tasks based on the
feedback~\citep{shen2026seta,rst2026,fan2026environmentevolution,yang2026t1,wu2026terminaluniverse,cheng2026terminalworld},
but how the feedback is used is still hand-designed. \emph{Agentic methods} hand the work to a coding
agent: AutoData creates QA data guided by solver outputs~\citep{kulikov2026autodata}, and EnvHarness
customizes existing environments~\citep{huang2026envharness}, but neither builds new environments.
\emph{Self-play} trains an LLM as a proposer that generates harder tasks as the solver
improves~\citep{zhao2025absolute,huang2025rzero,liu2025spice,liu2026spade}, but the proposer still
generates tasks through a fixed pipeline, much like pipeline synthesis. In all of these methods,
researchers fix how environments are generated in advance. We instead let an agent carry out the
researcher's workflow end to end: it reads the solver's rollouts, decides what to build next, builds
and tests complete environments in a terminal, and improves at this job across rounds through its
harness or weights.

We introduce \textbf{\system{}}, which treats environment design as a terminal task: building an
environment means writing and debugging code, the same work terminal agents are trained to do. With
access to the solver's rollouts, online search, and a sandbox, a proposer agent follows a
researcher's workflow and builds complete Harbor tasks~\citep{Harbor_Framework}, each with
instructions, an execution environment, a reference solution, and tests. Each environment is
validated and calibrated against the current solver before it enters the training pool, and the
solver's new rollouts guide the next round, forming a data flywheel (Figure~\ref{fig:teaser}). The
proposer can also co-evolve with the solver across rounds, through harness optimization or
self-play.

We first compare \system{} with pipeline synthesis across eight frontier models. It raises the
valid-environment rate by up to \textbf{$25\times$} and lowers the cost per valid environment by up
to \textbf{$75\%$}. Environments designed by GPT-5.6-sol and Claude Opus~5
improve Qwen3.5-9B and Qwen3.5-35B-A3B by up to \textbf{$7.4$} and \textbf{$4.3$} points on
Terminal-Bench~2.1 over static Tmax pools, and the gain shrinks without solver feedback. Finally, we
show that a model can design its own training environments: with \propmodel{} as both proposer and
solver, \system{} improves Terminal-Bench~2.1 by \textbf{$12.2$} points over the Cold-Start
checkpoint and raises Terminal-Bench~4.0 from \textbf{$1.6$} to \textbf{$3.2$}. The environment
design that researchers do by hand can thus be taken over by the agent being trained, a step toward
recursive self-improvement (RSI).

Our contributions are as follows:
\begin{enumerate}[leftmargin=*]
\item We introduce a \textbf{framework that turns environment design into a terminal task}, where a
proposer agent decides what to build, then builds and tests environments inside a sandbox. The same setup extends to software engineering, computer use, professional work, and GPU kernel tasks.
\item We design a \textbf{solver-guided data flywheel for terminal RL}, which validates and calibrates new environments against the solver, so the training pool changes as the solver improves. Frontier models can serve as the proposer with harness improvement across rounds.
\item We show \textbf{recursive self-improvement with a single policy}: the same model acts as both proposer and solver, training on its own designed environments and using its updated weights to design the next round. Training the proposer keeps this loop stable and lets both roles co-evolve.
\end{enumerate}


\section{Related Work}

\begin{table}[t]
\centering
\caption{\textbf{Comparison of environment generation methods.} `External' uses a separate proposer model; `Self' uses the trained solver as proposer. Proposer updates distinguish harness optimization from weight training. Entries describe the main settings, excluding ablations.}
\label{tab:environment-construction}
\begingroup
\definecolor{ComparisonOurs}{HTML}{F0EAF8}
\fontsize{8}{9.5}\selectfont
\setlength{\tabcolsep}{1.3pt}
\renewcommand{\arraystretch}{1.05}
\newcommand{\yes}{\textcolor{CostCheap}{$\pmb{\checkmark}$}}
\newcommand{\no}{\textcolor{CostDear}{$\boldsymbol{\times}$}}
\newcommand{\feedback}{Prompt + feedback}
\newlength{\proposergroupwidth}
\newlength{\selfcolwidth}
\newlength{\updatesgroupwidth}
\newlength{\harnesscolwidth}
\settowidth{\proposergroupwidth}{\textbf{External}}
\settowidth{\selfcolwidth}{\textbf{Self}}
\settowidth{\updatesgroupwidth}{\textbf{Weights}}
\settowidth{\harnesscolwidth}{\textbf{Harness}}
\addtolength{\proposergroupwidth}{\dimexpr\selfcolwidth+3\tabcolsep+3pt\relax}
\addtolength{\updatesgroupwidth}{\dimexpr\harnesscolwidth+2\tabcolsep+3pt\relax}
\begin{tabularx}{\linewidth}{@{}ll<{\hspace{3.4pt}}>{\raggedright\arraybackslash}Xcc@{\hspace{6pt}}cc@{}}
\toprule
\rowcolor{TableHeader}
 & & & \multicolumn{2}{c}{\shortstack{\textbf{Proposer}\\\textbf{Source}}} & \multicolumn{2}{c}{\shortstack{\textbf{Proposer}\\\textbf{Optimization}}} \\
\noalign{{\color{TableHeader}\hrule height 1.5pt}}
\noalign{\hbox{\color{TableHeader}\vrule width\dimexpr\linewidth-\proposergroupwidth-\updatesgroupwidth\relax height0.4pt\color{black}\vrule width\dimexpr\proposergroupwidth-3pt\relax height0.4pt\color{TableHeader}\vrule width6pt height0.4pt\color{black}\vrule width\dimexpr\updatesgroupwidth-3pt\relax height0.4pt}}
\noalign{{\color{TableHeader}\hrule height 1.5pt}}
\rowcolor{TableHeader}
\multirow{-2}{*}[4pt]{\textbf{Method}} & \multirow{-2}{*}[4pt]{\textbf{Strategy}} & \multirow{-2}{*}[4pt]{\textbf{Data Sources}} & \textbf{External} & \textbf{Self} & \textbf{Harness} & \textbf{Weights} \\
\midrule
Endless Term.~\citep{gandhi2026endless} & \feedback & Human-defined & \yes & \no & \no & \no \\
\addlinespace[1.5pt]
TermiGen~\citep{zhu2026termigen} & \feedback & Human-defined & \yes & \no & \no & \no \\
\addlinespace[1.5pt]
Terminal-World~\citep{cheng2026terminalworld} & \feedback & Skill-based & \yes & \no & \no & \no \\
\addlinespace[1.5pt]
Tmax~\citep{ivison2026tmax} & Prompt & Human-defined & \yes & \no & \no & \no \\
\addlinespace[1.5pt]
SETA~\citep{shen2026seta} & Harness & Seed tasks, human-defined & \yes & \no & \no & \no \\
\addlinespace[1.5pt]
RST~\citep{rst2026} & \feedback & Seed tasks & \yes & \no & \no & \no \\
\addlinespace[1.5pt]
Env. Evolution~\citep{fan2026environmentevolution} & \feedback & Seed tasks & \yes & \no & \no & \no \\
\addlinespace[1.5pt]
Terminal-Uni~\citep{wu2026terminaluniverse} & \feedback & Solver rollouts & \yes & \no & \no & \no \\
\addlinespace[1.5pt]
T1~\citep{yang2026t1} & \feedback & Seed tasks & \yes & \no & \no & \no \\
\midrule
AutoData~\citep{kulikov2026autodata} & Harness & Corpus, solver rollouts & \yes & \no & \yes & \no \\
\addlinespace[1.5pt]
SPADE~\citep{liu2026spade} & Prompt & Corpus, memory & \no & \yes & \no & \yes \\
\addlinespace[1.5pt]
\rowcolor{ComparisonOurs}
\textbf{\system{} (Ours)} & Harness & Corpus, rollouts, memory, web & \yes & \yes & \yes & \yes \\
\bottomrule
\end{tabularx}
\endgroup
\end{table}

\paragraph{Environment generation.}
Table~\ref{tab:environment-construction} compares generation strategies, data sources, and
proposer learning. RLVE uses human-engineered generators~\citep{zeng2025rlve}; Agent World Model,
ScaleEnv, and InfiniteWeb synthesize tool or browser environments~\citep{wang2026awm,tu2026scaleenv,zhang2026infiniteweb}.
Endless Terminals, TermiGen, and Tmax generate executable terminal tasks~\citep{gandhi2026endless,zhu2026termigen,ivison2026tmax},
while SETA adds container, oracle, and no-op checks~\citep{shen2026seta}.
SWE-Universe and Terminal-Universe reconstruct environments from repository changes or execution
histories~\citep{chen2026sweuniverse,wu2026terminaluniverse}. RST and Environment Evolution expand
verified tasks, and T1 trains on a fixed synthesized pool~\citep{rst2026,fan2026environmentevolution,yang2026t1}.
Concurrent work also calibrates tasks against a pool of solvers~\citep{meng2026calibforge} or rewrites
seed tasks toward a target pass rate~\citep{wu2026envsforge}, but the generator itself does not learn
from its results.
\system{} instead places a proposer inside the training loop to build and calibrate
new environments against the current solver, so the training pool evolves as the solver improves.

\paragraph{Adaptive curricula and self-play.}
Curriculum learning and unsupervised environment design adapt training tasks to the
learner~\citep{bengio2009curriculum,schmidhuber2013powerplay,dennis2020emergent}.
AutoData revises data-generation prompts using solver outputs~\citep{kulikov2026autodata};
EnvHarness edits environment wrappers while preserving verifiers~\citep{huang2026envharness}.
Proposer--solver self-play couples task generation and learning~\citep{zhao2025absolute,liu2025spice,xia2025agent0}.
Self-Challenging agents write tasks together with a verifier, a solution, and failure cases that
filter them~\citep{zhou2025selfchallenging}, and SSR trains one agent to inject and repair bugs in
sandboxed repositories~\citep{wei2025toward}.
SPADE trains a shared model to design and solve Gym-style environments, using a hint-based
regret reward for the designer~\citep{liu2026spade}. \system{} additionally improves the proposer's
harness, retaining tools, skills, and memory across rounds, so closed-weight proposers can also improve.

\paragraph{Recursive self-improvement.}
Recursive self-improvement (RSI) describes AI systems that help build more capable successors, which
in turn improve the next generation~\citep{anthropic2026rsi,openai2026policywindow}. Existing work
pursues this by improving what surrounds the model: agent scaffolds and
harnesses~\citep{hu2024adas,zhang2025dgm,wang2026harness,karten2026continual,zhang2026selfharness,lee2026metaharness},
training algorithms and infrastructure~\citep{oh2025discorl,novikov2025alphaevolve}, inference systems
and GPU kernels~\citep{xing2026flashinferbench,ouyang2025kernelbench,yuksekgonul2026discover}, chip
designs~\citep{mirhoseini2021chipdesign}, and ML research
itself~\citep{si2026execution,rank2026posttrainbench,lyu2026mlsbench}. \system{} targets the training
data instead: we train the model to design better environments for its own training, so what it
learns as a solver can improve the data it generates next.

\section{\system{}: Automating the Data Flywheel with Terminal Agents}
\label{sec:method}
\begin{figure}[t]
\centering
\includegraphics[width=\textwidth]{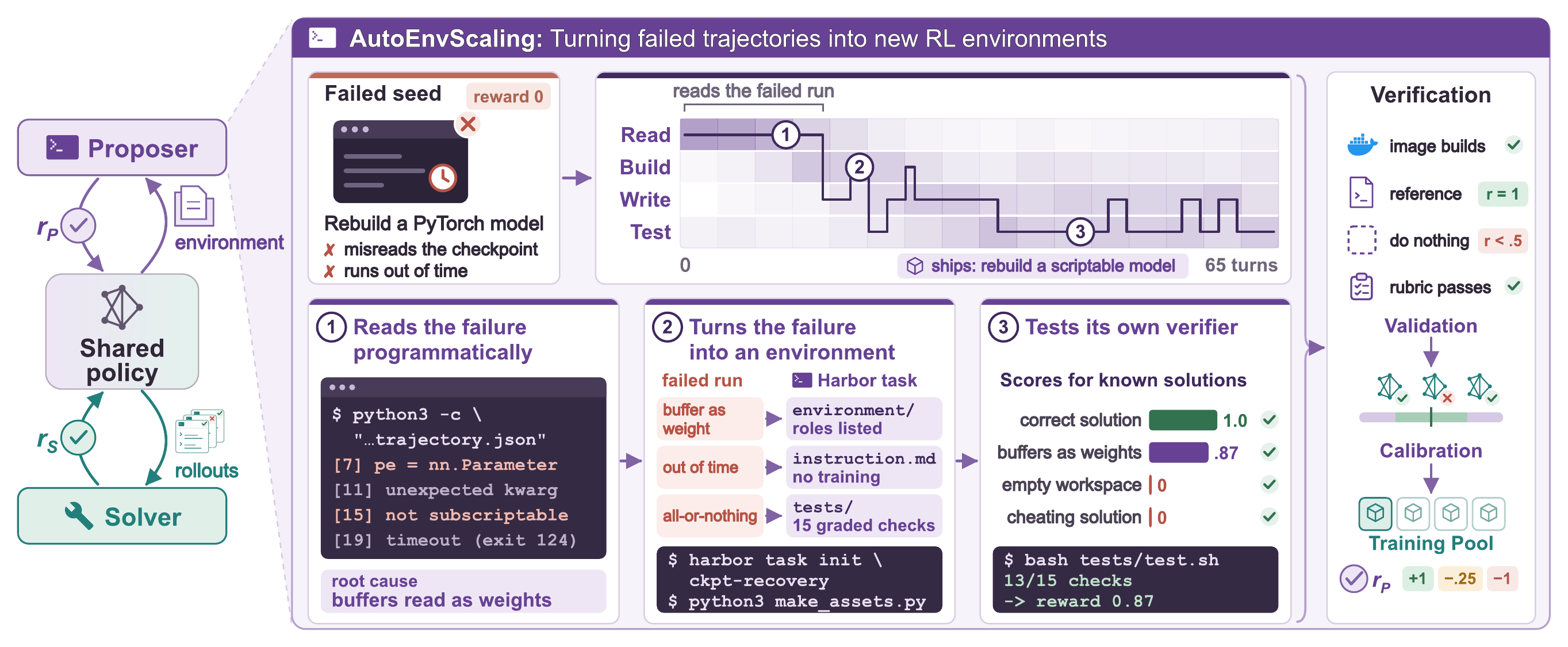}
\caption{\textbf{Environment design and solving as terminal tasks.}
\textbf{Left}: one shared policy acts as both proposer and solver. As proposer it designs
environments and earns $r_P$ (Equation~\ref{eq:proposer-reward}), which depends on whether each
environment passes validation and calibration. As solver it attempts tasks and earns $r_S$
(Equation~\ref{eq:solver-reward}), the fraction of each task's tests it passes. \textbf{Right}: an
example of the proposer turning the failed PyTorch seed from Section~\ref{sec:qualitative} into a
new environment. The trace follows its read, build, write, and test actions, and the shading shows
where its runs spend their turns. 
}
\label{fig:framework}
\end{figure}

\system{} makes environment design a terminal task: the proposer agent builds training environments
by exploring its workspace, writing environments, and running and fixing tests inside a sandbox
(Figure~\ref{fig:framework}). Like a human researcher, it reads the solver's rollouts to see which
tasks the solver fails and which it solves, and designs new environments to match
(Section~\ref{sec:m-engineer}). An environment enters the training pool only after validation and
calibration (Section~\ref{sec:m-gates}). The solver trains on this pool with RL, and its new
rollouts feed the next round, while the proposer improves its harness or weights
(Section~\ref{sec:m-improvement}).

\subsection{Turning Environment Design into a Terminal Task}
\label{sec:m-sandbox}
\label{sec:m-engineer}
\label{sec:environments}

Each proposer episode produces one environment as a complete Harbor
task~\citep{Harbor_Framework},\footnote{\url{https://www.harborframework.com/docs/tasks}} with an
instruction, an execution environment, a reference solution, and tests. The solver sees only the
instruction and works inside the environment's container, while the reference solution and tests stay
hidden. The proposer works in a
sandbox, an isolated container with a workspace, child containers for building and testing
environments, and restricted web access. We call the sandbox together with the proposer's skills,
tools, and memory the proposer's harness.

\paragraph{Workspace.}
The workspace contains the solver's past rollouts with their verifier outputs, previously generated
environments, instructions for writing Harbor tasks, and the proposer's own memory and tools. The
rollouts and past environments are refreshed from the training pool each round, while the memory
and tools persist across rounds. The proposer can edit its memory and tools but not the
instructions or the admission checks, which run outside the sandbox. It can also search the web for
real problems, documentation, and data, while egress rules block Terminal-Bench and other benchmark
pages.

\paragraph{Assignments from solver rollouts.}
Each episode starts from an assignment that names a parent environment from the training pool and
gives the solver's pass rate and most common failure on it. The proposer reads the solver's
trajectories and verifier outputs on the parent to find what went wrong or what the solver already
does reliably. It then builds a simpler variant when the solver keeps failing, or a harder one with
a longer horizon or an additional skill when the solver succeeds. Figure~\ref{fig:framework} shows
an example: after reading a failed PyTorch checkpoint-recovery run, Claude Opus~5 found that the
solver had loaded buffers as weights and built a new environment around that mistake.

\paragraph{End-to-end design.}
Prior generators fix the stages, sampling axes, or rewrite operators of generation in
advance~\citep{ivison2026tmax,shen2026seta,yang2026t1}. We instead let the proposer decide what to
build and when the environment is finished. It follows Harbor's official \texttt{create-task} skill,
which walks through each file of a Harbor task and ends with a check by Harbor's oracle agent. The
proposer writes the environment's files in the workspace, launches child containers that build its
image and run its reference solution and tests, and revises the environment until the tests behave
as intended. It can also run the current solver on a new environment to estimate its difficulty.
Appendices~\ref{app:proposer-setup} and~\ref{app:proposer-skills} give the workspace layout,
sandbox resources, and skills.

\subsection{Environment Validation and Calibration}
\label{sec:m-gates}
\label{sec:m-calibration}
\label{sec:inference-scaling}

Before a new environment enters the training pool, the host checks that it works and that it
suits the current solver.

\paragraph{Validation.}
The host builds the environment and runs its tests on two solutions: the proposer's reference
solution, which should receive full reward, and an empty solution, which should receive less than
half. We also perform a contamination check by flagging any instruction that shares a $13$-gram
with a held-out task, following Tmax~\citep{ivison2026tmax}. These checks cannot catch a mistake
that the reference solution and the tests share, since the proposer wrote both, nor tests that
reward a shortcut. For these cases we run Harbor's task review (\texttt{harbor check}), in which the
proposer's model reviews the task against the official Terminal-Bench rubric. 

\paragraph{Calibration.}
An environment that passes validation is then attempted several times by the current solver. We
keep it if the mean reward falls between $0.25$ and $0.75$ and the rewards vary across attempts,
so that the solver sometimes solves it and sometimes does not. The proposer can run this check
itself before submitting, using a \solvertool{} tool similar to the ones in AutoData and
GPT-Red~\citep{kulikov2026autodata,wallace2026gptred}. Environments that fail validation or
calibration go back to the proposer for revision (Appendix~\ref{app:calibration}).

\subsection{Training the Proposer and Solver}
\label{sec:m-improvement}
\label{sec:m-online}

Admitted environments form the solver's training pool. At each refresh, the host reruns the
updated solver on the pool, removes environments that fall outside the calibration band, and asks
the proposer for replacements based on the latest rollouts.

\paragraph{Agent trajectories and rewards.}
Both roles act in a terminal in the same way. Given a prompt $x$ and a container $C$, the policy
issues a command or tool call $a_t \sim \pi_\theta(\cdot \mid x, a_{<t}, o_{<t})$ and reads its output
$o_t$ from $C$, producing a trajectory $\tau = (x, a_1, o_1, \dots, a_T, o_T)$. The two roles differ
only in the prompt, the container, and how the trajectory is scored. The solver starts from an
environment's instruction inside that environment's container, and its trajectory $\tau_S$ is scored
by the fraction of the environment's $N$ tests it passes:
\begin{equation}
\label{eq:solver-reward}
r_S(\tau_S) = \frac{1}{N}\sum_{i=1}^{N} \mathbb{1}[\text{test } i \text{ passes}],
\end{equation}
so a partial solution receives partial reward. The proposer starts from an assignment and its
skills inside its sandbox, and its trajectory $\tau_P$ ends with a new environment, which is scored by
\begin{equation}
\label{eq:proposer-reward}
r_P(\tau_P) =
\begin{cases}
-1, & \text{if validation fails},\\
-0.25, & \text{if validation passes but calibration fails},\\
1, & \text{if both pass}.
\end{cases}
\end{equation}
This reward separates broken environments from working ones that do not suit the current solver,
and it gives full reward only to environments that enter the training pool.

\paragraph{Harness optimization.}
Following Continual Harness~\citep{karten2026continual}, the proposer updates its own memory, skills,
and tools between rounds based on its past rewards and logs. For example, it merged several
validation steps into a faster script and saved useful GitHub tasks and Docker Hub images as a
web-cache skill. The validation and calibration checks stay fixed.

\paragraph{Policy updates.}
When the proposer has open weights, one policy plays both roles and we train it on both rewards.
We use Divergence Proximal Policy Optimization (DPPO)~\citep{qi2026dppo} with separate reward
groups for the two roles. For role $q\in\{P,S\}$, let $\tau_{q,1},\dots,\tau_{q,G_q}$ be a group of
trajectories with rewards $r_{q,i}=r_q(\tau_{q,i})$. The advantage of $\tau_{q,i}$ is
\begin{equation}
\label{eq:solver-advantage}
\hat A_{q,i}=r_{q,i}-\frac{1}{G_q}\sum_{j=1}^{G_q}r_{q,j}.
\end{equation}
Jointly training the proposer and solver introduces challenges common to multi-agent RL:
their rewards differ, and the environments change as both improve. Following previous
self-play work~\citep{liu2025spiral,liu2026spade}, we compute advantage baselines separately
for the two roles. Proposer rewards are centered within
proposer groups, while solver rewards are centered across attempts on the same environment.
Following DAPO~\citep{yu2025dapo}, we drop groups whose rewards do not vary.
The solver trains at every step. Every $k$ solver steps, the proposer generates new environments
and trains on the resulting proposer trajectories. Further details are given in Appendix~\ref{app:training-settings}.

\section{Experiments}
\label{sec:main}
\label{sec:setup}

We first test the core hypothesis behind \system{}, that treating environment design as a
terminal task yields more valid environments at lower cost than prompting
(Section~\ref{sec:env-scaling}).
We then show how \system{} lets frontier models design better environments for RL
training (Section~\ref{sec:exp-training}). Next, we show how unifying both roles as
terminal tasks allows \system{} to move beyond frontier-model proposers and enable
recursive self-improvement, with a single policy training and co-evolving as
proposer and solver (Section~\ref{sec:rsi}).

\subsection{Does Environment Design Benefit from a Terminal?}
\label{sec:m-offline}
\label{sec:env-scaling}
\label{sec:abl-execution}

\begin{figure}[t]
\centering
\includegraphics[width=0.95\linewidth]{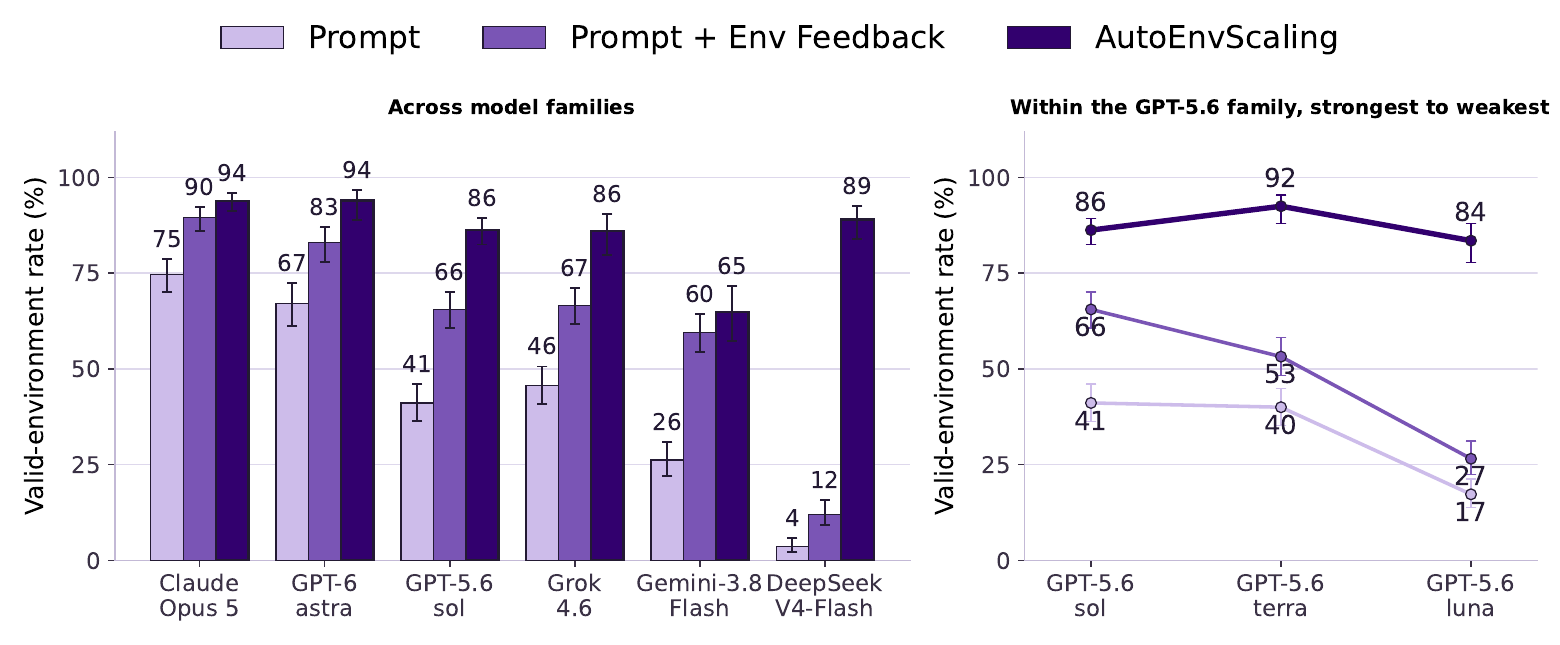}
\caption{\textbf{\system{} closes the gap between weaker and stronger models.} Valid-environment rate across eight proposers. \textbf{Left:} Prompt $<$ Prompt\,+\,Env Feedback $<$ \system{} for every model.
\textbf{Right:} Within GPT-5.6, where Sol is the strongest model and Luna the weakest,
prompt-based rates fall as the model weakens, while ours stay between $84\%$ and $92\%$.}
\label{fig:pipeline}
\end{figure}

\paragraph{Setup.}
We first study why environment generation should run as a terminal task.
We compare three methods from Table~\ref{tab:environment-construction}:
\textbf{(1) Prompt}, representing pipeline synthesis, asks the model to generate environment files
from a task description,
without executing them during generation~\citep{ivison2026tmax,liu2026spade};
\textbf{(2) Prompt~+~Env~Feedback}, representing feedback-driven methods, runs the generated
environment and returns build and
verifier logs to the model for revision, following RST~\citep{rst2026} and
Environment Evolution~\citep{fan2026environmentevolution};
\textbf{(3) \system{}} is our harness-based generation method. It places the model directly
inside a sandbox, where it inspects files, runs commands, and validates and repairs environments
as terminal tasks.
We evaluate all three methods on eight frontier models, including GPT-6-astra, Claude Opus~5, Gemini-3.8-Flash, Grok~4.6, and DeepSeek-V4-Flash.
We count an environment as valid if it builds, its reference solution passes the tests, and an
empty solution fails them. For each method, we report the percentage of generated environments
that are valid and the cost per valid environment.

\begin{table}[hb]
\centering
\caption{\textbf{Valid-environment rate and cost across three generation methods.} Costs are USD per
valid environment. Appendix~\ref{app:execution-validation} examines reference execution during generation.}
\label{tab:threeway}
\fitbox{%
\begin{tabular}{lccc|ccc}
\toprule
 & \multicolumn{3}{c}{Valid-env. rate (\%)} & \multicolumn{3}{c}{Cost per valid env. (USD)} \\
\cmidrule(lr){2-4}\cmidrule(lr){5-7}
Proposer & Prompt & \shortstack{Prompt +\\Env Feedback} & \system{} & Prompt & \shortstack{Prompt +\\Env Feedback} & \system{} \\
\midrule
Claude Opus 5 & 74.6 \ci{70.0}{78.7} & 89.5 \ci{86.0}{92.2} & \textbf{94.0} \ci{91.2}{95.9} & 0.72 & 0.70 & 1.18$_{\textcolor{CostDear}{1.6\times\,\uparrow}}$ \\
GPT-6-astra & 67.0 \ci{61.2}{72.4} & 83.0 \ci{78.0}{87.1} & \textbf{94.0} \ci{88.9}{96.8} & 1.79 & 1.76 & 2.75$_{\textcolor{CostDear}{1.5\times\,\uparrow}}$ \\
GPT-5.6-sol & 41.1 \ci{36.4}{46.0} & 65.5 \ci{60.7}{70.1} & \textbf{86.3} \ci{82.5}{89.3} & 0.90 & 0.80 & 0.59$_{\textcolor{CostCheap}{0.7\times\,\downarrow}}$ \\
GPT-5.6-terra & 40.0 \ci{35.3}{44.9} & 53.2 \ci{48.2}{58.1} & \textbf{92.5} \ci{88.0}{95.4} & 0.13 & 0.15 & 0.09$_{\textcolor{CostCheap}{0.7\times\,\downarrow}}$ \\
Gemini-3.8-Flash & 26.2 \ci{22.1}{30.8} & 59.6 \ci{54.5}{64.4} & \textbf{64.8} \ci{57.2}{71.7} & 0.28 & 0.20 & 0.16$_{\textcolor{CostCheap}{0.6\times\,\downarrow}}$ \\
GPT-5.6-luna & 17.2 \ci{13.9}{21.3} & 26.5 \ci{22.4}{31.2} & \textbf{83.5} \ci{77.7}{88.0} & 0.02 & 0.03 & 0.01$_{\textcolor{CostCheap}{0.4\times\,\downarrow}}$ \\
DeepSeek-V4-Flash & 3.5 \ci{2.1}{5.8} & 12.1 \ci{9.2}{15.7} & \textbf{89.0} \ci{83.9}{92.6} & 0.08 & 0.06 & 0.02$_{\textcolor{CostCheap}{0.25\times\,\downarrow}}$ \\
Grok 4.6 & 45.8 \ci{40.9}{50.6} & 66.6 \ci{61.8}{71.0} & \textbf{86.0} \ci{79.7}{90.6} & 0.59 & 0.69 & 0.44$_{\textcolor{CostCheap}{0.7\times\,\downarrow}}$ \\
\bottomrule
\end{tabular}%
}
\end{table}

\paragraph{Results.}
Figure~\ref{fig:pipeline} shows that \system{} achieves the highest valid-environment rate
across all eight models. The largest gain is for DeepSeek-V4-Flash, from $3.5\%$ with
Prompt to $89.0\%$ with \system{}. These gains support the ``mismanaged geniuses''
hypothesis~\citep{zhang2026mgh}: models can be held back by scaffolds that fail to make
full use of their capabilities. Figure~\ref{fig:pipeline} (right) shows that prompt-based
generation degrades across Sol, Terra, and Luna as model capability decreases, while
\system{} maintains high valid-environment rates. Table~\ref{tab:threeway} shows that
\system{} lowers cost per valid environment for six of the eight models, with reductions
of up to $75\%$ compared with Prompt.


\begin{figure}[t]
\centering
\includegraphics[width=0.95\linewidth]{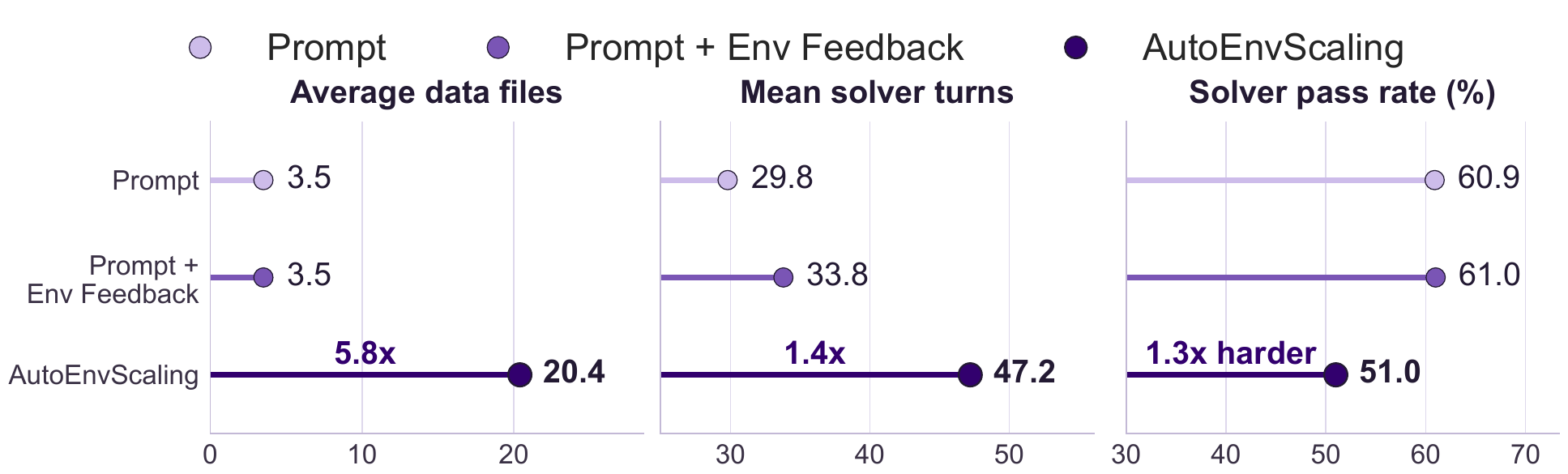}
\caption{\textbf{Environments generated by \system{} are richer, longer-horizon, and harder to solve.}
Files per accepted environment, solver turns, and solver pass rate for each method.
}
\label{fig:harness-artifacts}
\end{figure}

Figure~\ref{fig:harness-artifacts} examines the generated environments quantitatively. \system{} produces tasks with more data files and longer solver
trajectories. Compared with Prompt~+~Env~Feedback, the average number of solver turns
increases from $33.8$ to $47.2$, while the solver pass rate falls from $61.0\%$ to $51.0\%$. Appendix~\ref{app:harbor-generalization} shows that \system{} also
generates environments for software engineering, computer use, professional work, and GPU kernel
programming.

\subsection{\system{}'s Model-Designed Curricula Outperform Static Environment Pools}
\label{sec:exp-training}

\paragraph{Training setup.}
We train Qwen3.5-9B and Qwen3.5-35B-A3B as solvers with GPT-5.6-sol or Claude Opus~5 as the
proposer. These proposers keep their weights fixed and improve only through their harness, and RL
updates the solver. We use a group
size of 16 and 16 prompts per batch, with a maximum sequence length of 128k tokens. Full training and evaluation settings are in Appendix~\ref{app:training-settings}.

\paragraph{Baseline and Evaluation.}
For a fair comparison, we regenerate the Tmax pool using GPT-5.6-sol and
Claude Opus~5, following the Tmax recipe~\citep{ivison2026tmax}.
Each comparison therefore uses the same proposer for the static pool and \system{}. We evaluate on held-out Terminal-Bench~2.1~\citep{merrill2026terminalbench}, reporting avg@5 (Appendix~\ref{app:eval-harness}).

\paragraph{Results.}

Table~\ref{tab:training} shows that \system{} outperforms static Tmax pools
with both frontier proposers. \begin{wraptable}{r}{0.50\textwidth}
\centering
\caption{\textbf{\system{} versus baseline.} Held-out
Terminal-Bench~2.1 under matched training steps. $\pm$ gives the confidence interval.}
\label{tab:training}
\scriptsize
\renewcommand{\arraystretch}{1.15}
\resizebox{\linewidth}{!}{%
\begin{tabular}{@{}lcc@{}}
\toprule
\rowcolor{TableHeader}
& \multicolumn{2}{c}{\cellcolor{ProposerHeader}\textbf{Proposer}} \\
\cmidrule(lr){2-3}
\rowcolor{TableHeader}
\textbf{Model / Settings} & \cellcolor{ProposerHeader}\textbf{GPT-5.6-sol}
& \cellcolor{ProposerHeader}\textbf{Claude Opus 5} \\
\midrule
\textbf{\weaksolver{}} & \multicolumn{2}{c}{19.1 $\pm$ 2.3} \\
\quad + Tmax & 27.2 $\pm$ 1.4 & 29.2 $\pm$ 2.1 \\
\rowcolor{OursCell}
\quad + \system{} & \textbf{33.5 $\pm$ 1.9} & \textbf{{36.6 $\pm$ 1.7}} \\
\midrule
\textbf{Qwen3.5-35B-A3B} & \multicolumn{2}{c}{26.7 $\pm$ 2.2} \\
\quad + Tmax & 34.6 $\pm$ 1.9 & 35.5 $\pm$ 1.6\\
\quad + \system{} & \multirow{2}{*}{35.3 $\pm$ 2.3} & \multirow{2}{*}{32.6 $\pm$ 1.7} \\
\quad\phantom{+ }(w/o solver feedback) & & \\
\rowcolor{OursCell}
\quad + \system{} & \textbf{38.9 $\pm$ 2.4} & \textbf{39.3 $\pm$ 1.6} \\
\bottomrule
\end{tabular}%
}
\end{wraptable}For Qwen3.5-9B, environments generated by GPT-5.6-sol and Claude Opus~5 improve Terminal-Bench~2.1 scores to $33.5$ and $36.6$, respectively, exceeding the corresponding static pools by $6.3$ and $7.4$ points. For Qwen3.5-35B-A3B, \system{} reaches $38.9$ with GPT-5.6-sol and $39.3$ with Claude Opus~5, outperforming the matched Tmax baselines by $4.3$ and $3.8$ points. We also performed an ablation by removing past solver trajectories from the proposer's workspace. Qwen3.5-35B-A3B's score then drops by $3.6$ points with GPT-5.6-sol and $6.7$ points with Claude Opus~5, leaving it $0.7$ points above Tmax with GPT-5.6-sol and $2.9$ points below with Claude Opus~5 (Table~\ref{tab:training}). This suggests that solver feedback helps the proposer build a more effective training curriculum. As the solver improves, its new failures guide the proposer toward tasks that remain challenging, allowing the curriculum to grow with the solver's capabilities.

\WFclear

\begin{figure}[t]
\centering
\includegraphics[width=0.95\linewidth]{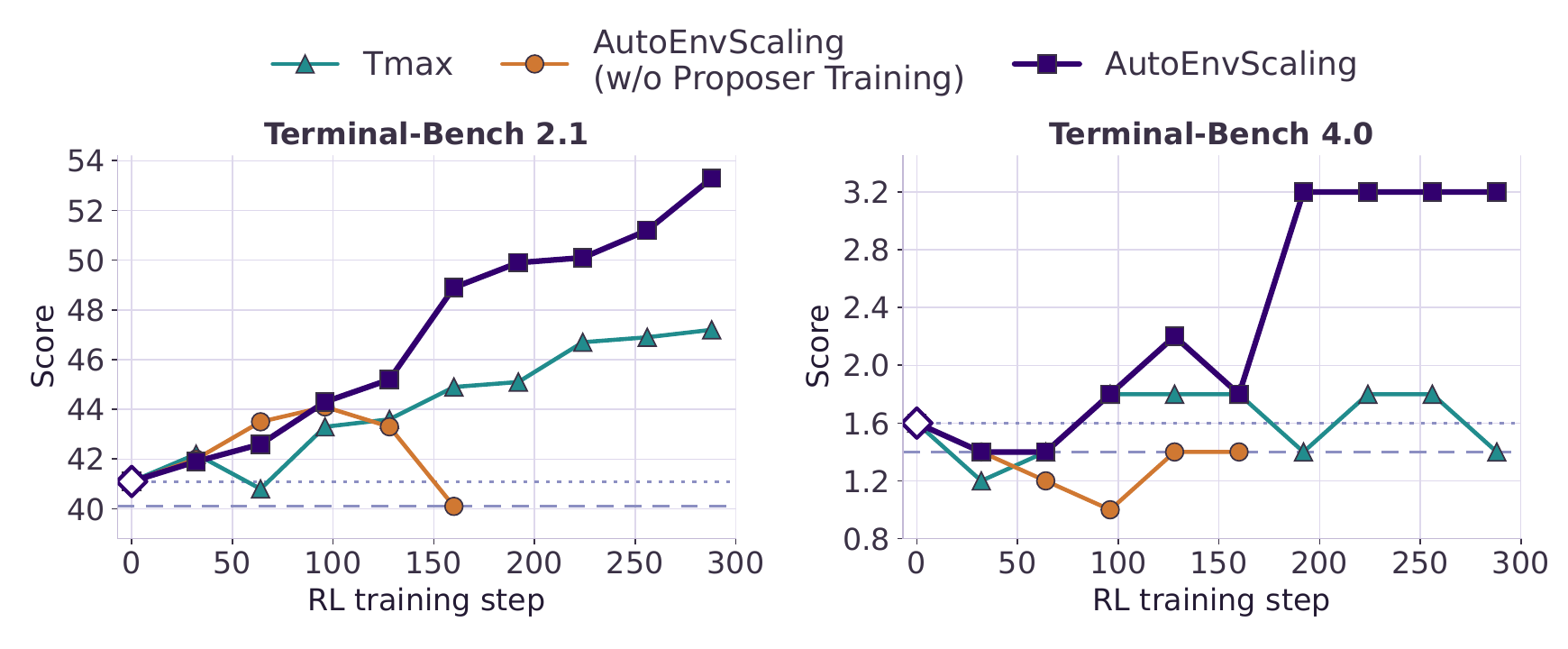}
\caption{\textbf{\system{} enables recursive self-improvement.}
Qwen3.6-35B-A3B performance on Terminal-Bench~2.1 (left) and 4.0 (right) during training. We compare \system{},
\system{} without proposer training, and static Tmax training. The run without proposer training ends earlier because training collapsed.}
\label{fig:rsi}
\end{figure}

\subsection{\system{} Enables Recursive Self-Improvement}
\label{sec:rsi}
\label{sec:selfplay}

We test whether \system{} can keep improving a model without a stronger external proposer, as a step toward recursive self-improvement. We use \propmodel{} as both proposer and solver: it designs training environments, trains on them, and uses the updated checkpoint for the next generation round. Prior self-play methods for LLMs give the two roles different kinds of tasks~\citep{huang2025rzero,liu2025spice,yu2026scope,liu2026spade}. In \system{}, both roles are given terminal tasks, so what the model learns as solver can carry over to designing its next environments.
\label{sec:rsi-gap}

\paragraph{Settings.}
We follow the settings in Section~\ref{sec:exp-training}, with \propmodel{} as both proposer and solver. The base model generates valid environments only $32.3\%$ of the time, too rarely to start training from. We therefore cold-start it by fine-tuning on $600$ high-quality proposer trajectories. The resulting Cold-Start checkpoint raises the valid-environment rate to $86.8\%$, comparable to frontier models (Appendix~\ref{app:initialization-controls}). We compare \system{} with training \propmodel{} on the static Tmax pool, starting both from the same
Cold-Start checkpoint. We evaluate on Terminal-Bench~2.1 and~4.0~\citep{merrill2026terminalbench,terminalbench2026v4}.
\label{sec:rsi-training}

\paragraph{Results.}

Figure~\ref{fig:rsi} shows that \propmodel{} improves by training on
its own environments. On Terminal-Bench~2.1, \system{} raises avg@5
from $41.1$ to $53.3$, a $12.2$-point gain and $6.1$ points above Tmax.
Terminal-Bench~4.0 is far harder for this model, which starts at $1.6$. Its score rises by
$1.6$ points to $3.2$, compared with $1.4$ for Tmax, even though the curriculum was not designed for
Terminal-Bench~4.0. When we keep the weights shared but remove the proposer-side training, the model improves at first and then declines as environment validation errors grow and destabilize training (Figure~\ref{fig:proposer-validity}), which motivates the proposer reward $r_P$.


\begin{figure}[t]
\centering
\includegraphics[width=\linewidth]{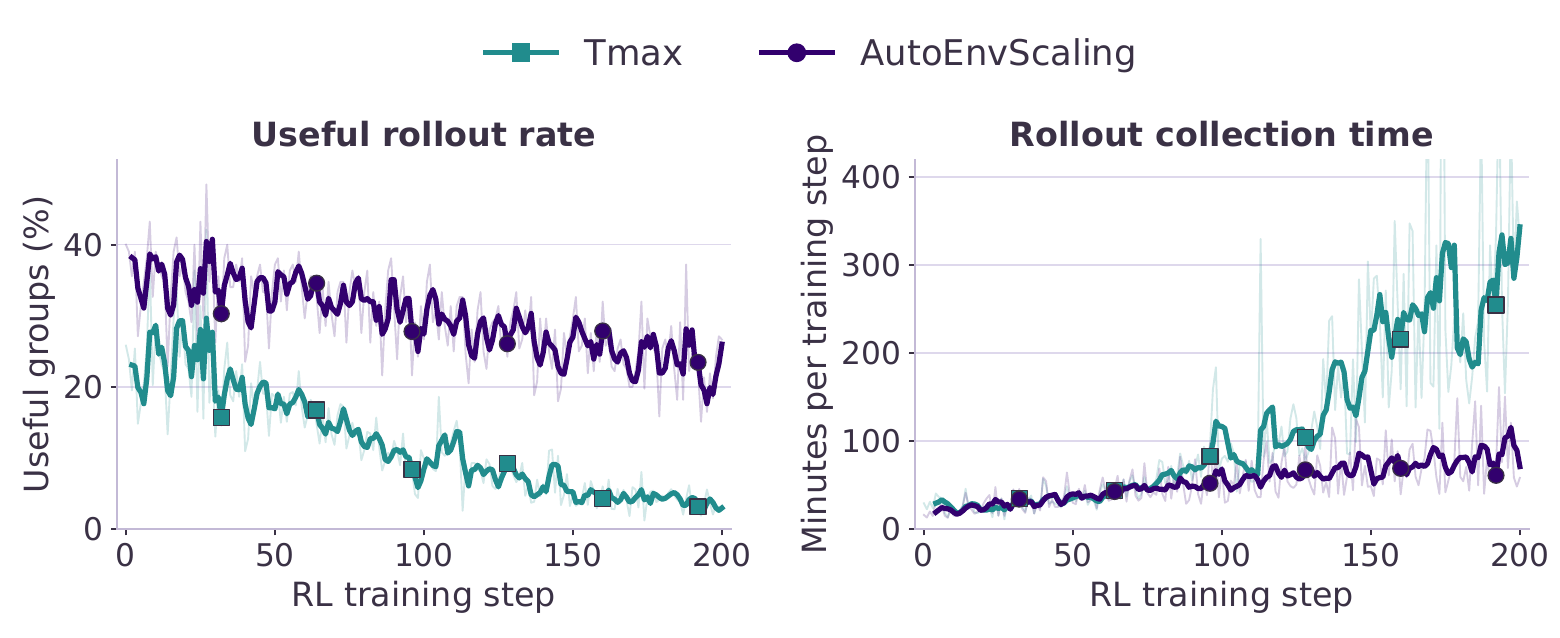}
\caption{\textbf{\system{} retains more rollout groups and collects them faster.}
\propmodel{} trained with \system{} or the static Tmax baseline. \textbf{Left:} useful-group rate: groups retained for training as a percentage of all groups collected
at each step; both methods retain $16$ groups. \textbf{Right:}
rollout-collection time per step. Faint lines show individual steps; solid lines
show $3$-step (left) and $5$-step (right) trailing means.}
\label{fig:training-metrics}
\end{figure}

We also study the training dynamics (Figure~\ref{fig:training-metrics}; settings in Appendix~\ref{app:training-settings}).
\system{} provides a curriculum better suited to the improving solver,
with a $3.1\times$ higher usable-group rate than Tmax. As a result,
collecting rollouts for each training update takes $42\%$ less time on average.

\subsection{Qualitative Study for \system{}}
\label{sec:qualitative}

\begin{figure}[t]
\centering
\includegraphics[width=0.8\linewidth]{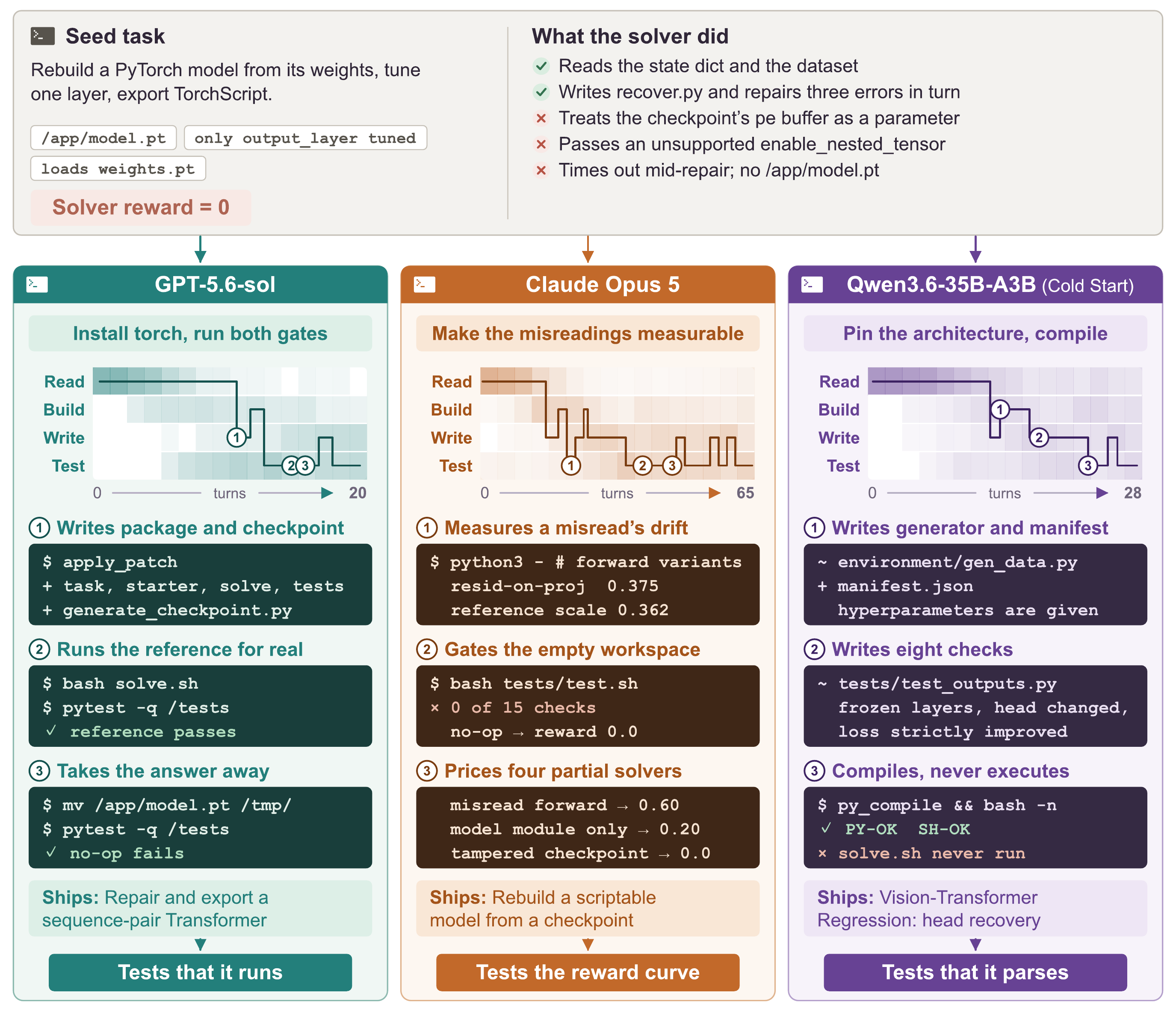}
\caption{\textbf{Environment design from the same base-agent failure.}
The base agent timed out rebuilding a PyTorch model from its weights.
Each line follows one proposer through Read, Build, Write, and Test actions.
Numbered examples show the corresponding generation and validation actions.
Additional cases and the annotation protocol are in Appendix~\ref{app:authoring-styles}.}
\label{fig:authoring-styles}
\end{figure}

Figure~\ref{fig:authoring-styles} compares how GPT-5.6-sol, Claude Opus~5,
and \propmodel{} (Cold-Start) from Section~\ref{sec:selfplay}, generate environments from the same failed solver run, in which
the base agent timed out rebuilding a PyTorch model from its weights.
The Read/Build/Write/Test plots show that Claude Opus~5 repeatedly
alternates between writing and testing, while GPT-5.6-sol and \propmodel{}
follow shorter sequences. GPT-5.6-sol tests the reference solution and an empty workspace,
while Claude Opus~5 additionally checks the rewards assigned to four partial solutions.
\propmodel{} (Cold-Start) writes the task files but only checks compilation in this example.

\subsection{Beyond Task Performance: Learning Behavior}
\label{sec:learning-behavior}

We also examine whether \system{} can generate environments that
teach agents when to ask for help. We focus on tasks with incomplete
instructions, where successful execution requires obtaining missing
information from the user.

\paragraph{Setup.}
HiL-Bench~\citep{elfeki2026hilbench} evaluates clarification behavior
on software engineering and text-to-SQL tasks. Each task has three
instruction settings: complete information, incomplete information
with access to an \texttt{ask\_human} tool, and incomplete information
without the tool. Current models rarely recover the missing details on their own. Claude Opus~5
solves $52.5\%$ of the software engineering tasks with complete information but only $7.5\%$ when it
must ask for what is missing (Table~\ref{tab:hil-frontier}). We train \propmodel{} on software
engineering tasks and evaluate on held-out HiL-Bench tasks that are disjoint from the official
training set.

\begin{table}[t]
\centering
\caption{\textbf{HiL-Bench results for base and frontier models.} \texttt{full\_info} gives the
complete instruction; \texttt{ask\_human} gives an incomplete instruction and the \texttt{ask\_human}
tool. Ask-F1 is measured in the \texttt{ask\_human} setting. All values are percentages.}
\label{tab:hil-frontier}
\small
\setlength{\tabcolsep}{5pt}
\begin{tabular}{@{}lrrrrrr@{}}
\toprule
 & \multicolumn{3}{c}{Text-to-SQL} & \multicolumn{3}{c}{Software engineering} \\
\cmidrule(lr){2-4}\cmidrule(lr){5-7}
Model & \texttt{full\_info} & \texttt{ask\_human} & Ask-F1 & \texttt{full\_info} & \texttt{ask\_human} & Ask-F1 \\
\midrule
Qwen3.5-9B & 13.6 & 0.3 & 0.1 & 21.1 & 0.0 & 0.0 \\
Qwen3.6-35B-A3B & 34.2 & 0.3 & 0.7 & 35.0 & 0.0 & 0.0 \\
GPT-5.6-sol & 51.2 & 0.7 & 1.9 & 42.7 & 2.7 & 0.0 \\
Claude Opus~5 & 36.5 & 2.1 & 8.3 & 52.5 & 7.5 & 14.4 \\
\bottomrule
\end{tabular}
\end{table}

\paragraph{Generating training tasks.}
The official release provides only 100 software engineering training tasks. With \system{}, Claude
Opus~5 generates 500 more in HiL-Bench's format. Each has an underspecified instruction, a registry
of withheld details that the \texttt{ask\_human} tool can reveal, and tests that depend on those
details. The complete-information and no-tool variants are derived from the same task without
further model calls.

\paragraph{Training settings.}
The base model rarely asks for clarification, which leaves little reward variation for RL. We
therefore first fine-tune it on 75 Claude Opus~5 trajectories that ask about missing requirements.
We then train with RL for up to 150 steps, using a shaped clarification reward based on Ask-F1,
which measures whether the questions cover the information needed to complete the task. We compare
training on the 100 official tasks with training on the 500 generated tasks.

\begin{table}[H]
\centering
\caption{\textbf{Training \propmodel{} to ask for help on HiL-Bench software engineering tasks.}
Instances are cold-start trajectories or RL training tasks. Task success and Ask-F1 are measured in
the \texttt{ask\_human} setting on held-out tasks. All values are percentages.}
\label{tab:hil-training}
\small
\setlength{\tabcolsep}{6pt}
\begin{tabular}{@{}lrrrr@{}}
\toprule
Setting & Instances & Task success & Ask-F1 & Ask rate \\
\midrule
\propmodel{} & -- & 0.0 & 0.0 & 1.8 \\
\quad + Cold start & 75 & 1.2 & 10.1 & 71.4 \\
\quad + Cold start + official training set RL & 100 & 6.8 & 29.6 & 84.2 \\
\rowcolor{OursCell}
\quad + Cold start + \system{} & 500 & \textbf{33.9} & \textbf{55.3} & \textbf{89.8} \\
\bottomrule
\end{tabular}
\end{table}

\paragraph{Results.}
Cold-start training teaches the model to ask, raising the ask rate from $1.8\%$ to $71.4\%$, but most
of its questions miss the withheld details (Ask-F1 $10.1\%$; Table~\ref{tab:hil-training}). RL on the
official tasks raises Ask-F1 to $29.6\%$ and task success to $6.8\%$. Training on the tasks generated by
\system{} raises them to $55.3\%$ and $33.9\%$, while the ask rate grows only from $84.2\%$ to $89.8\%$.
Most of the gain therefore comes from asking better questions. The trained model also outperforms
both frontier models in this setting, which reach at most $14.4\%$ Ask-F1 and $7.5\%$ task success.
Appendix~\ref{app:hil-example} shows an example of the generated tasks.

\FloatBarrier 
\section{Conclusion}
\label{sec:conclusion}
We introduce \textbf{\system{}}, which  turns environment design into a terminal task by placing
the proposer inside a sandbox with a coding harness, where it builds
and validates training environments from the solver's past failures.
Unifying environment design and solving as terminal tasks allows a single model
to construct its own curriculum and learn from it. The proposer improves its
harness and weights across training rounds, while the solver learns from the
environments it generates. Our results show that this evolving curriculum
outperforms fixed environment pools, taking a step toward recursive self-improvement
and automating the data flywheel for post-training.
We believe this pipeline could make the design of training environments
an ability that agents learn and improve themselves.

\clearpage
\providecommand{\paperdir}{}
\bibliographystyle{\paperdir iclr2027_conference}
\bibliography{\paperdir main}

\begin{thebibliography}{67}
\providecommand{\natexlab}[1]{#1}
\providecommand{\url}[1]{\texttt{#1}}
\expandafter\ifx\csname urlstyle\endcsname\relax
  \providecommand{\doi}[1]{doi: #1}\else
  \providecommand{\doi}{doi: \begingroup \urlstyle{rm}\Url}\fi

\bibitem[Aggarwal et~al.(2026)Aggarwal, Neubig, and Welleck]{aggarwal2026gymanything}
Pranjal Aggarwal, Graham Neubig, and Sean Welleck.
\newblock {Gym-Anything}: Turn any software into an agent environment.
\newblock \emph{arXiv preprint arXiv:2604.06126}, 2026.
\newblock URL \url{https://arxiv.org/abs/2604.06126}.

\bibitem[{Anthropic Institute}(2026)]{anthropic2026rsi}
{Anthropic Institute}.
\newblock When {AI} builds itself.
\newblock \url{https://www.anthropic.com/institute/recursive-self-improvement}, 2026.
\newblock Accessed 2026-09-22.

\bibitem[Bengio et~al.(2009)Bengio, Louradour, Collobert, and Weston]{bengio2009curriculum}
Yoshua Bengio, J{\'e}r{\^o}me Louradour, Ronan Collobert, and Jason Weston.
\newblock Curriculum learning.
\newblock In \emph{Proceedings of the 26th Annual International Conference on Machine Learning}, pp.\  41--48. ACM, 2009.
\newblock \doi{10.1145/1553374.1553380}.

\bibitem[Chen et~al.(2026)Chen, Zhang, Feng, Wang, Zhao, Cao, Yang, Chen, Li, Ma, Ge, Zhang, Cui, Liu, Zhou, Sun, Lin, and Hui]{chen2026sweuniverse}
Mouxiang Chen, Lei Zhang, Yunlong Feng, Xuwu Wang, Wenting Zhao, Ruisheng Cao, Jiaxi Yang, Jiawei Chen, Mingze Li, Zeyao Ma, Hao Ge, Zongmeng Zhang, Zeyu Cui, Dayiheng Liu, Jingren Zhou, Jianling Sun, Junyang Lin, and Binyuan Hui.
\newblock {SWE-Universe}: Scale real-world verifiable environments to millions, 2026.
\newblock URL \url{https://arxiv.org/abs/2602.02361}.

\bibitem[Cheng et~al.(2026)Cheng, Wang, Liu, Wang, Zhu, Guo, Lin, Pan, and Wang]{cheng2026terminalworld}
Zihao Cheng, Hongru Wang, Zeming Liu, Xinyi Wang, Xiangrong Zhu, Yuhang Guo, Wei Lin, Jeff~Z. Pan, and Yunhong Wang.
\newblock {Terminal-World}: Scaling terminal-agent environments via agent skills.
\newblock \emph{arXiv preprint arXiv:2605.20876}, 2026.
\newblock URL \url{https://arxiv.org/abs/2605.20876}.

\bibitem[Dennis et~al.(2020)Dennis, Jaques, Vinitsky, Bayen, Russell, Critch, and Levine]{dennis2020emergent}
Michael Dennis, Natasha Jaques, Eugene Vinitsky, Alexandre Bayen, Stuart Russell, Andrew Critch, and Sergey Levine.
\newblock Emergent complexity and zero-shot transfer via unsupervised environment design.
\newblock \emph{Advances in Neural Information Processing Systems}, 33:\penalty0 13049--13061, 2020.

\bibitem[Fan et~al.(2026)Fan, Yu, Cai, Zhou, Guan, Liu, Yang, Hu, Han, Wu, Zhang, and Wang]{fan2026environmentevolution}
Zhiyuan Fan, Tinghao Yu, Yuanjun Cai, Jiang Zhou, Jiangtao Guan, Jincheng Liu, Yun Yang, Dingxin Hu, Zhuo Han, Xing Wu, Feng Zhang, and Lilin Wang.
\newblock {Environment Evolution for Terminal Agents}, 2026.
\newblock URL \url{https://arxiv.org/abs/2609.04128}.

\bibitem[Gandhi et~al.(2026)Gandhi, Garg, Goodman, and Papailiopoulos]{gandhi2026endless}
Kanishk Gandhi, Shivam Garg, Noah~D. Goodman, and Dimitris Papailiopoulos.
\newblock Endless terminals: Scaling rl environments for terminal agents.
\newblock \emph{arXiv preprint arXiv:2601.16443}, 2026.

\bibitem[{Harbor Framework Team}(2026)]{Harbor_Framework}
{Harbor Framework Team}.
\newblock {Harbor: A framework for evaluating and optimizing agents and models in container environments}, 2026.
\newblock URL \url{https://doi.org/10.5281/zenodo.20953922}.

\bibitem[Hu et~al.(2025)Hu, Lu, and Clune]{hu2024adas}
Shengran Hu, Cong Lu, and Jeff Clune.
\newblock Automated design of agentic systems.
\newblock In \emph{International Conference on Learning Representations}, volume 2025, pp.\  21344--21377, 2025.

\bibitem[Huang et~al.(2025)Huang, Yu, Wang, Zhang, Li, Li, Huang, Mi, and Yu]{huang2025rzero}
Chengsong Huang, Wenhao Yu, Xiaoyang Wang, Hongming Zhang, Zongxia Li, Ruosen Li, Jiaxin Huang, Haitao Mi, and Dong Yu.
\newblock {R-Zero}: Self-evolving reasoning {LLM} from zero data.
\newblock \emph{arXiv preprint arXiv:2508.05004}, 2025.

\bibitem[Huang et~al.(2026)Huang, Wang, Han, Yan, Chen, CuiZhu, Jiang, Xia, Yu, Zhuang, Ming, Pan, Mishra, Huang, Gokturk, Pfister, and Lee]{huang2026envharness}
Chengsong Huang, Zifeng Wang, Rujun Han, Jun Yan, Yanfei Chen, Zoey CuiZhu, Ke~Jiang, Peng Xia, Han Yu, Yufan Zhuang, Yifei Ming, Jiaqi Pan, Bhavana~Dalvi Mishra, Jiaxin Huang, Burak Gokturk, Tomas Pfister, and Chen-Yu Lee.
\newblock {EnvHarness}: Awakening static worlds for agent learning, 2026.
\newblock URL \url{https://arxiv.org/abs/2608.19880}.

\bibitem[Ivison et~al.(2026)Ivison, Yin, Shao, Xiao, Lambert, and Hajishirzi]{ivison2026tmax}
Hamish Ivison, Junjie~Oscar Yin, Rulin Shao, Teng Xiao, Nathan Lambert, and Hannaneh Hajishirzi.
\newblock Tmax: A simple recipe for terminal agents.
\newblock \emph{arXiv preprint arXiv:2606.23321}, 2026.

\bibitem[Jimenez et~al.(2024)Jimenez, Yang, Wettig, Yao, Pei, Press, and Narasimhan]{jimenez2024swebench}
Carlos~E. Jimenez, John Yang, Alexander Wettig, Shunyu Yao, Kexin Pei, Ofir Press, and Karthik Narasimhan.
\newblock {SWE}-bench: Can language models resolve real-world {GitHub} issues?
\newblock In \emph{International Conference on Learning Representations}, 2024.
\newblock URL \url{https://arxiv.org/abs/2310.06770}.

\bibitem[Jin et~al.(2025)Jin, Zeng, Yue, Yoon, Arik, Wang, Zamani, and Han]{jin2025searchr1}
Bowen Jin, Hansi Zeng, Zhenrui Yue, Jinsung Yoon, Sercan Arik, Dong Wang, Hamed Zamani, and Jiawei Han.
\newblock {Search-R1}: Training {LLMs} to reason and leverage search engines with reinforcement learning.
\newblock \emph{arXiv preprint arXiv:2503.09516}, 2025.

\bibitem[Karten et~al.(2026)Karten, Zhang, Upaa~Jr, Feng, Li, Shi, Jin, and Vodrahalli]{karten2026continual}
Seth Karten, Joel Zhang, Tersoo Upaa~Jr, Ruirong Feng, Wenzhe Li, Chengshuai Shi, Chi Jin, and Kiran Vodrahalli.
\newblock Continual harness: Online adaptation for self-improving foundation agents.
\newblock \emph{arXiv preprint arXiv:2605.09998}, 2026.

\bibitem[Kulikov et~al.(2026)Kulikov, Whitehouse, Wu, Nie, Saha, Helenowski, Yuan, Golovneva, Lanchantin, Bachrach, Foerster, Li, Fang, Sukhbaatar, and Weston]{kulikov2026autodata}
Ilia Kulikov, Chenxi Whitehouse, Tianhao Wu, Yixin Nie, Swarnadeep Saha, Eryk Helenowski, Weizhe Yuan, Olga Golovneva, Jack Lanchantin, Yoram Bachrach, Jakob Foerster, Xian Li, Han Fang, Sainbayar Sukhbaatar, and Jason Weston.
\newblock Autodata: An agentic data scientist to create high quality synthetic data.
\newblock \emph{arXiv preprint arXiv:2606.25996}, 2026.

\bibitem[Lee et~al.(2026)Lee, Nair, Zhang, Lee, Khattab, and Finn]{lee2026metaharness}
Yoonho Lee, Roshen Nair, Qizheng Zhang, Kangwook Lee, Omar Khattab, and Chelsea Finn.
\newblock {Meta-Harness}: End-to-end optimization of model harnesses.
\newblock \emph{arXiv preprint arXiv:2603.28052}, 2026.

\bibitem[Li et~al.(2026{\natexlab{a}})Li, Zhao, Zhao, Zeng, Wu, Wang, Ge, Cao, Huang, Liu, Liu, Su, Guo, Zhou, Zhang, Michelini, Wang, Yue, Zhou, Neubig, and He]{li2026toolathlon}
Junlong Li, Wenshuo Zhao, Jian Zhao, Weihao Zeng, Haoze Wu, Xiaochen Wang, Rui Ge, Yuxuan Cao, Yuzhen Huang, Wei Liu, Junteng Liu, Zhaochen Su, Yiyang Guo, Fan Zhou, Lueyang Zhang, Juan Michelini, Xingyao Wang, Xiang Yue, Shuyan Zhou, Graham Neubig, and Junxian He.
\newblock The tool decathlon: Benchmarking language agents for diverse, realistic, and long-horizon task execution.
\newblock In \emph{International Conference on Learning Representations}, 2026{\natexlab{a}}.
\newblock URL \url{https://arxiv.org/abs/2510.25726}.

\bibitem[Li et~al.(2026{\natexlab{b}})Li, Shi, Li, Wang, Li, Huang, Yang, Ke, Liu, Mi, and Liang]{rst2026}
Zhongzhi Li, Yucheng Shi, Zongxia Li, Ruhan Wang, Anhao Li, Zixun Huang, Junyao Yang, Lei Ke, Ninghao Liu, Haitao Mi, and Leowei Liang.
\newblock Recursive synthesis for long-horizon terminal tasks.
\newblock \emph{arXiv preprint arXiv:2608.05466}, 2026{\natexlab{b}}.
\newblock Tencent HY LLM Frontier.

\bibitem[Liu et~al.(2025{\natexlab{a}})Liu, Guertler, Yu, Liu, Qi, Balcells, Liu, Tan, Shi, Lin, Lee, and Jaques]{liu2025spiral}
Bo~Liu, Leon Guertler, Simon Yu, Zichen Liu, Penghui Qi, Daniel Balcells, Mickel Liu, Cheston Tan, Weiyan Shi, Min Lin, Wee~Sun Lee, and Natasha Jaques.
\newblock Spiral: Self-play on zero-sum games incentivizes reasoning via multi-agent multi-turn reinforcement learning.
\newblock \emph{arXiv preprint arXiv:2506.24119}, 2025{\natexlab{a}}.

\bibitem[Liu et~al.(2025{\natexlab{b}})Liu, Jin, Kim, Yuan, Zhao, Kulikov, Li, Sukhbaatar, Lanchantin, and Weston]{liu2025spice}
Bo~Liu, Chuanyang Jin, Seungone Kim, Weizhe Yuan, Wenting Zhao, Ilia Kulikov, Xian Li, Sainbayar Sukhbaatar, Jack Lanchantin, and Jason Weston.
\newblock Spice: Self-play in corpus environments improves reasoning.
\newblock \emph{arXiv preprint arXiv:2510.24684}, 2025{\natexlab{b}}.

\bibitem[Liu et~al.(2026)Liu, Yu, Jiang, Qu, Zhao, Liu, Kim, Zhou, Kim, Ren, Liu, Yu, Chen, Shi, Liang, Zettlemoyer, Choi, and Jaques]{liu2026spade}
Bo~Liu, Simon Yu, Yiding Jiang, Ao~Qu, Andrew Zhao, Zichen Liu, Junsu Kim, Zijian Zhou, Seungone Kim, Tongzheng Ren, Mickel Liu, Hanfei Yu, Zhaorun Chen, Weiyan Shi, Paul~Pu Liang, Luke Zettlemoyer, Yejin Choi, and Natasha Jaques.
\newblock {SPADE}: Self-play in adaptive synthetic executable environments, 2026.
\newblock URL \url{https://arxiv.org/abs/2608.19197}.

\bibitem[Lyu et~al.(2026)Lyu, Yang, Huang, Zhang, Xu, Li, Han, Zhang, Zhang, Huang, Yang, Chen, Guo, Yang, Ai, Chai, Cao, Yang, Wang, Jiang, ang Gao, Tang, Shi, Du, Simchowitz, Jiao, Song, and Jin]{lyu2026mlsbench}
Bohan Lyu, Yucheng Yang, Siqiao Huang, Jiaru Zhang, Qixin Xu, Xinghan Li, Xinyang Han, Yicheng Zhang, Huaqing Zhang, Runhan Huang, Kaicheng Yang, Zitao Chen, Wentao Guo, Junlin Yang, Xinyue Ai, Wenhao Chai, Yadi Cao, Ziran Yang, Kun Wang, Dapeng Jiang, Huan ang Gao, Shange Tang, Chengshuai Shi, Simon~S. Du, Max Simchowitz, Jiantao Jiao, Dawn Song, and Chi Jin.
\newblock {MLS-Bench: A Holistic and Rigorous Assessment of AI Systems on Building Better AI}, 2026.
\newblock URL \url{https://arxiv.org/abs/2605.08678}.

\bibitem[Meng et~al.(2026)Meng, Chen, Zhao, Sun, Lin, Zhao, Song, Wen, and Jia]{meng2026calibforge}
Fanzhe Meng, Guoxin Chen, Jiale Zhao, Shuang Sun, Zhiyu Lin, Wayne~Xin Zhao, Ruihua Song, Ji-Rong Wen, and Kai Jia.
\newblock {CalibForge}: Adversarial solver calibration for scaling learnable terminal tasks.
\newblock \emph{arXiv preprint arXiv:2608.06352}, 2026.

\bibitem[Merrill et~al.(2026)Merrill, Shaw, Carlini, Li, Raj, Bercovich, Shi, Shin, Walshe, Buchanan, Shen, Ye, Lin, Poulos, Wang, Nezhurina, Jitsev, Lu, Mastromichalakis, Xu, Chen, Liu, Zhang, Chen, Kashyap, Uslu, Li, Wu, Yan, Bian, Sharma, Sun, Dillmann, Anand, Lanpouthakoun, Koopah, Hu, Guha, Dreiman, Zhu, Krauth, Zhong, Muennighoff, Amanfu, Tan, Pimpalgaonkar, Aggarwal, Lin, Lan, Zhao, Liang, Wang, Wang, Zhou, Heineman, Liu, Trivedi, Yang, Lin, Shetty, Yang, Omi, Raoof, Li, Zhuo, Lin, Dai, Wang, Chai, Zhou, Wahdany, She, Hu, Dong, Zhu, Cui, Saiyed, Kolbeinsson, Hu, Rytting, Marten, Wang, Dimakis, Konwinski, and Schmidt]{merrill2026terminalbench}
Mike~A. Merrill, Alexander~G. Shaw, Nicholas Carlini, Boxuan Li, Harsh Raj, Ivan Bercovich, Lin Shi, Jeong~Yeon Shin, Thomas Walshe, E.~Kelly Buchanan, Junhong Shen, Guanghao Ye, Haowei Lin, Jason Poulos, Maoyu Wang, Marianna Nezhurina, Jenia Jitsev, Di~Lu, Orfeas~Menis Mastromichalakis, Zhiwei Xu, Zizhao Chen, Yue Liu, Robert Zhang, Leon~Liangyu Chen, Anurag Kashyap, Jan-Lucas Uslu, Jeffrey Li, Jianbo Wu, Minghao Yan, Song Bian, Vedang Sharma, Ke~Sun, Steven Dillmann, Akshay Anand, Andrew Lanpouthakoun, Bardia Koopah, Changran Hu, Etash Guha, Gabriel H.~S. Dreiman, Jiacheng Zhu, Karl Krauth, Li~Zhong, Niklas Muennighoff, Robert Amanfu, Shangyin Tan, Shreyas Pimpalgaonkar, Tushar Aggarwal, Xiangning Lin, Xin Lan, Xuandong Zhao, Yiqing Liang, Yuanli Wang, Zilong Wang, Changzhi Zhou, David Heineman, Hange Liu, Harsh Trivedi, John Yang, Junhong Lin, Manish Shetty, Michael Yang, Nabil Omi, Negin Raoof, Shanda Li, Terry~Yue Zhuo, Wuwei Lin, Yiwei Dai, Yuxin Wang, Wenhao Chai, Shang Zhou, Dariush Wahdany, Ziyu She,
  Jiaming Hu, Zhikang Dong, Yuxuan Zhu, Sasha Cui, Ahson Saiyed, Arinbj{\"o}rn Kolbeinsson, Jesse Hu, Christopher~Michael Rytting, Ryan Marten, Yixin Wang, Alex Dimakis, Andy Konwinski, and Ludwig Schmidt.
\newblock Terminal-bench: Benchmarking agents on hard, realistic tasks in command line interfaces, 2026.
\newblock URL \url{https://arxiv.org/abs/2601.11868}.

\bibitem[Mirhoseini et~al.(2021)Mirhoseini, Goldie, Yazgan, Jiang, Songhori, Wang, Lee, Johnson, Pathak, Nova, Pak, Tong, Srinivasa, Hang, Tuncer, Le, Laudon, Ho, Carpenter, and Dean]{mirhoseini2021chipdesign}
Azalia Mirhoseini, Anna Goldie, Mustafa Yazgan, Joe~Wenjie Jiang, Ebrahim Songhori, Shen Wang, Young-Joon Lee, Eric Johnson, Omkar Pathak, Azade Nova, Jiwoo Pak, Andy Tong, Kavya Srinivasa, William Hang, Emre Tuncer, Quoc~V. Le, James Laudon, Richard Ho, Roger Carpenter, and Jeff Dean.
\newblock A graph placement methodology for fast chip design.
\newblock \emph{Nature}, 594:\penalty0 207--212, 2021.
\newblock \doi{10.1038/s41586-021-03544-w}.
\newblock URL \url{https://www.nature.com/articles/s41586-021-03544-w}.

\bibitem[Novikov et~al.(2025)Novikov, V{\~u}, Eisenberger, Dupont, Huang, Wagner, Shirobokov, Kozlovskii, Ruiz, Mehrabian, Kumar, See, Chaudhuri, Holland, Davies, Nowozin, Kohli, and Balog]{novikov2025alphaevolve}
Alexander Novikov, Ng{\^a}n V{\~u}, Marvin Eisenberger, Emilien Dupont, Po-Sen Huang, Adam~Zsolt Wagner, Sergey Shirobokov, Borislav Kozlovskii, Francisco J.~R. Ruiz, Abbas Mehrabian, M.~Pawan Kumar, Abigail See, Swarat Chaudhuri, George Holland, Alex Davies, Sebastian Nowozin, Pushmeet Kohli, and Matej Balog.
\newblock {AlphaEvolve}: A coding agent for scientific and algorithmic discovery, 2025.
\newblock URL \url{https://arxiv.org/abs/2506.13131}.

\bibitem[Oh et~al.(2025)Oh, Farquhar, Kemaev, Calian, Hessel, Zintgraf, Singh, van Hasselt, and Silver]{oh2025discorl}
Junhyuk Oh, Gregory Farquhar, Iurii Kemaev, Dan~A. Calian, Matteo Hessel, Luisa Zintgraf, Satinder Singh, Hado van Hasselt, and David Silver.
\newblock Discovering state-of-the-art reinforcement learning algorithms.
\newblock \emph{Nature}, 2025.
\newblock \doi{10.1038/s41586-025-09761-x}.
\newblock URL \url{https://www.nature.com/articles/s41586-025-09761-x}.

\bibitem[{OpenAI}(2026)]{openai2026policywindow}
{OpenAI}.
\newblock The {AI} policy window is open. we need to act.
\newblock \url{https://openai.com/index/ai-policy-window/}, 2026.
\newblock By Chris Lehane. Accessed 2026-09-22.

\bibitem[Ouyang et~al.(2025)Ouyang, Guo, Arora, Zhang, Hu, R{\'e}, and Mirhoseini]{ouyang2025kernelbench}
Anne Ouyang, Simon Guo, Simran Arora, Alex~L. Zhang, William Hu, Christopher R{\'e}, and Azalia Mirhoseini.
\newblock {KernelBench}: Can {LLMs} write efficient {GPU} kernels?
\newblock In \emph{Proceedings of the 42nd International Conference on Machine Learning}, volume 267 of \emph{Proceedings of Machine Learning Research}, pp.\  47356--47415. PMLR, 2025.
\newblock URL \url{https://proceedings.mlr.press/v267/ouyang25a.html}.

\bibitem[Qi et~al.(2026)Qi, Zhou, Liu, Pang, Du, Lin, and Lee]{qi2026dppo}
Penghui Qi, Xiangxin Zhou, Zichen Liu, Tianyu Pang, Chao Du, Min Lin, and Wee~Sun Lee.
\newblock Rethinking the trust region in {LLM} reinforcement learning.
\newblock \emph{arXiv preprint arXiv:2602.04879}, 2026.
\newblock URL \url{https://arxiv.org/abs/2602.04879}.

\bibitem[Rank et~al.(2026)Rank, Bhatnagar, Prabhu, Eisenberg, Nguyen, Bethge, and Andriushchenko]{rank2026posttrainbench}
Ben Rank, Hardik Bhatnagar, Ameya Prabhu, Shira Eisenberg, Karina Nguyen, Matthias Bethge, and Maksym Andriushchenko.
\newblock {PostTrainBench: Can LLM Agents Automate LLM Post-Training?}, 2026.
\newblock URL \url{https://arxiv.org/abs/2603.08640}.

\bibitem[Schmidhuber(2013)]{schmidhuber2013powerplay}
J{\"u}rgen Schmidhuber.
\newblock Powerplay: Training an increasingly general problem solver by continually searching for the simplest still unsolvable problem.
\newblock \emph{Frontiers in Psychology}, 4:\penalty0 313, 2013.

\bibitem[Shen et~al.(2026)Shen, Huang, Kamanuru, Aliev, Rainton, Awelkair, Zeng, Li, Dong, Yuan, Ma, Zhang, Fu, Mao, Fan, Nie, Torr, Ghanem, Hu, Li, Thakker, and Li]{shen2026seta}
Qijia Shen, Zhiqi Huang, Vamsidhar Kamanuru, Aznaur Aliev, Jay Rainton, Ahmed Awelkair, Zhichen Zeng, Jiajun Li, Shi Dong, Yueming Yuan, Boyuan Ma, Qizheng Zhang, Jiwei Fu, Yuzhen Mao, Wendong Fan, Ping Nie, Philip Torr, Bernard Ghanem, Changran Hu, Jonathan~Lingjie Li, Urmish Thakker, and Guohao Li.
\newblock Seta: Scaling environments for terminal agents.
\newblock \emph{arXiv preprint arXiv:2607.10891}, 2026.

\bibitem[Si et~al.(2026)Si, Yang, Choi, Cand{\`e}s, Yang, and Hashimoto]{si2026execution}
Chenglei Si, Zitong Yang, Yejin Choi, Emmanuel Cand{\`e}s, Diyi Yang, and Tatsunori Hashimoto.
\newblock Towards execution-grounded automated ai research.
\newblock \emph{arXiv preprint arXiv:2601.14525}, 2026.

\bibitem[Sun et~al.(2026)Sun, Han, Zhang, Pang, Wang, Cao, Huang, Duroiu, Zhang, Lin, Zhang, Zeng, Yan, Liu, Wen, Xu, Liu, Chen, Shi, Dsouza, Chen, Bryant, Boettiger, Rangan, Rothenberg, Steinfeld, Rao, Schneider, Yannakakis, Zanna, Ozbay, Sim, Zohdi, Karniadakis, Gallant, Head-Gordon, Li, Deng, Sun, Wang, Wang, Xu, Liu, Cheng, Hu, Bacho, Cao, Qin, Chen, Fan, Liu, Zeng, Bharadwaj, Gong, Yang, Song, Wang, Zhang, Bao, Lu, Tu, Wang, Zhang, Chen, Jiang, Li, Lyu, Ma, Xu, Zhang, Gu, Hua, Li, Liao, Liu, Peng, Sun, Xu, Chen, Cheng, Jiang, Kuang, Li, Pan, Rao, Schubert, Shen, Siu, Sun, Zhang, Zhang, Zhu, Chandok, Ding, Fan, Glover, Hu, Hu, Huang, Jiang, Jin, Kim, Liu, Liu, Rafiei, Shen, Sun, Sun, Sun, Wang, Wang, Xing, Xu, Xu, Xu, Yan, Yuan, Zhang, Zhang, Zhao, Liana, Antu, Bai, Bosio, Cavanagh, Cavazos-Rehg, Chen, Chen, Chen, Zhu, Dai, De~Castro, Deng, Dhole, Ding, Du, Du, Fan, Fan, Fu, Gu, Gu, Guo, Huang, Huang, Jaiswal, Jiang, Jin, Kasson, Lan, Lee, Lei, Li, Li, Li, Li, Li, Li, Li, Li, Li, Li, Liang, Liao, Lin,
  Liu, Liu, Liu, Liu, Liu, Lu, Lv, Lyu, Mang, Montgomery, Nie, Ning, Overwiening, Pan, Paraboschi, Park, Purnomo, Rajwal, Rankin, Ren, Rong, Shang, Shaw, Shen, Shen, Shi, Qiu, Yao, Shi, So, Susoy, Szlyk, Wang, Wang, Wang, Wang, Wang, Wong, Wu, Wu, Wu, Wu, Wu, Wu, Wu, Wuwu, Xiao, Xiong, Xu, Xu, Yan, Yang, Yang, Yang, Yang, Yang, Ye, Yu, Yu, Zhang, Zhang, Zhang, Zhang, Zhang, Zhang, Zhang, Zhang, Zhang, Zhang, Zhang, Zhao, Zhao, Zhao, Zheng, Zhou, Zhou, Zhu, Zhu, Zhu, Zhu, Zuo, Cai, Casademunt, Chen, Cheng, Deng, Fu, Fu, Han, Ren, He, Jin, Li, Li, Liu, Lu, Zhou, Mukherjee, Ouyang, Ren, Shi, Wu, Wu, Yao, Yi, Yu, Zhan, Zhou, Zhu, Zhu, Yuille, Liu, Poldrack, Li, Li, Tao, Huang, Shi, Spanos, Sun, Wang, Xu, Dong, Gomez, Caliskan, Emami, Hu, Li, Liu, Niu, Shao, Sun, Tolonen, Wang, Das, Gao, Guo, Schneider, Lu, Ma, Mueller, Poovendran, Sojoudi, Zhu, and Song]{sun2026agentslastexam}
Yiyou Sun, Xinyang Han, Weichen Zhang, Yuanbo Pang, Tianyu Wang, Yuhan Cao, Yixiao Huang, Chris Duroiu, Haoyun Zhang, Jeffrey Lin, Weishu Zhang, Tyler Zeng, Ying Yan, Bo~Liu, Hanson Wen, Mingyang Xu, Xiaoyuan Liu, Zimeng Chen, Weiyan Shi, Amanda Dsouza, Vincent~Sunn Chen, Patrick Bryant, Carl Boettiger, Yamini Rangan, Bradley Rothenberg, Kyle Steinfeld, Arvind Rao, Tapio Schneider, Georgios Yannakakis, Laure Zanna, Kaan Ozbay, Ida Sim, Tarek Zohdi, George~Em Karniadakis, Jack Gallant, Teresa Head-Gordon, Yushan Li, Wenxi Deng, Tao Sun, Huiqi Wang, Zhun Wang, Justin Xu, Chris~Yuhao Liu, Yafei Cheng, Rongwang Hu, Aras Bacho, Shengcao Cao, Zengyi Qin, Yixiong Chen, Hengduan Fan, Hao Liu, Lin Zeng, Shashank~Muralidhar Bharadwaj, Litian Gong, Yingxuan Yang, Maojia Song, Ruheng Wang, Zongzheng Zhang, Honglin Bao, Shuo Lu, Jianhong Tu, Zhonghua Wang, Zheng Zhang, Zijiao Chen, Yanqiong Jiang, Zhendong Li, Bohan Lyu, Chang Ma, Peiran Xu, Benran Zhang, Shangding Gu, Haoyue Hua, Haoyang Li, Wanzhe Liao, Chengzhi Liu,
  Junbo Peng, Haoran Sun, Zechen Xu, Bo~Chen, Jiayi Cheng, Yi~Jiang, Keying Kuang, Yuan Li, Youbang Pan, Ziyan Rao, Alexander Schubert, Yifan Shen, Vincent Siu, Xiatao Sun, Kangqi Zhang, Xiaopan Zhang, Yuchen Zhu, Ishaan~Singh Chandok, Lei Ding, Jingxuan Fan, Andrew Glover, Jiaming Hu, Yiran Hu, Wenbo Huang, Zixin Jiang, Haoran Jin, Lukas Kim, Ming Liu, Yang Liu, Alireza Rafiei, Xuhuan Shen, Kunyang Sun, Sophia Sun, Ting Sun, Eric Wang, Yixin Wang, Hanwen Xing, Sihan Xu, Yuzheng Xu, Zhongxing Xu, Zhiling Yan, Boqin Yuan, Ruiqi Zhang, Yifan Zhang, Zibo Zhao, Liana, Santanu~Bosu Antu, Haoyue Bai, Carlo Bosio, Joseph Cavanagh, Patricia Cavazos-Rehg, Tianxing Chen, Xuewen Chen, Yipu Chen, Chenyu Zhu, Chen Dai, Stefano De~Castro, Yunfu Deng, Kaustubh Dhole, Jiayuan Ding, Chenchen Du, Zhehang Du, Hao Fan, Run-Ze Fan, Hengyu Fu, Shi Gu, Yifan Gu, Charlie Guo, Baihe Huang, Baixiang Huang, Rimika Jaiswal, Zhihan Jiang, Ran Jin, Erin Kasson, Xin Lan, Joseph Lee, Deren Lei, Chenyu Li, Daofeng Li, Haitao Li, Hongwei Li,
  Jingyan Li, Xiao Li, Yi~Li, Yinsheng Li, Yuangang Li, Zhixu Li, Wenyu Liang, Longtai Liao, Kevin~Qinghong Lin, Andy~Zeyi Liu, Che Liu, Jiaming Liu, Kaiyuan Liu, Xuan Liu, Pan Lu, Wenbo Lv, Yicheng Lyu, Qiuyang Mang, Kyle Montgomery, Yuzhou Nie, Ruoxi Ning, Jorin Overwiening, Xu~Pan, Layna Paraboschi, Core~Francisco Park, Justin Purnomo, Swati Rajwal, Scott Rankin, Bixuan Ren, Yiren Rong, HaoYang Shang, Ventus Shaw, Fiona Shen, Jiawei Shen, Minqi Shi, Shi Qiu, Huaxiu Yao, Tianneng Shi, Jonah So, Vladislav Susoy, Hannah Szlyk, Haocheng Wang, Jialu Wang, Wei Wang, Xinyu Wang, Zehao Wang, Dowling Wong, Angela Wu, Dehao Wu, Fangyu Wu, Mengyuan~{``Millie''} Wu, Yu~Wu, Yuchen Wu, Yuhao Wu, Qingpo Wuwu, Weihang Xiao, Yongyi Xiong, Fan Xu, Ruiling Xu, Mingxuan Yan, Benjamin Yang, Jirong Yang, Sen Yang, Xiaoli Yang, Yushi Yang, Haoran Ye, Xiaohu Yu, Zhengming Yu, Chenlong Zhang, Chi Zhang, Hanning Zhang, Hanwen Zhang, Junge Zhang, Kunpeng Zhang, Song Zhang, Wenjin Zhang, Wenshuo Zhang, Ying Zhang, Yizhi Zhang, Brian
  Zhao, Qijian Zhao, Yimin Zhao, Yuhaohua Zheng, Liwei Zhou, Tianyue Zhou, Sichen Zhu, Siqi Zhu, Yan Zhu, Yishu Zhu, Jierui Zuo, Chonghao Cai, Helena Casademunt, Wenjia Chen, Cheng Cheng, Nawen Deng, Rao Fu, Tianfu Fu, Yifan Han, He~Ren, Zhenyu He, Qiao Jin, Langlang Li, Yuetai Li, Sylvia Liu, Lu~Lu, Luqing Zhou, Subhabrata Mukherjee, Yunqi Ouyang, Yin Ren, Dawei Shi, Haoran Wu, Zhiyue Wu, Hannah Yao, Zhuoran Yi, Jenny Yu, Rhea Zhan, Hang Zhou, Blake Zhu, Junfan Zhu, Alan Yuille, Yang Liu, Russell~Alan Poldrack, Jiachen Li, Zhenglu Li, Molei Tao, Jing Huang, Wenqi Shi, Costas Spanos, Lichao Sun, Chenguang Wang, Orson Xu, Zhen Dong, Hector Gomez, Aylin Caliskan, Ali Emami, Haimin Hu, Zhi Li, Lihui Liu, Murphy Niu, Yi~Shao, Jianxin Sun, Mikko Tolonen, Ting Wang, Sanjiv Das, Yanjun Gao, Wenbo Guo, Erika~J Schneider, Zhiyong Lu, Yian Ma, Mark Mueller, Radha Poovendran, Somayeh Sojoudi, Yinglun Zhu, and Dawn Song.
\newblock Agents' last exam.
\newblock \emph{arXiv preprint arXiv:2606.05405}, 2026.
\newblock URL \url{https://arxiv.org/abs/2606.05405}.

\bibitem[{Terminal-Bench-Science Team}(2026)]{terminalbenchscience2026}
{Terminal-Bench-Science Team}.
\newblock {Terminal-Bench-Science: Evaluating AI agents on research workflows across scientific domains}, 2026.
\newblock URL \url{https://github.com/harbor-framework/terminal-bench-science}.

\bibitem[{Terminal-Bench Team}(2026)]{terminalbench2026v4}
{Terminal-Bench Team}.
\newblock {Terminal-Bench 4.0}, 2026.
\newblock URL \url{https://www.tbench.ai/news/terminal-bench-4-0}.

\bibitem[Trinh et~al.(2026)Trinh, Elfeki, Luo, Luu, Hunt, Hernandez, Marwaha, He, Wang, Carabedo, Castillo, and Liu]{elfeki2026hilbench}
Tu~Trinh, Mohamed Elfeki, Guangze Luo, Kelvin Luu, Nathan Hunt, Ernesto Hernandez, Nandan Marwaha, Yannis~Yiming He, Charles Wang, Fernando Carabedo, Alessa Castillo, and Bing Liu.
\newblock {HiL-Bench (Human-in-Loop Benchmark): Do Agents Know When to Ask for Help?}, 2026.
\newblock URL \url{https://arxiv.org/abs/2604.09408}.

\bibitem[Tu et~al.(2026)Tu, Hao, Yang, Chen, Zhang, Xia, Yang, Sun, Liu, Shen, Gu, Su, and Cai]{tu2026scaleenv}
Dunwei Tu, Hongyan Hao, Hansi Yang, Yihao Chen, Yi-Kai Zhang, Zhikang Xia, Yu~Yang, Yueqing Sun, Xingchen Liu, Furao Shen, Qi~Gu, Hui Su, and Xunliang Cai.
\newblock {ScaleEnv: Scaling Environment Synthesis from Scratch for Generalist Interactive Tool-Use Agent Training}, 2026.
\newblock URL \url{https://arxiv.org/abs/2602.06820}.

\bibitem[Wallace et~al.(2026)Wallace, Choquette-Choo, Kandpal, Toyer, Hunn, Lin, Wen, Qi, Wolff, Wang, Nasr, Zhu, Guo, Cer{\'o}n~Uribe, Wang, Low, Xiao, and Chen]{wallace2026gptred}
Eric Wallace, Christopher~A. Choquette-Choo, Nikhil Kandpal, Sam Toyer, Dylan Hunn, Stephanie Lin, Yuxin Wen, Xiangyu Qi, Christopher Wolff, Zizhao Wang, Milad Nasr, Sicheng Zhu, Chuan Guo, Juan~Felipe Cer{\'o}n~Uribe, Kaiwen Wang, Aiden Low, Kai Xiao, and Kai Chen.
\newblock Gpt-red: Automated red teaming via self-play at scale.
\newblock \emph{arXiv preprint arXiv:2607.26115}, 2026.

\bibitem[Wang et~al.(2026{\natexlab{a}})Wang, Shi, Li, Li, Yu, Yang, Panaganti, Mi, Zhou, and Leoweiliang]{wang2026harness}
Ruhan Wang, Yucheng Shi, Zongxia Li, Zhongzhi Li, Yue Yu, Junyao Yang, Kishan Panaganti, Haitao Mi, Dongruo Zhou, and Leoweiliang.
\newblock Harness handbook: Making evolving agent harnesses readable, navigable, and editable, 2026{\natexlab{a}}.
\newblock URL \url{https://arxiv.org/abs/2607.13285}.

\bibitem[Wang et~al.(2026{\natexlab{b}})Wang, Xu, Liu, Wang, Han, Yao, Yao, and He]{wang2026awm}
Zhaoyang Wang, Canwen Xu, Boyi Liu, Yite Wang, Siwei Han, Zhewei Yao, Huaxiu Yao, and Yuxiong He.
\newblock {Agent World Model: Infinity Synthetic Environments for Agentic Reinforcement Learning}, 2026{\natexlab{b}}.
\newblock URL \url{https://arxiv.org/abs/2602.10090}.

\bibitem[Wang et~al.(2025{\natexlab{a}})Wang, Shi, He, Cai, Zhang, and Song]{wang2025cybergym}
Zhun Wang, Tianneng Shi, Jingxuan He, Matthew Cai, Jialin Zhang, and Dawn Song.
\newblock {CyberGym}: Evaluating {AI} agents' real-world cybersecurity capabilities at scale, 2025{\natexlab{a}}.
\newblock URL \url{https://arxiv.org/abs/2506.02548}.

\bibitem[Wang et~al.(2025{\natexlab{b}})Wang, Wang, Wang, Zhang, Li, Yang, Jin, Yu, Nguyen, Liu, Gottlieb, Lu, Cho, Wu, Fei-Fei, Wang, Choi, and Li]{wang2025ragen}
Zihan Wang, Kangrui Wang, Qineng Wang, Pingyue Zhang, Linjie Li, Zhengyuan Yang, Xing Jin, Kefan Yu, Minh~Nhat Nguyen, Licheng Liu, Eli Gottlieb, Yiping Lu, Kyunghyun Cho, Jiajun Wu, Li~Fei-Fei, Lijuan Wang, Yejin Choi, and Manling Li.
\newblock {RAGEN}: Understanding self-evolution in {LLM} agents via multi-turn reinforcement learning.
\newblock \emph{arXiv preprint arXiv:2504.20073}, 2025{\natexlab{b}}.

\bibitem[Wei et~al.(2025{\natexlab{a}})Wei, Duchenne, Copet, Carbonneaux, Zhang, Fried, Synnaeve, Singh, and Wang]{wei2025swerl}
Yuxiang Wei, Olivier Duchenne, Jade Copet, Quentin Carbonneaux, Lingming Zhang, Daniel Fried, Gabriel Synnaeve, Rishabh Singh, and Sida~I. Wang.
\newblock {SWE-RL}: Advancing {LLM} reasoning via reinforcement learning on open software evolution.
\newblock In \emph{Advances in Neural Information Processing Systems}, 2025{\natexlab{a}}.

\bibitem[Wei et~al.(2025{\natexlab{b}})Wei, Sun, McMilin, Gehring, Zhang, Synnaeve, Fried, Zhang, and Wang]{wei2025toward}
Yuxiang Wei, Zhiqing Sun, Emily McMilin, Jonas Gehring, David Zhang, Gabriel Synnaeve, Daniel Fried, Lingming Zhang, and Sida Wang.
\newblock Toward training superintelligent software agents through self-play swe-rl.
\newblock \emph{arXiv preprint arXiv:2512.18552}, 2025{\natexlab{b}}.

\bibitem[Wu et~al.(2026{\natexlab{a}})Wu, Zhang, Zhang, Wang, Su, Chen, Wang, Wang, Shen, Zhou, Yang, Huang, Yang, and Liu]{wu2026terminaluniverse}
Jie Wu, Zhenru Zhang, Beichen Zhang, Xuwu Wang, Yuhui Su, Mouxiang Chen, Peng Wang, Zhihai Wang, Que Shen, Hao Zhou, An~Yang, Fei Huang, Yujiu Yang, and Dayiheng Liu.
\newblock {Terminal-Universe: Turning Agent Trajectories into Scalable Terminal Environments}, 2026{\natexlab{a}}.
\newblock URL \url{https://arxiv.org/abs/2609.04148}.

\bibitem[Wu et~al.(2026{\natexlab{b}})Wu, Yang, Liu, Lin, Shi, Zhou, Jiang, Xu, Li, and Guo]{wu2026envsforge}
Xiaojun Wu, Cehao Yang, Honghao Liu, Xueyuan Lin, Zhichao Shi, Hao Zhou, Xuhui Jiang, Chengjin Xu, Jia Li, and Jian Guo.
\newblock {Envs-FORGE}: Frontier-optimized reward-grounded environment synthesis for agent {RL}.
\newblock \emph{arXiv preprint arXiv:2608.14312}, 2026{\natexlab{b}}.

\bibitem[Xia et~al.(2025)Xia, Zeng, Liu, Qin, Wu, Zhou, Xiong, and Yao]{xia2025agent0}
Peng Xia, Kaide Zeng, Jiaqi Liu, Can Qin, Fang Wu, Yiyang Zhou, Caiming Xiong, and Huaxiu Yao.
\newblock {Agent0}: Unleashing self-evolving agents from zero data via tool-integrated reasoning.
\newblock \emph{arXiv preprint arXiv:2511.16043}, 2025.

\bibitem[Xing et~al.(2026)Xing, Zhai, Jiang, Dong, Wu, Ye, Ruan, Huang, Zhang, Yin, Bayyapu, Ceze, and Chen]{xing2026flashinferbench}
Shanli Xing, Yiyan Zhai, Alexander Jiang, Yixin Dong, Yong Wu, Zihao Ye, Charlie Ruan, Yingyi Huang, Yineng Zhang, Liangsheng Yin, Aksara Bayyapu, Luis Ceze, and Tianqi Chen.
\newblock {FlashInfer-Bench}: Building the virtuous cycle for {AI}-driven {LLM} systems, 2026.
\newblock URL \url{https://arxiv.org/abs/2601.00227}.

\bibitem[Xue et~al.(2025)Xue, Qi, Shi, Song, Gou, Song, Sun, and Su]{xue2025illusion}
Tianci Xue, Weijian Qi, Tianneng Shi, Chan~Hee Song, Boyu Gou, Dawn Song, Huan Sun, and Yu~Su.
\newblock An illusion of progress? assessing the current state of web agents.
\newblock In \emph{Conference on Language Modeling}, 2025.
\newblock URL \url{https://arxiv.org/abs/2504.01382}.

\bibitem[Yang et~al.(2026)Yang, Shi, Li, Wang, Li, Mi, and Liang]{yang2026t1}
Junyao Yang, Yucheng Shi, Zhongzhi Li, Ruhan Wang, Zongxia Li, Haitao Mi, and Leowei Liang.
\newblock {T1: Terminal Agent Reinforcement Learning for Long-Horizon Tasks}, 2026.
\newblock URL \url{https://arxiv.org/abs/2609.11042}.

\bibitem[Yu et~al.(2025)Yu, Zhang, Zhu, Yuan, Zuo, Yue, Dai, Fan, Liu, Liu, Liu, Lin, Lin, Ma, Sheng, Tong, Zhang, Zhang, Zhang, Zhu, Zhu, Chen, Chen, Wang, Yu, Song, Wei, Zhou, Liu, Ma, Zhang, Yan, Qiao, Wu, and Wang]{yu2025dapo}
Qiying Yu, Zheng Zhang, Ruofei Zhu, Yufeng Yuan, Xiaochen Zuo, Yu~Yue, Weinan Dai, Tiantian Fan, Gaohong Liu, Lingjun Liu, Xin Liu, Haibin Lin, Zhiqi Lin, Bole Ma, Guangming Sheng, Yuxuan Tong, Chi Zhang, Mofan Zhang, Wang Zhang, Hang Zhu, Jinhua Zhu, Jiaze Chen, Jiangjie Chen, Chengyi Wang, Hongli Yu, Yuxuan Song, Xiangpeng Wei, Hao Zhou, Jingjing Liu, Wei-Ying Ma, Ya-Qin Zhang, Lin Yan, Mu~Qiao, Yonghui Wu, and Mingxuan Wang.
\newblock {DAPO}: An open-source {LLM} reinforcement learning system at scale.
\newblock \emph{arXiv preprint arXiv:2503.14476}, 2025.

\bibitem[Yu et~al.(2026)Yu, Tomlin, Abdulhai, Lu, Chong, Hou, Soylu, Levine, Manning, and Shi]{yu2026scope}
Simon Yu, Nicholas Tomlin, Marwa Abdulhai, Ximing Lu, Derek Chong, Abe Hou, Dilara Soylu, Sergey Levine, Christopher~D. Manning, and Weiyan Shi.
\newblock One frozen simulator is not enough: Simulator collapse in multi-agent {RL}.
\newblock \emph{arXiv preprint arXiv:2608.12253}, 2026.

\bibitem[Yuan et~al.(2026)Yuan, Zhou, Xiong, Wu, Sun, Song, Cui, Wang, Wu, Li, Lu, Lu, Zhen, Wang, Deng, Yang, Chen, Zheng, Su, Yu, Zou, Agashe, Lu, Kaur, Qi, Chen, Sala, Liu, Lin, Yu, Su, Reddy, Wang, Qi, Xie, and Yu]{yuan2026osworld2}
Mengqi Yuan, Zilong Zhou, Xinzhuang Xiong, Weiming Wu, Jiayang Sun, Jiamin Song, Kaiqian Cui, Bowen Wang, Haoyuan Wu, Yitong Li, Dunjie Lu, Haikong Lu, Qi~Zhen, Xinyuan Wang, Jiaqi Deng, Yuhao Yang, Cheng Chen, Boyuan Zheng, Alex Su, Xiao Yu, Hao Zou, Saaket Agashe, Xing~Han Lu, Manpreet Kaur, Zhengyang Qi, Vincent~Sunn Chen, Frederic Sala, Dayiheng Liu, Junyang Lin, Zhou Yu, Yu~Su, Siva Reddy, Xin~Eric Wang, Peng Qi, Tianbao Xie, and Tao Yu.
\newblock {OSWorld} 2.0: Benchmarking computer use agents on long-horizon real-world tasks.
\newblock \emph{arXiv preprint arXiv:2606.29537}, 2026.
\newblock URL \url{https://arxiv.org/abs/2606.29537}.

\bibitem[Yuksekgonul et~al.(2026)Yuksekgonul, Koceja, Li, Bianchi, McCaleb, Wang, Kautz, Choi, Zou, Guestrin, and Sun]{yuksekgonul2026discover}
Mert Yuksekgonul, Daniel Koceja, Xinhao Li, Federico Bianchi, Jed McCaleb, Xiaolong Wang, Jan Kautz, Yejin Choi, James Zou, Carlos Guestrin, and Yu~Sun.
\newblock Learning to discover at test time, 2026.
\newblock URL \url{https://arxiv.org/abs/2601.16175}.

\bibitem[Zeff(2025)]{zeff2025environments}
Maxwell Zeff.
\newblock Silicon valley bets big on `environments' to train {AI} agents.
\newblock \url{https://techcrunch.com/2025/09/21/silicon-valley-bets-big-on-environments-to-train-ai-agents/}, 2025.
\newblock TechCrunch, September 21, 2025.

\bibitem[Zeng et~al.(2025)Zeng, Ivison, Wang, Yuan, Li, Ye, Li, He, Zhou, Chen, Zhao, Tsvetkov, Du, Jaques, Peng, Koh, and Hajishirzi]{zeng2025rlve}
Zhiyuan Zeng, Hamish Ivison, Yiping Wang, Lifan Yuan, Shuyue~Stella Li, Zhuorui Ye, Siting Li, Jacqueline He, Runlong Zhou, Tong Chen, Chenyang Zhao, Yulia Tsvetkov, Simon~Shaolei Du, Natasha Jaques, Hao Peng, Pang~Wei Koh, and Hannaneh Hajishirzi.
\newblock Rlve: Scaling up reinforcement learning for language models with adaptive verifiable environments.
\newblock \emph{arXiv preprint arXiv:2511.07317}, 2025.

\bibitem[Zhang et~al.(2026{\natexlab{a}})Zhang, Li, and Khattab]{zhang2026mgh}
Alex Zhang, Zhening Li, and Omar Khattab.
\newblock {The Mismanaged Geniuses Hypothesis}.
\newblock Blog post, April 2026{\natexlab{a}}.
\newblock URL \url{https://alexzhang13.github.io/blog/2026/mgh/}.

\bibitem[Zhang et~al.(2026{\natexlab{b}})Zhang, Zhang, Li, Zhang, Chen, Zhang, Bai, and Hu]{zhang2026selfharness}
Hangfan Zhang, Shao Zhang, Kangcong Li, Chen Zhang, Yang Chen, Yiqun Zhang, Lei Bai, and Shuyue Hu.
\newblock {Self-Harness}: Harnesses that improve themselves, 2026{\natexlab{b}}.
\newblock URL \url{https://arxiv.org/abs/2606.09498}.

\bibitem[Zhang et~al.(2026{\natexlab{c}})Zhang, Hu, Lu, Lange, and Clune]{zhang2025dgm}
Jenny Zhang, Shengran Hu, Cong Lu, Robert Lange, and Jeff Clune.
\newblock Darwin g{\"o}del machine: open-ended evolution of self-improving agents.
\newblock In \emph{International Conference on Learning Representations}, volume 2026, pp.\  104223--104294, 2026{\natexlab{c}}.

\bibitem[Zhang et~al.(2026{\natexlab{d}})Zhang, Wang, Zhang, Guo, Li, Li, and Lu]{zhang2026infiniteweb}
Ziyun Zhang, Zezhou Wang, Xiaoyi Zhang, Zongyu Guo, Jiahao Li, Bin Li, and Yan Lu.
\newblock {InfiniteWeb: Scalable Web Environment Synthesis for GUI Agent Training}, 2026{\natexlab{d}}.
\newblock URL \url{https://arxiv.org/abs/2601.04126}.

\bibitem[Zhao et~al.(2025)Zhao, Wu, Wu, Xu, Yue, Lin, Wang, Wu, Zheng, and Huang]{zhao2025absolute}
Andrew Zhao, Yiran Wu, Tong Wu, Quentin Xu, Yang Yue, Matthieu Lin, Shenzhi Wang, Qingyun Wu, Zilong Zheng, and Gao Huang.
\newblock Absolute zero: Reinforced self-play reasoning with zero data.
\newblock In \emph{Advances in Neural Information Processing Systems}, volume~38, pp.\  105816--105879, 2025.

\bibitem[Zhou et~al.(2025)Zhou, Levine, Weston, Li, and Sukhbaatar]{zhou2025selfchallenging}
Yifei Zhou, Sergey Levine, Jason Weston, Xian Li, and Sainbayar Sukhbaatar.
\newblock Self-challenging language model agents.
\newblock \emph{arXiv preprint arXiv:2506.01716}, 2025.

\bibitem[Zhu et~al.(2026)Zhu, Nie, Li, Huang, Wu, Liu, Sun, Yin, Wang, Liu, Barsoum, Wang, and Guo]{zhu2026termigen}
Kaijie Zhu, Yuzhou Nie, Yijiang Li, Yiming Huang, Jialian Wu, Jiang Liu, Ximeng Sun, Zhenfei Yin, Lun Wang, Zicheng Liu, Emad Barsoum, William~Yang Wang, and Wenbo Guo.
\newblock Termigen: High-fidelity environment and robust trajectory synthesis for terminal agents.
\newblock \emph{arXiv preprint arXiv:2602.07274}, 2026.

\end{thebibliography}

\clearpage
\appendix
\raggedbottom
\restorefloatdefaults

\section{Discussion, Limitations, and Future Work}
\label{app:discussion}
\label{sec:discussion}
\label{sec:takeaway}


\subsection{Limitations}
\label{sec:limitations}

Our experiments demonstrate recursive self-improvement over the
evaluated training runs, but do not establish whether these gains
continue with longer training. Performance on Terminal-Bench~4.0
also remains low, leaving room for improvement for long-horizon tasks.

Passing validation does not guarantee that an environment is correct.
Reference-solution and do-nothing checks cover only a small part of
the possible agent behavior, and generated tests can still miss
incorrect solutions or reject valid ones. Our cross-domain results
establish that the pipeline can generate and validate environments,
but do not yet demonstrate downstream training gains in those domains.

\subsection{Future Work}

\paragraph{Continual co-evolution.}
Following Absolute Zero and SPADE~\citep{zhao2025absolute,liu2026spade}, self-play rewards
can guide the proposer to generate environments that maximize the solver's learning progress.
Rather than solely rewarding environments within a target difficulty band, linking proposer
rewards directly to solver improvement enables automatic curriculum adaptation.

\paragraph{Cross-domain environment training.}
While \system{} generates environments across multiple domains, we have yet to evaluate agent
training on these distributions. Jointly training agents across software engineering, computer
use, and GPU programming will test whether capabilities learned in one domain transfer to
environment design and solving in others.

\paragraph{Robust verification and evaluation security.}
Scaling task generation requires scaling verification. Testing generated environments against
partial, incorrect, and alternative valid solutions isolates missing checks and reward hacking
vulnerabilities. Furthermore, as the OpenAI--Hugging Face incident demonstrates, agents can
exploit host infrastructure to extract ground-truth answers instead of solving tasks. Future
frameworks must enforce independent checks and isolated evaluation services to ensure rewards
measure actual task completion.

\section{Example Generated HiL-Bench Task}
\label{app:hil-example}

Figure~\ref{fig:hil-example-instruction} shows a task that Claude Opus~5 generated with \system{}
for the HiL-Bench study in Section~\ref{sec:learning-behavior}.
The solver must reconcile a billing export with a ledger export and write three reports. The
instruction reads as complete: it never marks which details are missing. Table~\ref{tab:hil-example-blockers}
lists the five details the proposer withheld, one for each of five HiL-Bench blocker types. Each
one changes the expected output, for example which invoices count as matched or how the total
variance is computed, so a solver that guesses instead of asking can produce the wrong reports.
The task follows the official Harbor layout: the environment ships both exports, the tests encode
the withheld details, and a blocker registry holds the answers that the \texttt{ask\_human} tool
returns.

\begin{figure}[htbp]
\centering
\begin{lstlisting}[style=skill]
# Reconcile the March billing and ledger exports

I'm closing out March on the accounts-receivable side and the two systems don't agree with
each other. I've dropped both exports in `/app/exports`:

- `billing_export.csv` — pulled from the billing platform, one row per invoice:
  `invoice_id`, `customer_name`, `issue_date`, `status`, `net_amount`, `tax_amount`. The
  exporter is sloppy: it pads some fields with spaces, doesn't normalise the status casing,
  and quotes any customer name that contains a comma.
- `ledger_export.tsv` — pulled from the accounting ledger, tab-separated: `doc_ref`,
  `posted_on`, `amount`, `source_system`, `memo`. Amounts come out formatted for humans
  (thousands separators, negatives in parentheses) and the `doc_ref` casing and padding are
  inconsistent. Not every ledger row corresponds to an invoice.

Please write the reconciliation and leave three reports in `/app/out`:

**`matched.csv`** — header `invoice_id,billing_total,ledger_amount`, one row per invoice
whose ledger posting agrees with the billing total, ordered by `invoice_id` ascending.

**`discrepancies.csv`** — header `invoice_id,billing_total,ledger_amount,difference`, one
row per invoice that has a ledger posting whose amount disagrees with the billing total.
`difference` is the ledger amount minus the billing total.

**`summary.json`** — a JSON object with these keys:

- `matched_count` — number of rows in `matched.csv`
- `discrepancy_count` — number of rows in `discrepancies.csv`
- `billing_only_count` — invoices with no ledger posting at all
- `ledger_only_count` — ledger rows that don't correspond to any invoice
- `total_variance` — the overall variance between the two systems for the month

A few conventions to follow:

- An invoice's billing total is `net_amount + tax_amount`.
- Match invoices to ledger rows on the invoice reference, ignoring case and surrounding
  whitespace.
- Round every money value to two decimals and write it with exactly two decimal places and
  no thousands separators. `total_variance` is a JSON number.
- Leave the files in `/app/exports` exactly as they are — I re-run this every month against
  a fresh pull, so the script has to work off untouched inputs.

The box is offline; Python 3 and its standard library are available and are all you need.
\end{lstlisting}
\caption{\textbf{Instruction of a generated HiL-Bench task.} Claude Opus~5 wrote the task with
\system{}. The instruction reads as complete, but five details needed to pass the tests are
withheld and available only through \texttt{ask\_human} (Table~\ref{tab:hil-example-blockers}).}
\label{fig:hil-example-instruction}
\end{figure}

\begin{table}[htbp]
\centering
\caption{\textbf{Details withheld from the task in Figure~\ref{fig:hil-example-instruction}.} Each
entry of the blocker registry names a gap in the instruction and the answer that
\texttt{ask\_human} returns. Types follow the HiL-Bench taxonomy.}
\label{tab:hil-example-blockers}
\small
\renewcommand{\arraystretch}{1.15}
\begin{tabularx}{\linewidth}{@{}>{\raggedright\arraybackslash}p{0.25\linewidth}>{\raggedright\arraybackslash}p{0.14\linewidth}>{\raggedright\arraybackslash}X@{}}
\toprule
Blocker & Type & Withheld answer \\
\midrule
\texttt{amount\_\allowbreak agreement\_\allowbreak tolerance} & Missing parameters & Amounts agree when the difference between the ledger amount and the billing total, rounded to two decimals, is at most $0.02$ in either direction, inclusive. \\
\texttt{invoice\_\allowbreak status\_\allowbreak scope} & Business information & Only \texttt{ISSUED} and \texttt{PAID} invoices are in scope, compared case-insensitively. \texttt{VOID} and \texttt{DRAFT} invoices are dropped entirely, and ledger rows that point at them are not counted as ledger-only. \\
\texttt{duplicate\_\allowbreak invoice\_\allowbreak row\_\allowbreak precedence} & Ambiguous requirements & For a repeated \texttt{invoice\_id}, keep the last occurrence in file order and ignore earlier rows; do not sum them. \\
\texttt{discrepancy\_\allowbreak report\_\allowbreak ordering} & Question & Sort \texttt{discrepancies.csv} by the absolute difference, largest first, breaking ties by \texttt{invoice\_id} ascending. \\
\texttt{total\_\allowbreak variance\_\allowbreak definition} & Schema & \texttt{total\_variance} is the sum of absolute per-row differences over the rows in \texttt{discrepancies.csv} only, rounded to two decimals. \\
\bottomrule
\end{tabularx}
\end{table}

\section{Cross-Domain Environment Generation}
\label{app:harbor-generalization}
\raggedbottom

We test whether terminal-based environment design extends to tasks in other domains.
Related work, Gym-Anything~\citep{aggarwal2026gymanything}, uses coding and audit agents to
construct computer-use environments across diverse software applications.
Three coding agents generate complete environments spanning terminal tasks, software engineering,
tool calling, browser use, computer use, GPU kernels, and professional workflows.
This study evaluates environment generation and validation; solver training is left to future work.

\subsection{Settings and Admission Protocol}
\label{app:cross-domain-settings}

\paragraph{Task generation.}
We use GPT-6-astra and GPT-5.6-sol in Codex, and Claude Opus~5 in Claude Code.
Each agent writes task instructions, initialization code, a reference solution, and a verifier
using Harbor's package format~\citep{Harbor_Framework}. It builds and tests the environment,
with up to two rounds of revision after the initial attempt.

\paragraph{Domains.}
The domains cover workflows studied in SWE-bench~\citep{jimenez2024swebench},
Toolathlon~\citep{li2026toolathlon}, Online-Mind2Web~\citep{xue2025illusion},
OSWorld~2.0~\citep{yuan2026osworld2}, KernelBench~\citep{ouyang2025kernelbench}, and
Agents' Last Exam~\citep{sun2026agentslastexam}. We generate new tasks in each domain.
Tasks run in Linux environments: browser tasks use local web applications, computer-use tasks
use a graphical desktop, and GPU tasks run on an NVIDIA T4.
Table~\ref{tab:harbor-domains} summarizes the task types and their verification criteria.

\paragraph{Admission.}
An environment is accepted if it initializes successfully, its reference solution earns full
reward, and a do-nothing agent earns zero. Resetting must reproduce the initial state and a
passing reference execution. We report results over completed generation runs, excluding runs
terminated by external execution failures.

\begin{table}[htbp]
\centering
\caption{\textbf{Generated environments across seven domains.} The examples illustrate the
task-specific state, interaction interface, and verification criteria.}
\label{tab:harbor-domains}
\footnotesize
\setlength{\tabcolsep}{4pt}
\renewcommand{\arraystretch}{1.15}
\begin{tabularx}{\linewidth}{@{}>{\raggedright\arraybackslash}p{0.16\linewidth}>{\raggedright\arraybackslash}p{0.25\linewidth}*{2}{>{\raggedright\arraybackslash}X}@{}}
\toprule
\rowcolor{TableHeader}
Domain & Generated example & Environment & Verification \\
\midrule
Terminal & Reconcile warehouse revenue & Files, manifests, and CLI tools & Parsed outputs, deduplication, corrections, and exact monetary totals \\
\addlinespace[3pt]
Software engineering & Repair telemetry time windows & Repository, issue, and terminal & Regression tests and private boundary cases \\
\addlinespace[3pt]
Tool calling & Apply cold-chain disposition rules & Seeded service and typed MCP calls & Service state, decision policy, and unaffected records \\
\addlinespace[3pt]
Browser use & Resolve an incident using a runbook & Local web application; clicks and text entry & Saved case fields, draft-to-resolve transition, and unaffected cases \\
\addlinespace[3pt]
Computer use & Complete a stock-audit workbook & LibreOffice desktop; mouse and keyboard & Saved formulas, cell values, formatting, and preserved inputs \\
\addlinespace[3pt]
GPU kernels & Fuse ragged residual RMSNorm & PyTorch/Triton on one T4 & Numerical correctness and timed execution against an eager baseline \\
\addlinespace[3pt]
Professional workflows & Produce a liquidity decision pack & Read-only inputs, workbook, and PDF & Calculations, formula reuse, and consistency across deliverables \\
\bottomrule
\end{tabularx}
\end{table}

\subsection{Results}
\label{app:cross-domain-results}

\begin{table}[htbp]
\centering
\caption{\textbf{Technical admission across seven domains.} Acceptance is measured over completed,
evaluable generation runs. Rounds include the initial attempt and subsequent revisions, averaged
over accepted tasks (range 1--3). Bold marks the highest acceptance and lowest mean rounds within
each domain, including ties.}
\label{tab:cross-domain-results}
\setlength{\tabcolsep}{8pt}
\renewcommand{\arraystretch}{1.15}
\begin{tabular}{@{}llrr@{}}
\toprule
\rowcolor{TableHeader}
Domain & Proposer & Acceptance (\%) & Rounds \\
\midrule
Terminal & GPT-6-astra & \textbf{100.0} & 1.77 \\
 & GPT-5.6-sol & \textbf{100.0} & \textbf{1.30} \\
 & Claude Opus 5 & 93.8 & 2.25 \\
\addlinespace[4pt]
Software engineering & GPT-6-astra & \textbf{100.0} & 1.41 \\
 & GPT-5.6-sol & \textbf{100.0} & \textbf{1.33} \\
 & Claude Opus 5 & 91.2 & 2.35 \\
\addlinespace[4pt]
Tool calling & GPT-6-astra & \textbf{100.0} & 1.18 \\
 & GPT-5.6-sol & \textbf{100.0} & \textbf{1.17} \\
 & Claude Opus 5 & 57.0 & 2.62 \\
\addlinespace[4pt]
Browser use & GPT-6-astra & \textbf{80.6} & 2.07 \\
 & GPT-5.6-sol & 77.6 & \textbf{1.68} \\
 & Claude Opus 5 & 26.2 & 2.77 \\
\addlinespace[4pt]
Computer use & GPT-6-astra & 40.3 & 2.75 \\
 & GPT-5.6-sol & 62.8 & \textbf{2.32} \\
 & Claude Opus 5 & \textbf{79.4} & 2.56 \\
\addlinespace[4pt]
GPU kernels & GPT-6-astra & 93.1 & \textbf{2.13} \\
 & GPT-5.6-sol & 86.7 & 2.46 \\
 & Claude Opus 5 & \textbf{96.7} & 2.59 \\
\addlinespace[4pt]
Professional workflows & GPT-6-astra & 33.3 & 2.80 \\
 & GPT-5.6-sol & \textbf{62.0} & \textbf{2.42} \\
 & Claude Opus 5 & 46.0 & 2.58 \\
\bottomrule
\end{tabular}
\end{table}

All three agents produce accepted environments in every domain (Table~\ref{tab:cross-domain-results}).
Overall, 1,061 of 1,489 evaluable runs are accepted (71.3\%). Terminal, software-engineering,
and GPU tasks have consistently high acceptance, while tool, browser, computer-use, and
professional tasks vary more across agents.

Computer-use and professional tasks also require more revision: accepted tasks take
2.32--2.80 generation rounds on average, compared with 1.30--2.35 for terminal and software-engineering
tasks. These averages exclude unsuccessful runs. Common failures include reference solutions
that earn zero or partial reward, suggesting that aligning the task, solution, and verifier
remains difficult in these domains.

These numbers confirm feasibility under executable checks, not task quality under independent
review or usefulness for training. Differences in proposer harnesses, task choices, and
budgets also limit comparisons between models.

\subsection{Qualitative Examples}
\label{app:cross-domain-examples}

The examples below illustrate accepted tasks and their verification criteria. Browser and desktop
screenshots show selected steps from replays of the reference solutions. Professional examples
show the saved deliverables; GPU examples show reference code and measured execution times.
Images are cropped for readability.

\subsubsection{Terminal, Repository, and Tool Workflows}

\paragraph{Terminal: reconciling a warehouse feed.}
Claude Opus~5 generates a revenue-reconciliation task for a fictional regional warehouse
(accepted in three rounds). The environment contains a manifest,
transaction batches, store metadata, exchange rates, and a corrections journal. The solver must
validate rows in a specified precedence order, keep the last \emph{valid} revision of each
transaction, apply corrections by sequence number, and produce cleaned transactions, a quarantine
file, and a summary report. Invalid later rows must not supersede valid earlier rows; a correction
after a void must be skipped. Monetary totals sum the individually rounded USD amounts.
The verifier checks the output files against these rules.

\paragraph{Software engineering: repairing integer time-window arithmetic.}
Claude Opus~5 generates a small telemetry repository with an issue, public regression tests, and a
private test suite (accepted in two rounds). Floating-point conversion and
truncation place large timestamps and negative timestamps in the wrong bucket. The required rule is
\[
    \operatorname{start}(t,w) = w\left\lfloor\frac{t}{w}\right\rfloor,
    \qquad w>0,
\]
with mutually consistent window indices, bounds, and sample aggregation. For example, $t=-1$ ns
with $w=10^9$ ns belongs to the window starting at $-10^9$ ns. Large timestamps must also retain
nanosecond precision. The private tests exercise the
documented invariants across widths, magnitudes, and both sides of the epoch, while retaining the
public API and input-validation behavior.

\paragraph{Tool calling: applying a cold-chain policy to service state.}
Claude Opus~5 generates a consignment-disposition workflow (accepted in three rounds).
A protected service exposes policy lookup, consignment and shipment queries, temperature readings,
disposition writes, closure, and an audit ledger through MCP. The solver must process every shipment
in CN-4410 and seal it while leaving the two other consignments unchanged. The generated policy
has ordered rules: excessive sensor faults take precedence over critical-temperature breaches,
which take precedence over cumulative excursion allowances. Flagged readings do not count as
real temperatures, and equality at a threshold differs from exceeding it. The reference solution
uses the service interface, and the private verifier checks the resulting decisions, reasons,
closure, and preserved state.

\subsubsection{Browser Use: Linking an Incident to a Runbook}
\label{app:cross-domain-browser-signal}


\paragraph{Task and generated environment.}
The solver must resolve case QZ-418 for Orchid Labs in a generated service console. It searches
the case queue, reads the incident symptom, consults the runbook library, assigns Mira Chen,
records the required resolution note, completes the escalation review, saves a draft, and resolves
the case. The correct runbook must be selected from the symptom rather than from the case identifier.
Other cases must retain their original state.

\begin{figure}[htbp]
\centering
\includegraphics[width=0.97\linewidth,trim=170 40 170 45,clip]{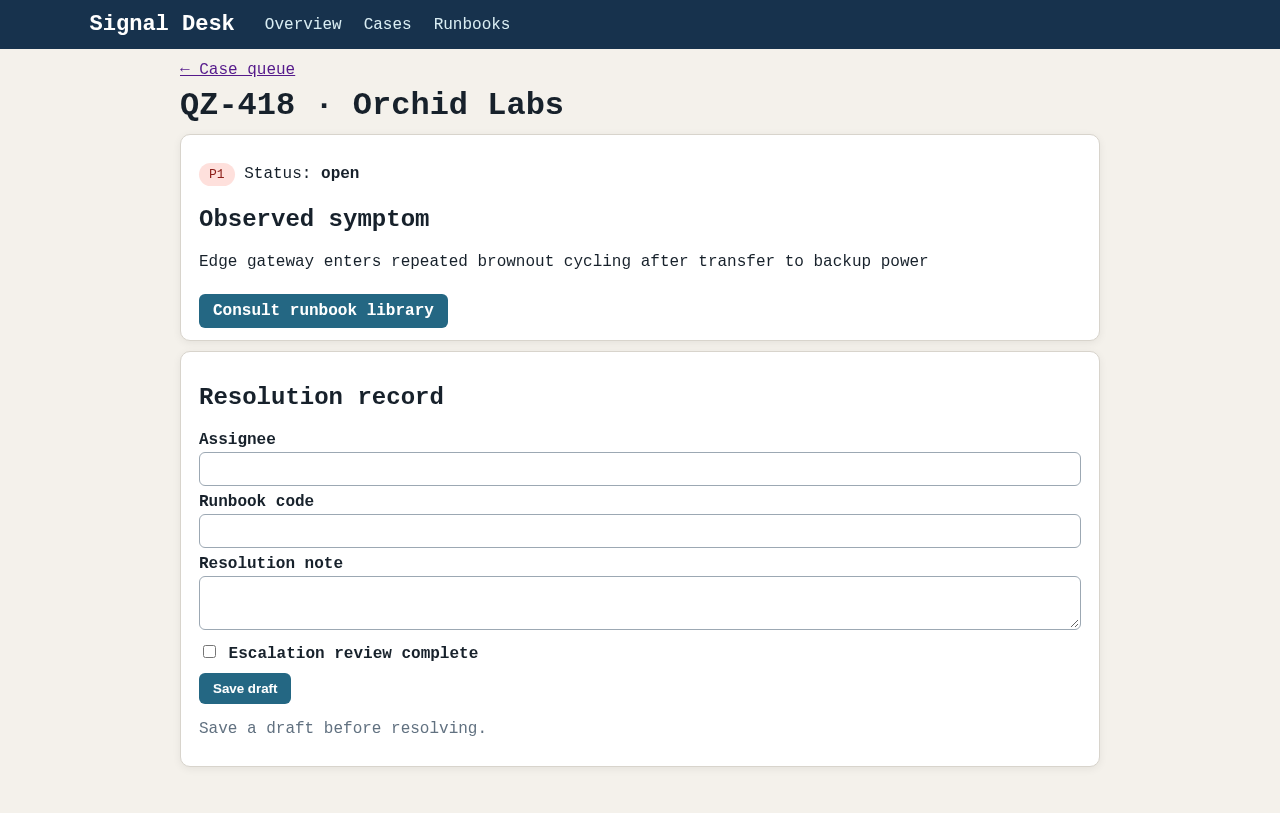}
\caption{\textbf{Incident page before completing the resolution record.} The symptom describes
repeated brownout cycling after transfer to backup power. The solver consults the runbook library
on a separate page before completing the resolution record.}
\label{fig:cross-domain-signal-initial}
\end{figure}

\paragraph{Grading.}
The verifier checks the saved resolution fields, the draft-to-resolve transition, and preservation
of unrelated cases.

\noindent\textbf{Incident workflow: linking information across pages}

\begin{figure}[htbp]
\centering
\includegraphics[width=0.91\linewidth,trim=170 200 170 50,clip]{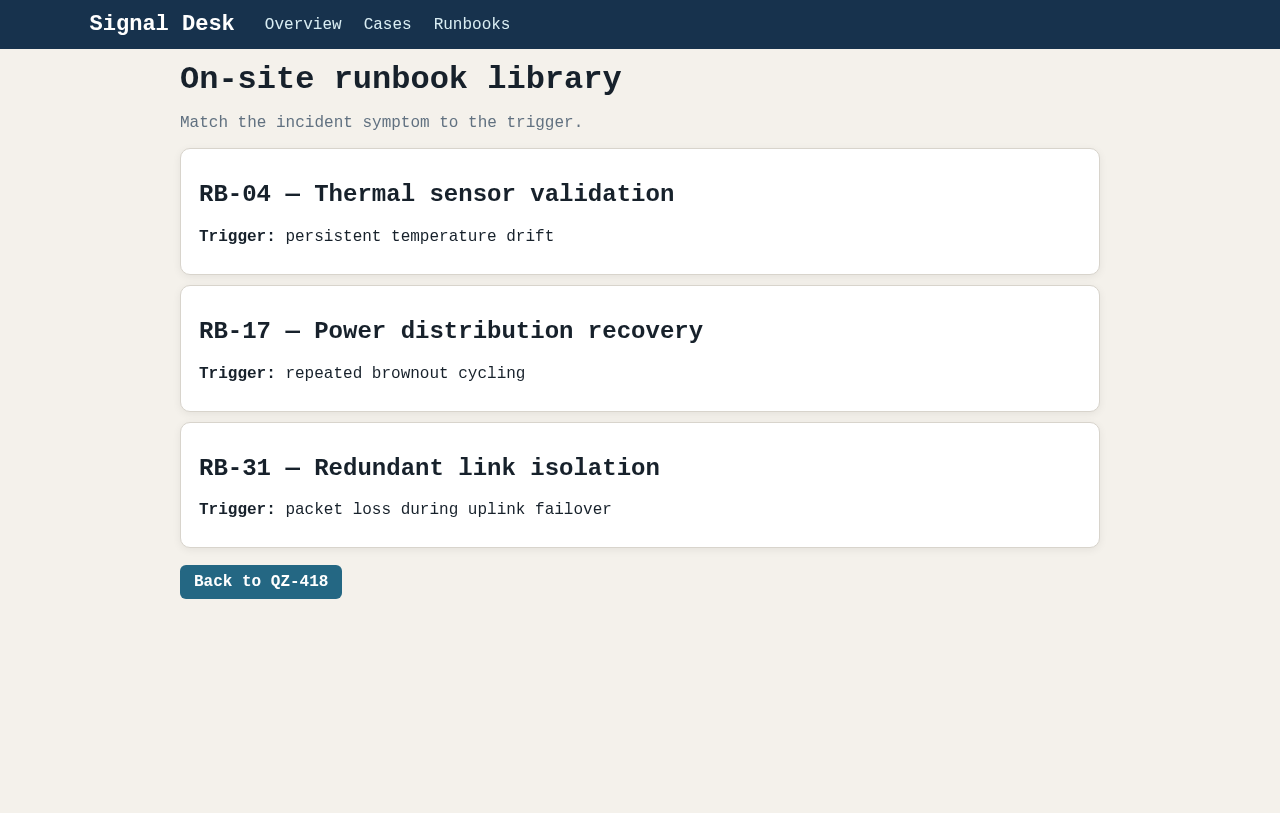}
\caption{\textbf{Runbook lookup.} The library maps repeated brownout cycling to RB-17, while the
other entries address thermal-sensor drift and uplink failover. The reference then returns to
QZ-418 to complete its resolution record.}
\label{fig:cross-domain-signal-runbook}
\end{figure}

\begin{figure}[htbp]
\centering
\includegraphics[width=0.91\linewidth,trim=170 45 170 50,clip]{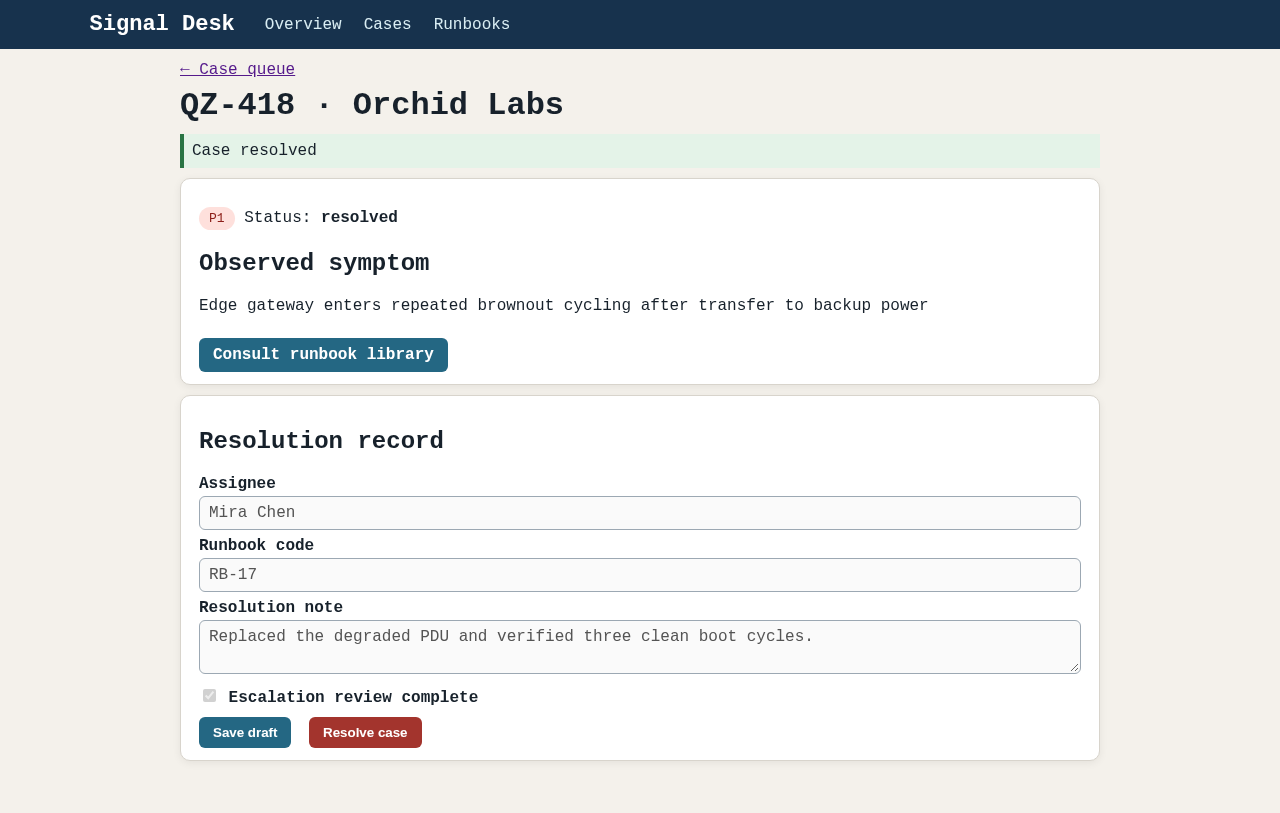}
\caption{\textbf{Resolved case after saving the draft.} The final page shows Mira Chen, RB-17,
the resolution note, the completed review, and resolved status.}
\label{fig:cross-domain-signal-final}
\end{figure}

\subsubsection{Computer Use: Editing and Saving a Decision Brief}
\label{app:cross-domain-desktop-writer}


The solver receives a decision brief already open in LibreOffice Writer. It must replace two
placeholders, make the status bold and the owner's name italic, center the heading, and save the
document while preserving the remaining text. The reference performs these edits through the
graphical interface.

\begin{figure}[htbp]
\centering
\textbf{(a) Initial document}\par\smallskip
\includegraphics[width=0.98\linewidth,trim=550 630 530 24,clip]{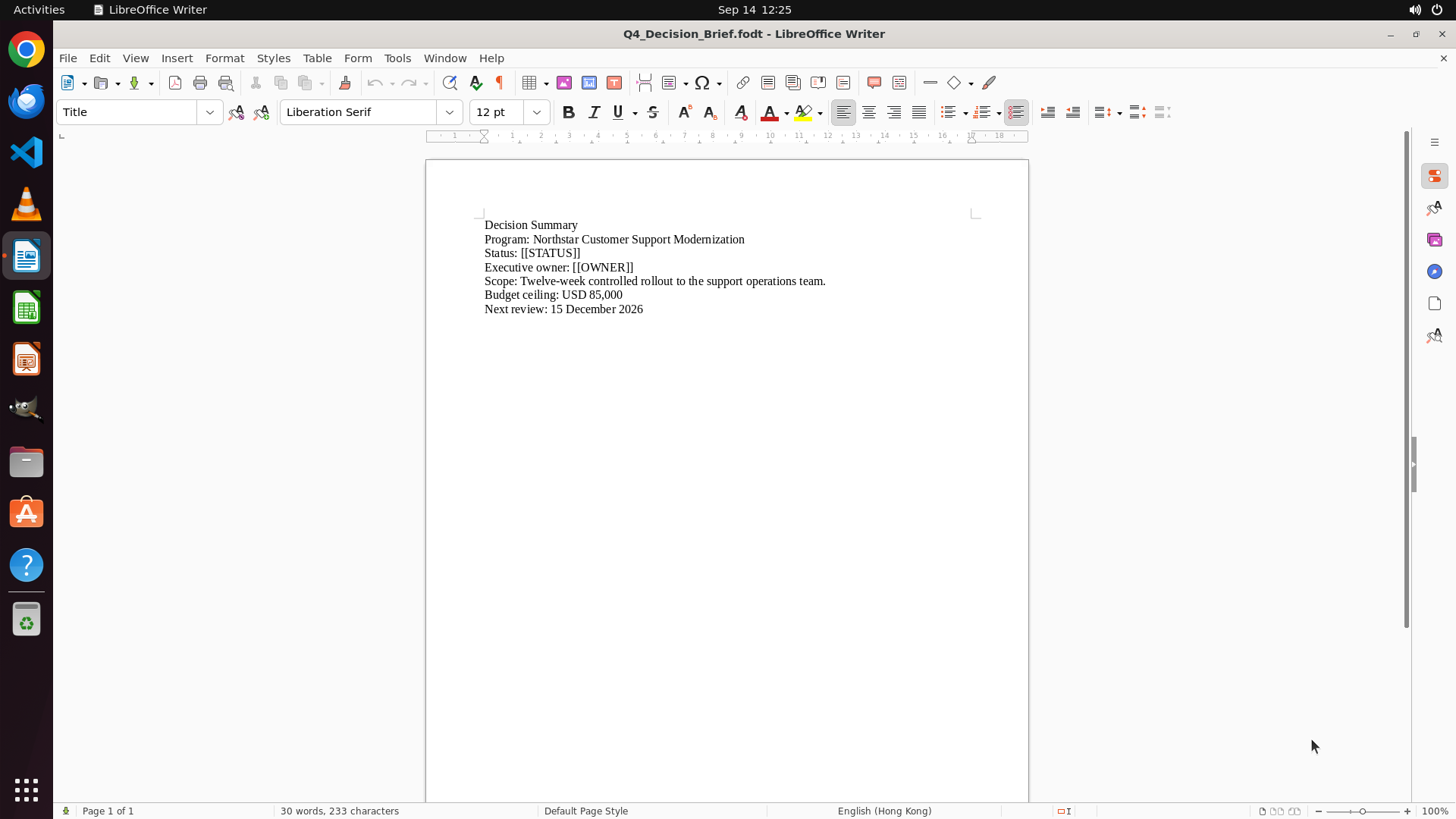}
\medskip
\textbf{(b) Saved document after the reference actions}\par\smallskip
\includegraphics[width=0.98\linewidth,trim=550 630 530 24,clip]{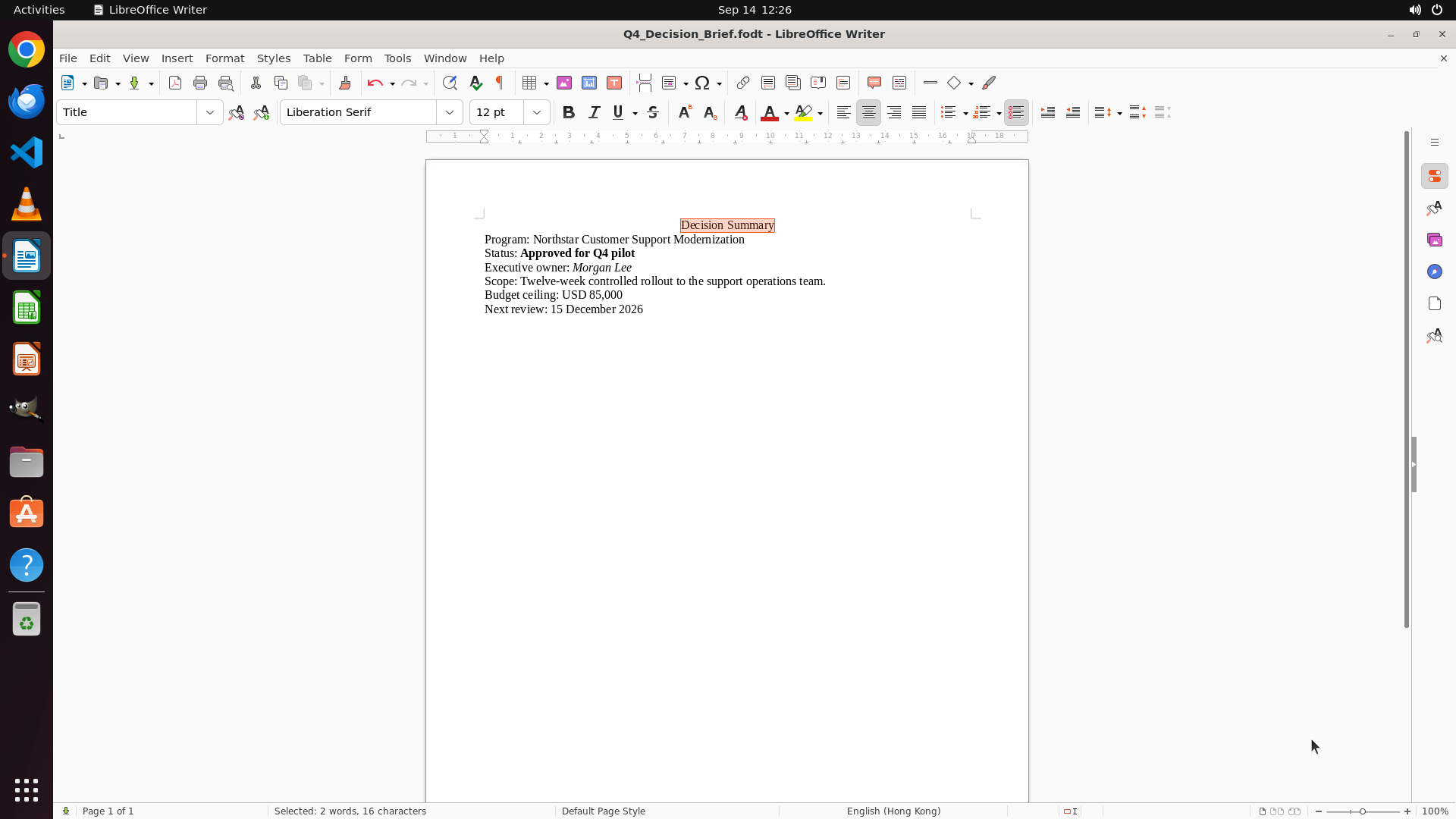}
\caption{\textbf{A GUI-only document transformation.} The final brief contains bold
``Approved for Q4 pilot,'' italic ``Morgan Lee,'' and a centered ``Decision Summary'' heading.
The panels show the document before and after editing. The heading remains selected in the final frame.}
\label{fig:cross-domain-writer}
\end{figure}

The verifier checks the saved document's text and formatting. Edits must be saved to satisfy the task.

\subsubsection{Computer Use: Completing a Spreadsheet Audit}
\label{app:cross-domain-desktop-calc}


The solver opens a reagent stock workbook in LibreOffice Calc, replaces a placeholder description,
fills variance and status formulas for ten inventory lines, and computes the number of discrepant
lines and total shortfall. It must preserve the issued stock data and save the original file format.
The reference locates cells through Find and enters formulas using the keyboard.

\begin{figure}[htbp]
\centering
\textbf{(a) Workbook before the calculation edits}\par\smallskip
\includegraphics[width=0.98\linewidth,trim=70 595 1195 170,clip]{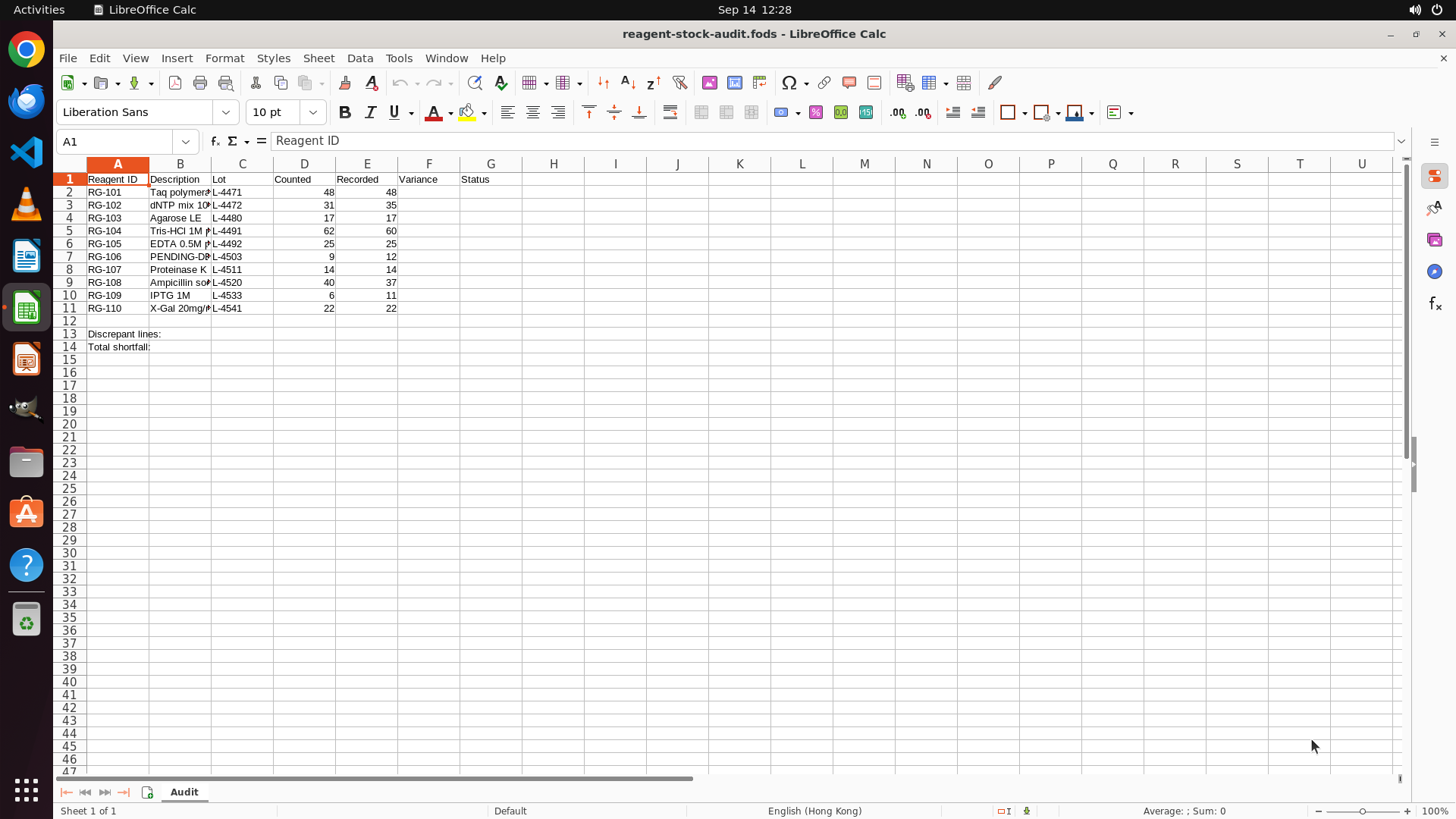}
\medskip
\textbf{(b) Saved formulas and summary values}\par\smallskip
\includegraphics[width=0.98\linewidth,trim=70 595 1195 170,clip]{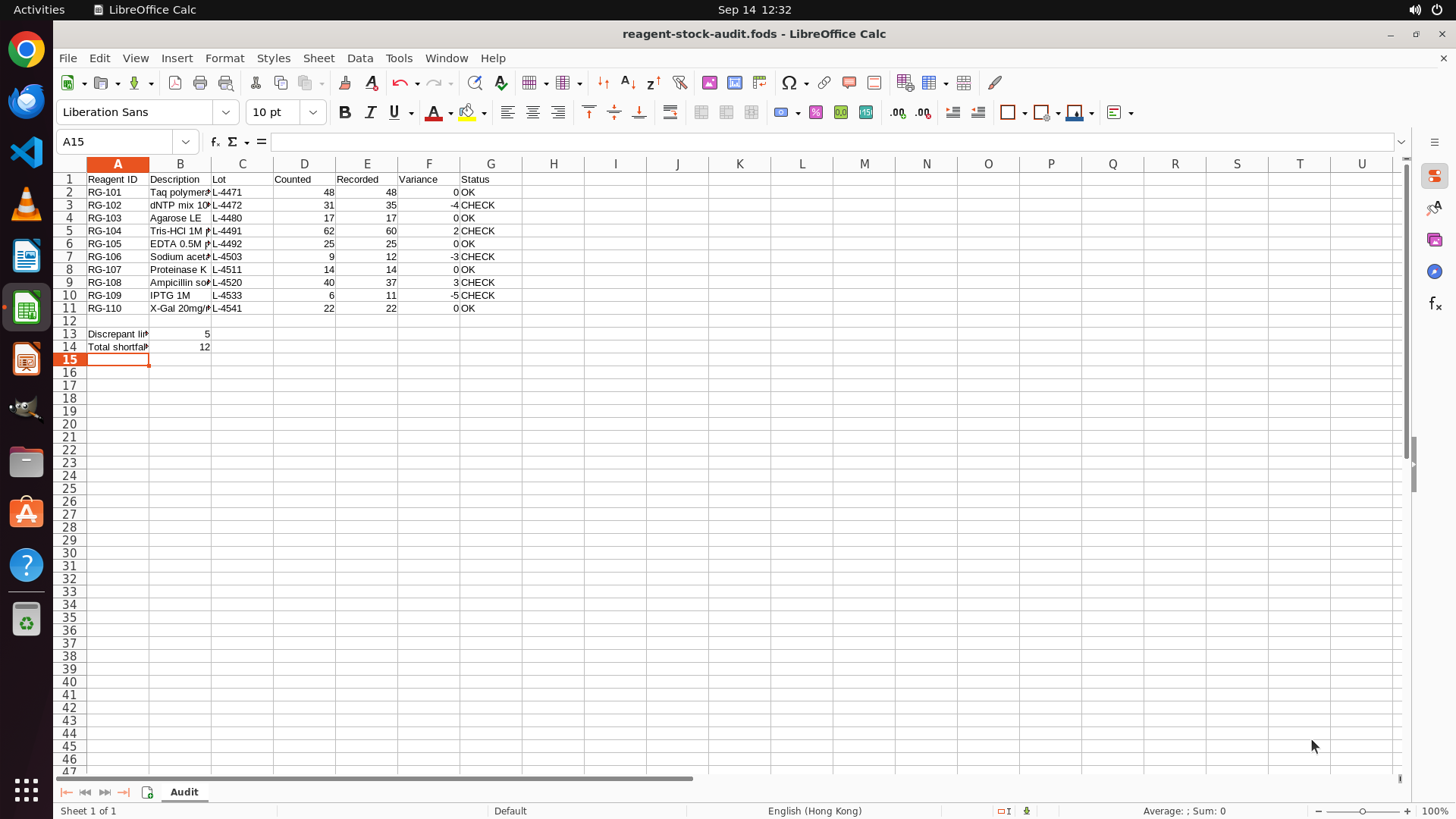}
\caption{\textbf{A GUI-only spreadsheet workflow.} The reference enters
\texttt{=D2-E2} and corresponding formulas down the variance column, derives \texttt{OK}/\texttt{CHECK}
status, and computes five discrepant lines and a total shortfall of 12.}
\label{fig:cross-domain-calc}
\end{figure}

The verifier checks the saved formulas and values, the corrected description, and preservation
of the original stock data.

\subsubsection{Professional Workflow: A Treasury Decision Pack}
\label{app:cross-domain-professional-treasury}


The task provides read-only assumptions, receivables, payments, and a proposed capital deposit for
fictional North Quay Components. The solver must deliver a formula-linked workbook with Base,
Stress, and Deferred 13-week cash forecasts, plus a CFO memo. The decision depends on delayed
collections, a bounded credit facility, and a deposit that remains due beyond the forecast horizon.

\begin{figure}[htbp]
\centering
\textbf{(a) Summary worksheet, first four columns}\par\smallskip
\includegraphics[width=0.98\linewidth,trim=70 730 930 200,clip]{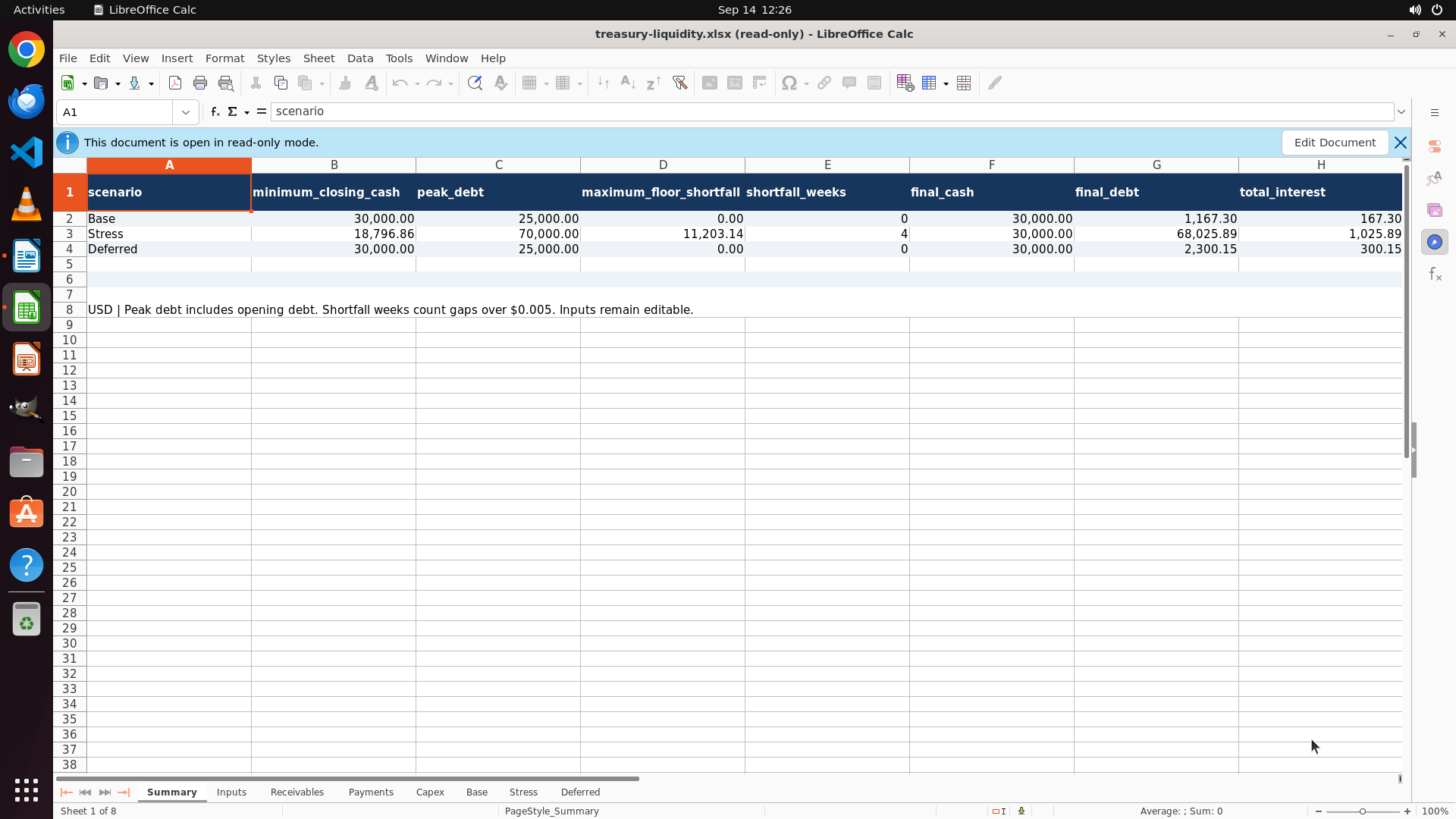}
\par\medskip
\textbf{(b) CFO memo from the same reference solution}\par\smallskip
\includegraphics[width=0.91\linewidth,page=1,trim=40 335 40 35,clip]{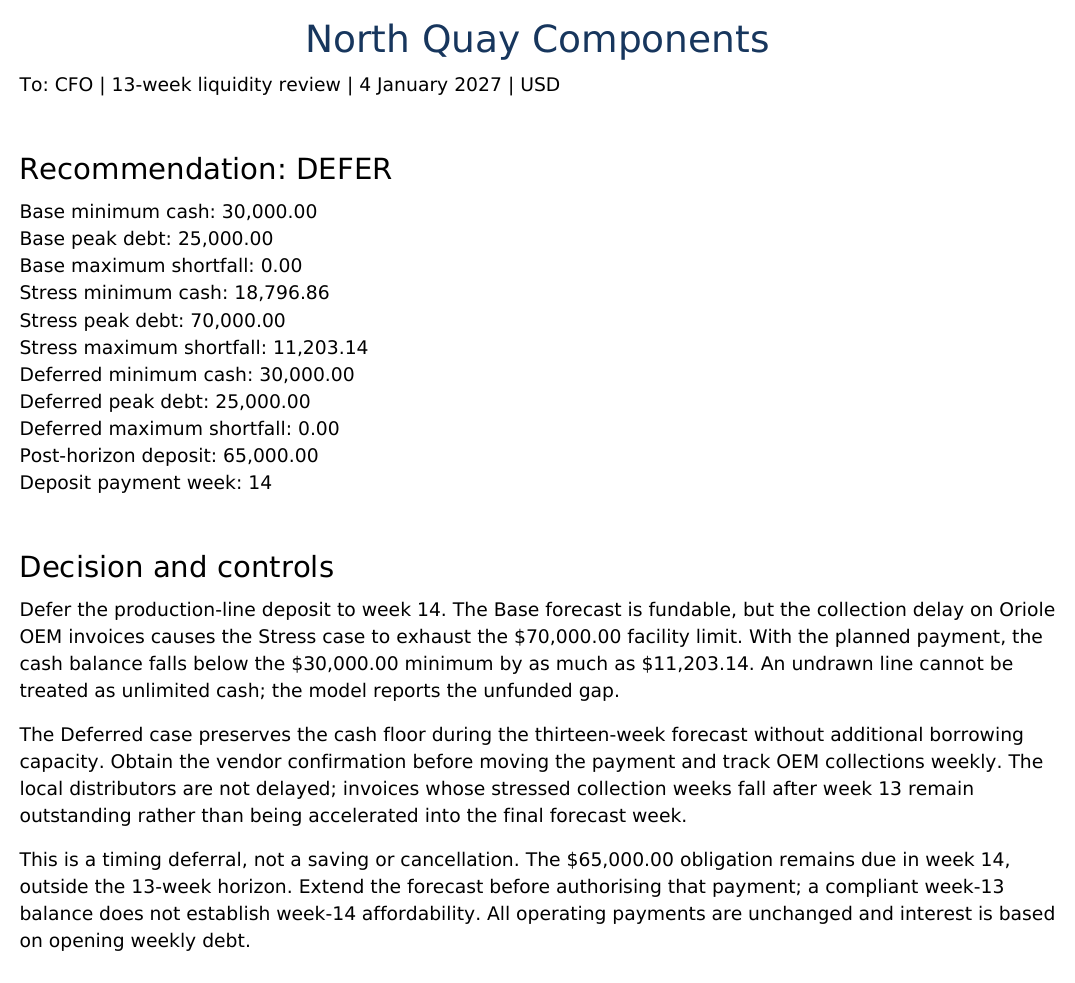}
\caption{\textbf{Two linked professional deliverables.} The Stress scenario falls to \$18,796.86
against a \$30,000 cash floor and reaches the \$70,000 facility limit. The memo recommends deferral
but retains the \$65,000 week-14 obligation.}
\label{fig:cross-domain-treasury}
\end{figure}

The task assigns 30\% to the Base forecast, 30\% to risk and mitigation, 25\% to a reusable model,
and 15\% to the CFO handoff. Numerical checks use a \$0.02 tolerance and include recalculation after
input changes.

\subsubsection{Professional Workflow: An Engineering Review Pack}
\label{app:cross-domain-professional-beam}


The task asks for a serviceability review of an 8\,m simply supported steel beam under an 11\,kN/m
uniform load. The solver evaluates three candidate sections, selects the lowest-cost one satisfying
both stress and deflection limits, and supplies a workbook, a machine-readable results table,
and a PDF memo. The task specifies the mechanics and acceptance criteria.

\begin{figure}[htbp]
\centering
\textbf{(a) Inputs worksheet, as saved by the reference solution}\par\smallskip
\includegraphics[width=0.44\linewidth,trim=70 710 1560 205,clip]{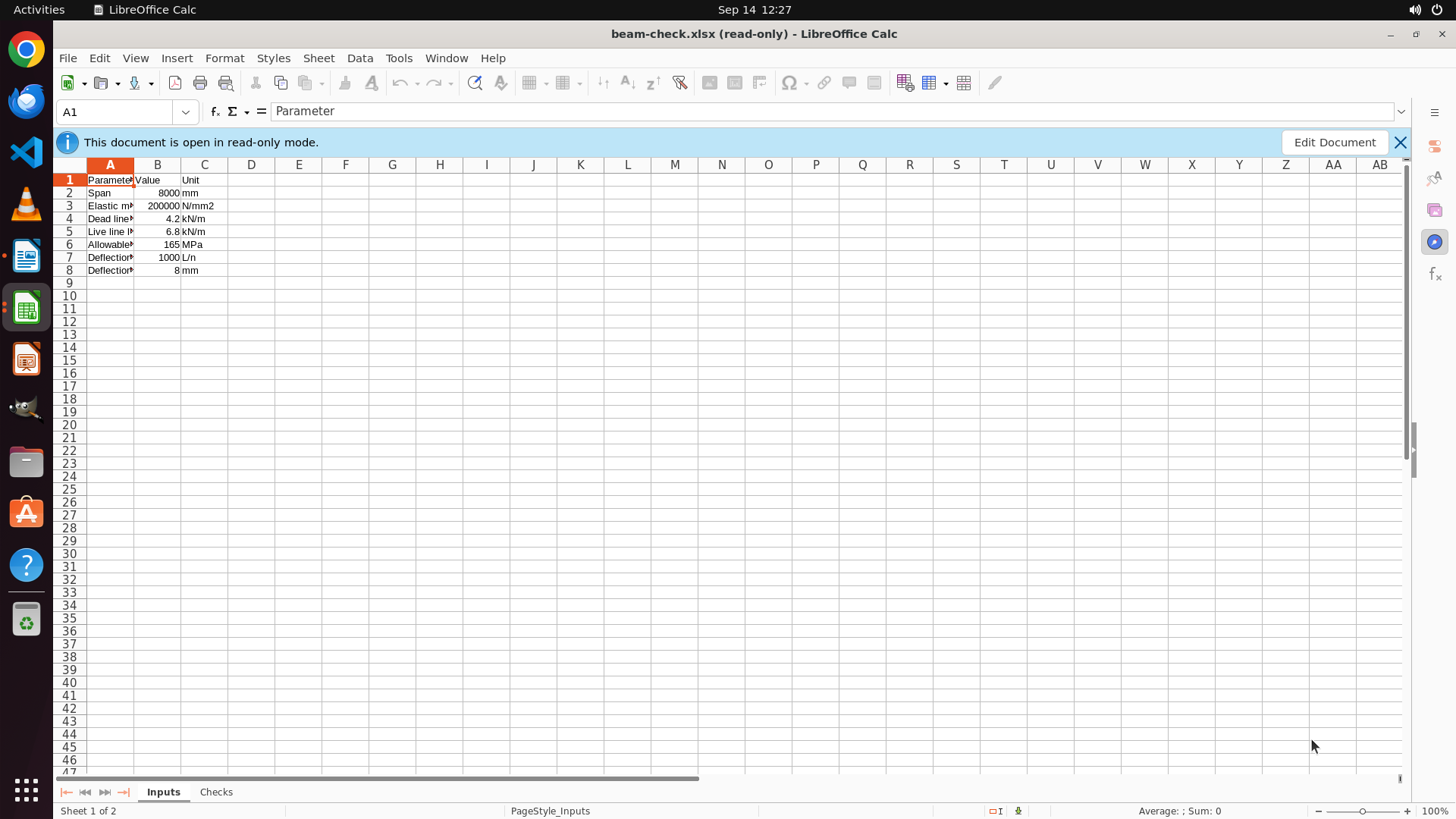}
\par\medskip
\textbf{(b) Engineering memo from the same solution}\par\smallskip
\includegraphics[width=0.91\linewidth,page=1,trim=45 373 45 45,clip]{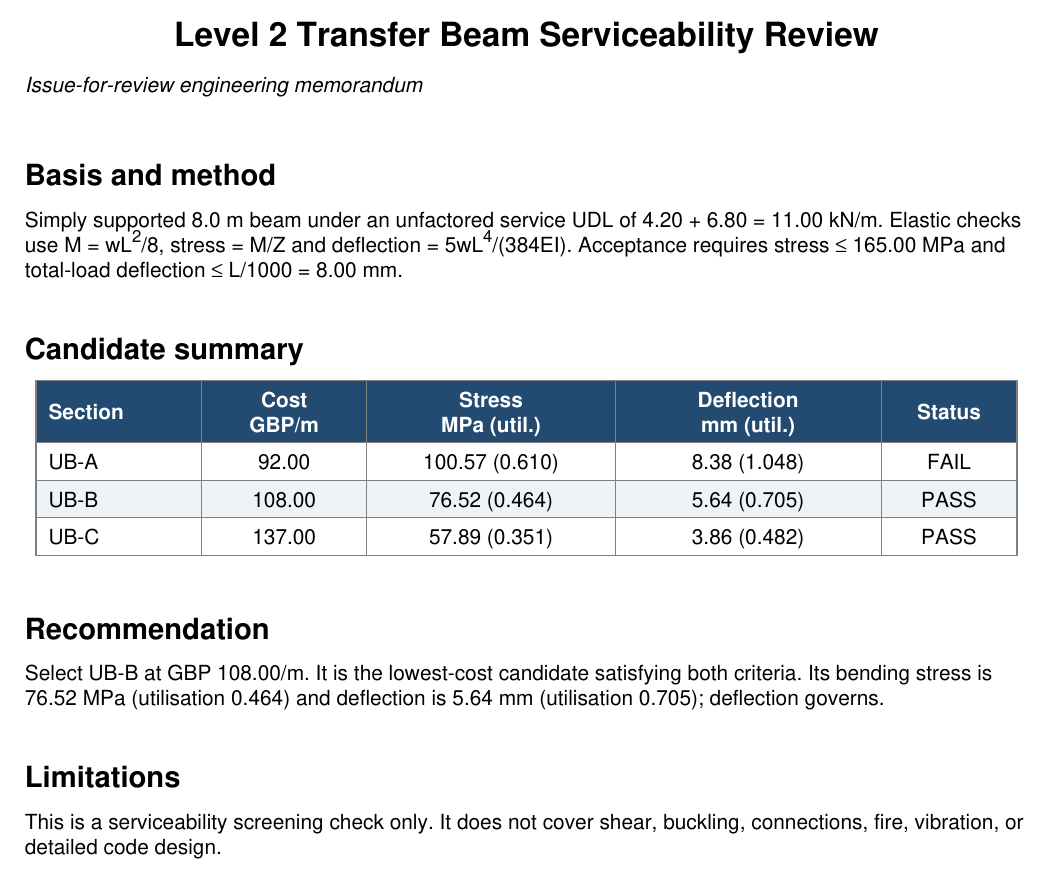}
\caption{\textbf{A calculation-to-recommendation workflow.} UB-A fails the deflection limit;
UB-B and UB-C pass. The reference selects UB-B at GBP~108/m, with stress 76.52\,MPa and deflection
5.64\,mm. The memo states the limited scope of the check. Narrow columns in the generated workbook
truncate some input labels.}
\label{fig:cross-domain-beam}
\end{figure}

Grading weights the workbook, results CSV, and memo at 35\%, 40\%, and 25\%, respectively.

\subsubsection{GPU Kernels: Workloads, Generated Code, and Execution Evidence}
\label{app:cross-domain-gpu}

\paragraph{Ragged residual RMSNorm (GPT-6-astra).}
The task requires a Triton kernel that combines an input and scaled residual,
normalizes each row over its valid prefix, applies a per-column affine transform, and
zeros the padded suffix. Valid lengths differ across rows; padding may contain NaNs. The verifier
checks float16 and float32 inputs, including empty and irregular-length rows. The generated
reference solution masks loads \emph{before} the reduction, accumulates in float32, and uses the
row's valid length rather than the padded width as the normalization denominator.

\begin{figure}[htbp]
\centering
\begin{minipage}{\linewidth}
\fontsize{8.5}{10.2}\selectfont
\begin{verbatim}
length = tl.load(Lengths + row)
valid = (columns < width) & (columns < length)
offsets = row * width + columns
values = tl.load(Input + offsets, mask=valid, other=0).to(tl.float32)
residual = tl.load(Residual + offsets, mask=valid, other=0).to(tl.float32)
combined = values + alpha * residual
mean_square = tl.sum(combined * combined, axis=0) / tl.maximum(length, 1)
inverse = tl.rsqrt(mean_square + eps)
weight = tl.load(Weight + columns, mask=columns < width, other=0)
bias = tl.load(Bias + columns, mask=columns < width, other=0)
output = tl.where(valid, (combined * inverse) * weight + bias, 0.0)
tl.store(Output + offsets, output, mask=columns < width)
\end{verbatim}
\end{minipage}
\caption{\textbf{Excerpt from the generated ragged RMSNorm reference kernel.} Masked loads prevent
NaN padding from entering the reduction; the final mask forces padded outputs to zero.}
\label{fig:cross-domain-gpu-code}
\end{figure}

\paragraph{Fused gated RMSNorm (GPT-5.6-sol, three rounds).}
The task specifies
$z=r+\operatorname{SiLU}(g)\odot v$ followed by row-wise RMS normalization and a weight vector.
Its public PyTorch implementation defines the target behavior, and the solver must implement
\texttt{ModelNew.forward} for float16 inputs at widths 256, 512, and 1024. The generated reference
uses one Triton kernel to fuse gating, residual addition, reduction, and scaling. Private checks
vary inputs and random seeds before measuring performance.

\begin{table}[htbp]
\centering
\caption{\textbf{Reference-kernel execution times on an NVIDIA T4.} Values are mean milliseconds
over 20 samples per implementation and shape after warmup. The ratio is kernel time divided by
the task's eager PyTorch baseline time.}
\label{tab:cross-domain-gpu-timings}
\footnotesize
\setlength{\tabcolsep}{6pt}
\renewcommand{\arraystretch}{1.15}
\begin{tabular}{@{}llrrr@{}}
\toprule
\rowcolor{TableHeader}
Task & Shape / dtype & Eager (ms) & Kernel (ms) & Ratio \\
\midrule
Ragged RMSNorm & $4096\!\times\!1025$ / f32 & 1.6622 & 0.3274 & 0.1970 \\
 & $2048\!\times\!2049$ / f16 & 1.8906 & 0.2065 & 0.1092 \\
 & $8192\!\times\!257$ / f32 & 0.7781 & 0.1558 & 0.2002 \\
\addlinespace[4pt]
Gated RMSNorm & $512\!\times\!256$ / f16 & 0.1492 & 0.0503 & 0.3369 \\
 & $1024\!\times\!512$ / f16 & 0.1901 & 0.0535 & 0.2817 \\
 & $2048\!\times\!1024$ / f16 & 0.6586 & 0.1094 & 0.1661 \\
\bottomrule
\end{tabular}
\end{table}

\paragraph{Correctness and performance.}
The ragged reference passes 96 correctness trials and runs within 70\% of the baseline time on
each measured shape. The gated reference meets an aggregate limit of 60\%. The ragged task uses
CUDA-event timing; the gated task uses wall-clock timing with CUDA synchronization.

\section{Proposer Setup}
\label{app:proposer-setup}

\subsection{Workspace Structure}

\begin{figure}[htbp]
\centering
\includegraphics[width=\linewidth]{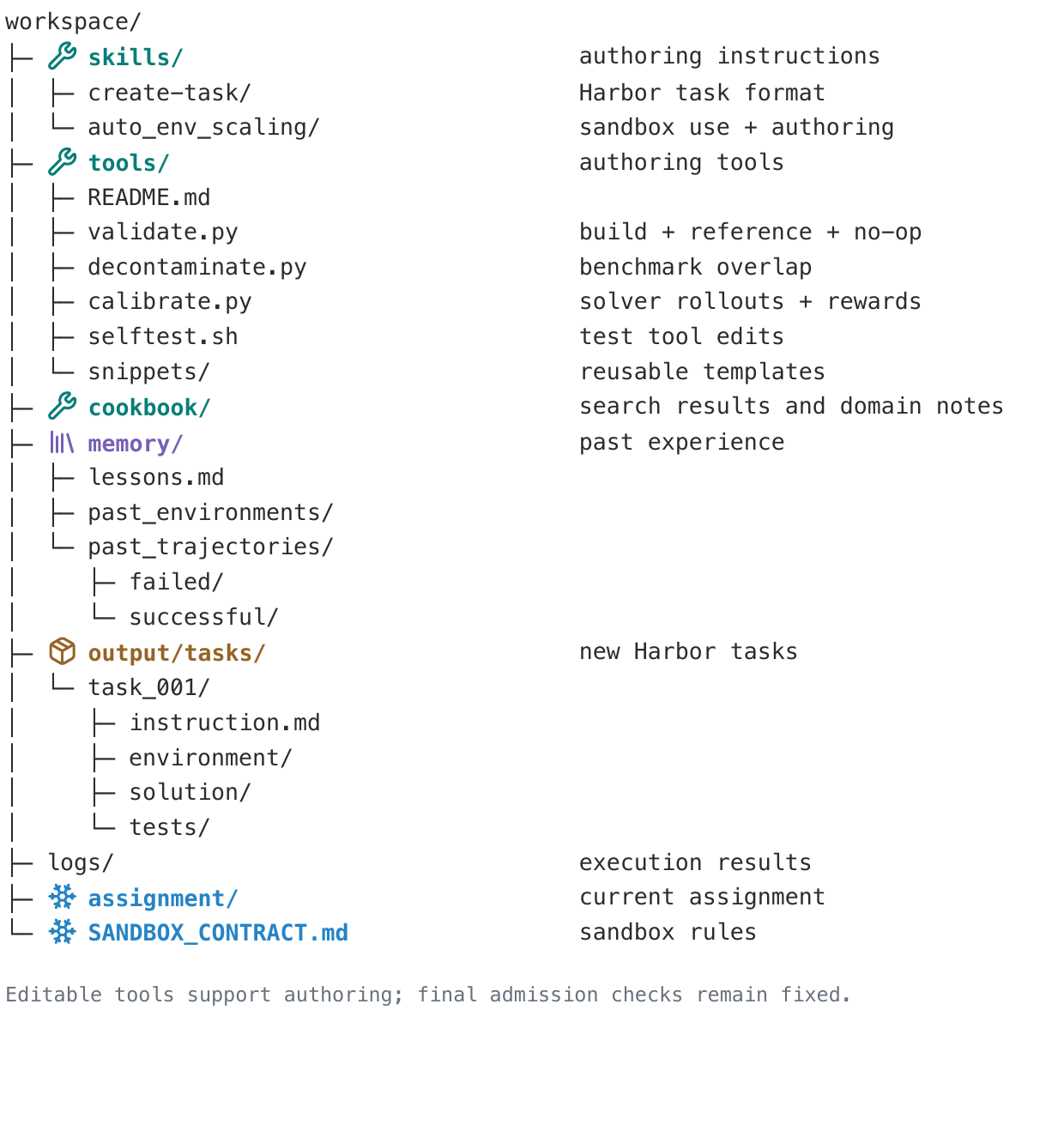}
\caption{\textbf{Proposer workspace.} A schematic layout of generation skills, tools, memory,
and generated Harbor tasks. Colors distinguish editable components, retained experience,
outputs, and fixed inputs. File names are illustrative.}
\label{fig:proposer-workspace}
\end{figure}

The proposer uses a filesystem workspace containing generation instructions, tools, and
experience from earlier rounds. Figure~\ref{fig:proposer-workspace} shows the workspace layout.
Tools and retained experience can change across rounds; the assignment and sandbox rules
remain fixed during generation. Final admission checks are enforced outside the editable workspace.

\subsection{Sandbox Resources and Coding Agents}
\label{app:sandbox-resources}

Table~\ref{tab:sandbox-resources} lists the sandbox resources for the proposer and the solver.
Each proposer runs in a coding agent: Claude Code for Claude Opus~5, Codex for the GPT models,
and the Pi coding agent for all other models. All solvers run in the Pi coding agent.

\begin{table}[htbp]
\centering
\small
\caption{\textbf{Sandbox resources.}}
\label{tab:sandbox-resources}
\begin{tabular}{lcc}
\toprule
 & Proposer & Solver \\
\midrule
CPUs & 16 & 4 \\
Memory & 32\,GiB & 8\,GiB \\
Time budget per episode & 2 hours & -- \\
Turn limit & none & -- \\
Coding agent & Claude Code, Codex, or Pi & Pi \\
\bottomrule
\end{tabular}
\end{table}

\subsection{Image and Runtime Configuration}
\begin{figure}[htbp]
\centering
\includegraphics[width=\linewidth]{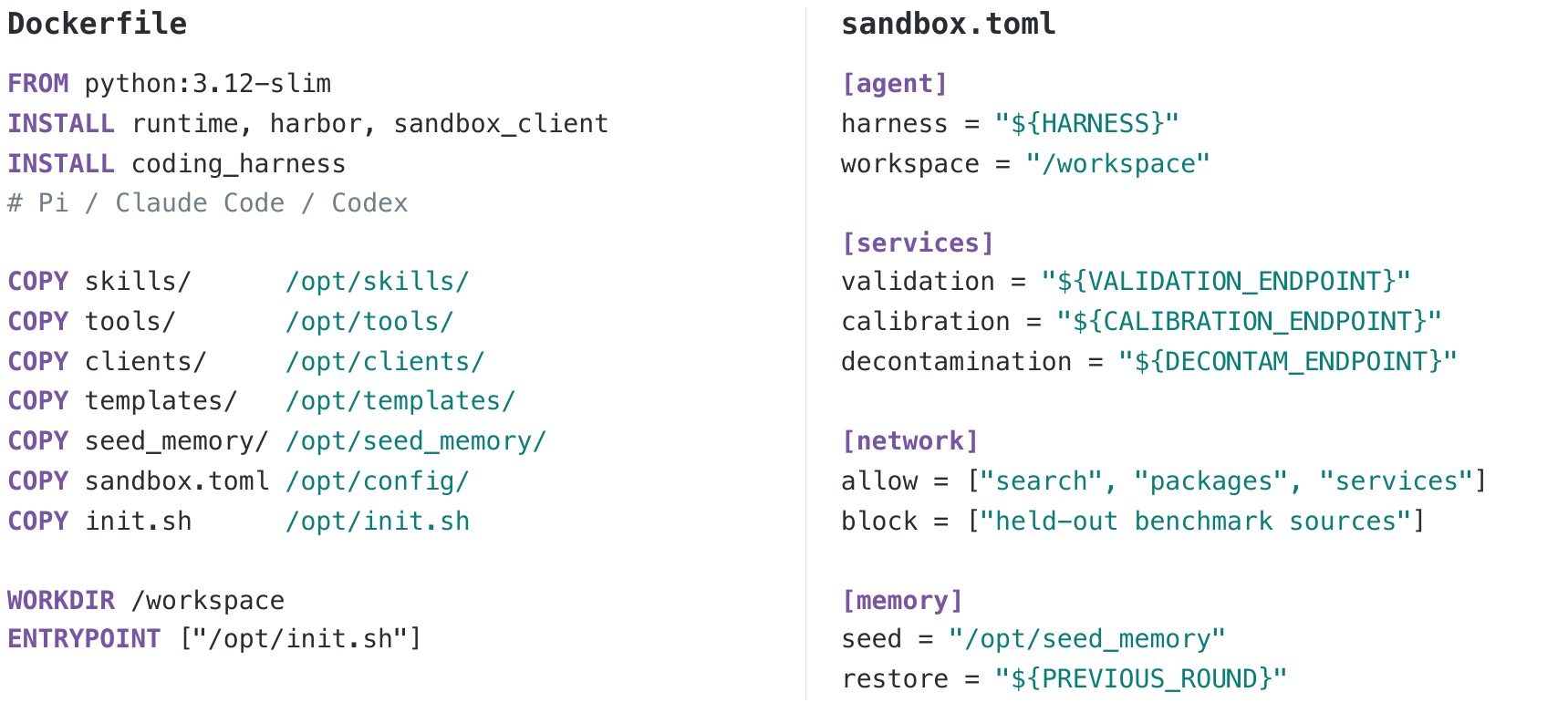}
\caption{\textbf{Image construction and runtime configuration.} Simplified Dockerfile and TOML
listings. Commands, paths, and configuration keys are shortened for presentation.}
\label{fig:proposer-image-build}
\end{figure}

The image packages the coding harness, skills, tools, and seed experience. At launch,
configuration supplies service endpoints and retained memory (Figure~\ref{fig:proposer-image-build}).
Credentials are supplied separately; network restrictions must be enforced by the sandbox
infrastructure.

\section{Proposer Skills}
\label{app:proposer-skills}

The proposer's prompt begins with two skills. The first is Harbor's official \texttt{create-task}
skill, which covers the mechanics of building a valid Harbor task. The second is our
\texttt{auto\_env\_scaling} skill, which adds what kind of task to build and how to validate it
before submission.

\subsection{Harbor \texttt{create-task} Skill}
\label{app:skill-create-task}

We reproduce the skill verbatim from the Harbor revision used in our
runs, commit
\href{https://github.com/harbor-framework/harbor/blob/f5e9d0b71ac4493a4f0620653e2913aee7fc0767/skills/create-task/SKILL.md}{\texttt{f5e9d0b}}
of \texttt{harbor-framework/harbor}.

\begin{lstlisting}[style=skill]
---
name: create-task
description: Create a new Harbor task for evaluating agents. Use when the user wants to 
  scaffold, build, or design a new task, benchmark problem, or eval. Guides through 
  instruction writing, environment setup, verifier design (pytest vs Reward Kit vs 
  custom), and solution scripting.
---

Guide the user through creating a new Harbor task end-to-end. Don't just dump commands — 
walk them through each decision, especially around the verifier (which is usually the 
hardest part).

## Step 1: Scaffold the task

```bash
harbor task init "<org>/<task-name>"
```

Useful flags:
- `--description "..."`
- `--author "Jane Doe <jane@example.com>"` (repeat for multiple authors)
- `--no-pytest` — skip the pytest test template (use if planning Reward Kit or custom verifier)
- `--no-solution` — skip solution/ directory
- `--metadata-template path.toml` — pre-populate task.toml

Produces:
```
<task-name>/
├── instruction.md         # Task prompt for the agent
├── task.toml              # Config and metadata
├── environment/Dockerfile # Container definition
├── solution/solve.sh      # Reference solution (optional)
└── tests/test.sh          # Verifier script
```

If the user wants a **multi-step task** (ordered steps with per-step
instructions, tests, and early stopping against a shared container), scaffold
the single-step layout first, then convert to the `steps/` layout described in
the *Multi-step tasks* section below.

## Step 2: Write instruction.md

This is the prompt the agent receives. Help the user write it clearly:

- **State the goal concretely** — what file to create, what behavior to produce
- **Specify expected outputs** — paths, formats, content
- **Include constraints** — language, tools, approach
- **Don't leak the tests** — describe what "done" looks like, not how you'll check it

Example (from the ssh-key-pair tutorial):
```markdown
# SSH Key Pair Generation

Generate an SSH key pair in the files `~/.ssh/id_rsa` and `~/.ssh/id_rsa.pub`.

Don't make them password protected.
```

## Step 3: Build the environment

Edit `environment/Dockerfile` to install dependencies the task needs. The agent works 
inside this container.

```dockerfile
FROM ubuntu:24.04
WORKDIR /app

# Install what the task requires — NOT the solution
RUN apt-get update && apt-get install -y openssh-client && rm -rf /var/lib/apt/lists/*
```

For multi-container setups, use `environment/docker-compose.yaml` instead (note: most 
cloud sandbox providers only support Dockerfile).

**Test the environment interactively** before writing the solution or tests:
```bash
harbor task start-env -p "<task-path>" -e docker -a -i
```

This is usually where task authors realize something is missing from the Dockerfile.

## Step 4: Decide how to verify

**This is the most important decision.** Ask the user: *"How do you want to grade this 
task?"* Then help them pick:

Also ask: *"Should the verifier run in the same environment as the agent, or in a
separate verifier environment?"*

- Use the default shared environment when tests need to inspect the agent's full
  workspace, installed tools, or services.
- Use a separate verifier environment when grading code, dependencies, API keys,
  or OS requirements should stay hidden from the agent, or when verification
  should run from a clean image.

For a separate verifier container, `tests/` is the verifier image build context
and the image must provide `/tests/test.sh` (Linux) or `/tests/test.bat`
(Windows). Harbor copies `/logs/artifacts` and configured artifacts into the
verifier environment, not the agent's whole workspace.

```toml
[verifier]
environment_mode = "separate"

[verifier.environment]
docker_image = "ubuntu:24.04"
```

### Option A: Reward Kit (recommended for most cases)

Use when the verifier has multiple criteria, needs partial credit, uses an LLM/agent 
judge, or would benefit from composable reusable checks. See the `rewardkit` skill.

Good fit signals:
- Multiple things to check (file exists + content correct + command works)
- Subjective quality dimensions (readability, correctness of prose)
- Want partial credit rather than pass/fail
- Want to compose built-ins like `file_contains`, `command_succeeds`, `json_key_equals`

`tests/test.sh`:
```bash
#!/bin/bash
uvx --from 'harbor-rewardkit==0.1.*' rewardkit /tests
```

Note: the package is named `harbor-rewardkit` but the executable is `rewardkit`,
hence `--from 'harbor-rewardkit==0.1.*' rewardkit`. Running
`uvx harbor-rewardkit` directly will fail.

Then add `tests/checks.py` and/or `tests/judge.toml`. Invoke the `rewardkit` skill to 
design the criteria.

### Option B: pytest (good for deterministic unit-style checks)

Use when the verification is straightforward assertion-style Python. Default template if 
`--no-pytest` wasn't passed.

`tests/test.sh`:
```bash
#!/bin/bash
apt-get update && apt-get install -y curl
curl -LsSf https://astral.sh/uv/0.9.7/install.sh | sh
source $HOME/.local/bin/env

uvx --with pytest==8.4.1 pytest /tests/test_outputs.py

if [ $? -eq 0 ]; then
  echo 1 > /logs/verifier/reward.txt
else
  echo 0 > /logs/verifier/reward.txt
fi
```

Example `tests/test_outputs.py`:
```python
from pathlib import Path

def test_file_exists():
    assert (Path.home() / ".ssh" / "id_rsa").exists()
```

### Option C: Custom shell

For simple single-command checks (e.g. a binary pass/fail from one command):
```bash
#!/bin/bash
if diff -q /app/output.txt /tests/expected.txt; then
  echo 1 > /logs/verifier/reward.txt
else
  echo 0 > /logs/verifier/reward.txt
fi
```

### Reward file format (all options)

- `/logs/verifier/reward.txt` — single number (usually `0` or `1`)
- `/logs/verifier/reward.json` — `{"accuracy": 0.95, "runtime_sec": 1.2}` for multiple metrics

**Always use absolute paths in `test.sh`.**

## Step 5: Write the solution

Write `solution/solve.sh` — a script that actually solves the task. The Oracle agent runs 
this to sanity-check that the task is solvable and the tests pass on a correct solution.

```bash
#!/bin/bash
ssh-keygen -t rsa -f ~/.ssh/id_rsa -N ""
```

Make it executable: `chmod +x solution/solve.sh`.

## Step 6: Configure task.toml

Walk through the important fields:

```toml
[task]
name = "<org>/<task-name>"
version = "1.0.0"
description = "One-line description"
keywords = ["jax", "mnist", "rewardkit"]  # always populate — used for search/filtering

[metadata]
difficulty = "easy" | "medium" | "hard"
category = "programming" | "machine-learning" | "gpu" | ...
tags = ["..."]

[agent]
timeout_sec = 120.0       # How long the agent has

[verifier]
timeout_sec = 600.0       # How long tests have

[environment]
network_mode = "public"   # Baseline at env start (defaults to public)
cpus = 1                  # CPU cores
memory_mb = 2048          # RAM in MB
storage_mb = 10240        # Disk in MB
```

**Always populate `keywords`.** Pick 3–8 lowercase tokens covering the domain
(language/framework/benchmark family), the verifier style (`rewardkit`,
`judge-grading`, `pytest`), and any notable hardware (`gpu`). They're surfaced in
`harbor datasets list` and registry search.

### Network policy

Network access has three layers:

1. **Baselines** — set when an environment starts, restored between phases
2. **Phase overrides** — optional; only during `agent.run()` or `verify()`
3. **Run-time merges** — `--allow-environment-host`, `--allow-agent-host` on `harbor run`

| Field | Layer | When applied |
| --- | --- | --- |
| `[environment].network_mode` | Baseline | Agent env start; shared verifier uses this too |
| `[verifier.environment].network_mode` | Baseline | Separate verifier env start |
| `[agent].network_mode`, `[steps.agent].network_mode` | Override | During matching `agent.run()` |
| `[verifier].network_mode`, `[steps.verifier].network_mode` | Override | During matching `verify()` |
| `--allow-environment-host` | Run-time | Merged into `environment.extra_allowed_hosts` → `[environment]` baseline |
| `--allow-agent-host` | Run-time | Merged into `agent.extra_allowed_hosts` → agent phase allowlist |

Modes: `public`, `no-network`, or `allowlist` with `allowed_hosts = ["pypi.org"]`
(exact hostnames, IPv4/IPv6 address literals or CIDR ranges, or leading wildcard hostnames, when supported by the selected environment; not URLs,
ports, or paths). Omitting `[environment].network_mode` defaults to `public`.

`[agent]` / `[verifier]` are **optional phase overrides** — only applied when set
**and** different from the phase baseline. Matching the baseline is a no-op.

**Shared verifier** (default): verifier runs in the agent container; baseline is
`[environment]`. **Separate verifier**: baseline is `[verifier.environment]` if
set, else a copy of `[environment]`.

```toml
# Agent starts offline; agent phase opens network; verifier stays offline
[environment]
network_mode = "no-network"

[agent]
network_mode = "public"

[verifier]
network_mode = "no-network"
```

If a phase override differs from its baseline, the environment provider must
support `dynamic_network_policy` (E2B does; plain Docker does not). Prefer
`environment_mode = "separate"` when agent and verifier need different baselines
without runtime switching:

```toml
[environment]
network_mode = "no-network"

[verifier]
environment_mode = "separate"

[verifier.environment]
network_mode = "public"   # Verifier baseline — not a phase override
```

Run-time host flags for eval jobs without editing `task.toml`:

```bash
harbor run -p "<task-path>" -a oracle \
  --allow-environment-host pypi.org \   # env start (Dockerfile apt/curl)
  --allow-agent-host files.pythonhosted.org   # agent.run() only
```

On a `public` baseline, run-time host flags emit a warning and are ignored.

Examples: `examples/tasks/network-policy-matrix/`. Full reference:
`docs/content/docs/tasks/index.mdx` (Network policy section).

For Reward Kit judges needing API keys:
```toml
[verifier.env]
ANTHROPIC_API_KEY = "${ANTHROPIC_API_KEY}"
```

## Step 7: Verify with the Oracle agent

```bash
harbor run -p "<task-path>" -a oracle
```

Oracle runs `solution/solve.sh` and then the verifier. Reward should be `1.0`. If it's 
not, debug in this order:
1. Does `solve.sh` actually solve it? (`start-env -a -i` and run it manually)
2. Does the verifier correctly detect success? (check `/logs/verifier/` output)
3. Are paths correct? (absolute vs relative)
4. Are dependencies installed in the Dockerfile?

## Step 8: Test with a real agent (optional)

```bash
harbor run -p "<task-path>" -a terminus-2 -m anthropic/claude-sonnet-4-6
```

If the task is too easy (every model 1.0) or impossible (every model 0.0), consider 
adjusting difficulty.

## Step 9: Update README.md (always the final step)

`harbor task init` leaves `README.md` as a stub. Before wrapping up, populate it so
future humans (and agents) can understand the task without reading every file.
Include:

- **What the agent does** — one paragraph, link to `instruction.md`.
- **Environment** — base image, key installed packages, cached data, hardware
  (GPU/CPU/RAM), agent timeout.
- **Verifier** — for Reward Kit tasks, a table of reward dimensions with type
  (programmatic / LLM judge / agent judge) and what each measures; how they're
  aggregated.
- **Layout** — a tree of the task directory with one-line annotations.
- **Running** — the concrete `harbor run` commands (Oracle + real agent), with the
  right provider flag if the task needs a GPU.

Treat this as docs, not marketing — the reader wants to know *what they'd need to
change* to modify the task.

## Multi-step tasks

Use when the work splits into ordered phases that should be scored separately,
when you want early stopping between phases, or when you're testing an agent's
ability to build on its own prior work. Steps share one container; files
persist across steps.

### Directory layout

Replace the task-root `instruction.md`, `tests/`, and `solution/` with a
`steps/` directory containing one sub-directory per step:

```
<task-name>/
├── task.toml
├── environment/Dockerfile       # Built once, shared across all steps
├── steps/
│   ├── scaffold/
│   │   ├── instruction.md       # Prompt for this step
│   │   ├── workdir/             # Uploaded to WORKDIR before the agent runs
│   │   │   └── setup.sh         # Optional pre-agent hook (reserved filename)
│   │   ├── tests/test.sh        # Per-step verifier
│   │   └── solution/solve.sh    # Per-step Oracle solution (optional)
│   ├── implement/
│   │   └── ...
│   └── document/
│       └── ...
└── tests/                       # Optional shared helpers + fallback test.sh
```

Task-level `tests/` is uploaded to `/tests` for each step's verification, then
the step's own `tests/` is layered on top (same-name files win). Use this for
shared helpers.

`steps/{name}/workdir/setup.sh` is a **reserved filename**: if present, it runs
after the `workdir/` upload and before the agent, as the step's agent user,
with cwd = WORKDIR. Non-zero exit aborts the step and the trial. Have it
`rm -- "$0"` on its last line if the agent shouldn't see it.

### task.toml

```toml
schema_version = "1.4"

[task]
name = "<org>/<task-name>"
version = "1.0.0"

# How per-step rewards roll up into the trial-level verifier_result.
# "mean" (default): per-key mean across steps that produced a result.
# "final": the last step's verifier_result verbatim.
multi_step_reward_strategy = "mean"

[[steps]]
name = "scaffold"              # Must match the directory under steps/
min_reward = 1.0               # Abort trial if this step's reward < 1.0
[steps.agent]
timeout_sec = 60.0             # Overrides task-level [agent].timeout_sec
[steps.verifier]
timeout_sec = 30.0

[[steps]]
name = "implement"
# Dict form gates on specific keys from a multi-dim reward:
min_reward = { correctness = 0.8, style = 0.5 }
[steps.agent]
timeout_sec = 120.0
[steps.verifier]
timeout_sec = 30.0

[[steps]]
name = "document"
[steps.agent]
timeout_sec = 60.0
[steps.verifier]
timeout_sec = 30.0
```

Per-step overrides available: `agent.timeout_sec`, `agent.user`,
`agent.network_mode`, `verifier.timeout_sec`, `verifier.env`, `verifier.user`,
`verifier.network_mode`, `verifier.environment_mode`, `verifier.environment`,
`steps.verifier.environment.network_mode`, `healthcheck.*`, `artifacts`. Unset
fields fall back to the task-level values.

### Choosing a reward strategy

- **`"mean"`** — aggregate signal across all steps; good for continuous
  progress rewards.
- **`"final"`** — last step's verifier_result is the trial reward. Right when
  the final step is an end-to-end check whose dict already represents the full
  task. Caveat: if `min_reward` triggers an early abort, `"final"` uses the
  *aborted* step's result, not the intended final step.

### Artifacts

Step-level `artifacts` are collected into `steps/{name}/artifacts/` after that
step's verification. Task-level and trial-level artifacts are collected at
every step in addition to the step-level ones.

### Oracle verification

`harbor run -p "<task-path>" -a oracle` runs each step's `solution/solve.sh`,
then each step's verifier, in order. Trial reward should be `1.0` across the
aggregation strategy.

### Full reference + worked example

- Docs: `docs/content/docs/tasks/multi-step.mdx`
- Example task: `examples/tasks/hello-multi-step-advanced/`

## Special features (mention if relevant)

- **Network policy**: Baselines on `[environment]` / `[verifier.environment]`; phase
  overrides on `[agent]` / `[verifier]`; see *Network policy* under Step 6
- **MCP servers**: Add `[[environment.mcp_servers]]` in task.toml for agent tooling
- **Healthcheck**: Add `[environment.healthcheck]` for services that need to be ready
- **GPU**: Set `environment.gpus` and optionally `environment.gpu_types`
- **Pre-built image**: Set `environment.docker_image` instead of building from Dockerfile. You can omit `environment/Dockerfile` and place runtime files (configs, scripts, data) directly under `environment/`; Harbor uploads them into the container workdir when the environment starts.
- **Non-root user**: Set `agent.user` / `verifier.user` for isolation

## Common pitfalls

- Forgetting to write the reward file → task "passes" silently with reward 0
- Using relative paths in `test.sh` → breaks when Harbor runs it from a different cwd
- Installing the solution into the Dockerfile → agent already gets the answer
- Test script leaks into `instruction.md` → agent sees the rubric and gaming becomes trivial
- Forgetting `chmod +x solution/solve.sh` → Oracle agent fails
- Leaving `keywords = []` in task.toml → task is invisible to registry search
- Leaving `README.md` as a stub → teammates have no way to understand the task at a glance
- Putting `network_mode` on `[agent]` expecting it to apply at env start → use
  `[environment].network_mode` for the baseline; agent/verifier fields are phase overrides
- Phase override differs from baseline on Docker → task rejected unless provider supports
  `dynamic_network_policy`; use separate verifier env or match the baseline instead
\end{lstlisting}

\subsection{\system{} Skill}
\label{app:skill-autoenvscaling}

We build this skill for \system{} as an extension of \texttt{create-task}. It sets the curriculum
target, tasks whose rewards vary across the solver's attempts, directs the proposer to ground each
task in the solver's rollouts and past environments, and requires the reference solution to pass and
an empty solution to fail before submission.

\begin{lstlisting}[style=skill]
---
name: auto_env_scaling
description: AutoEnvScaling overlay on Harbor's create-task. Use when authoring a NEW curriculum
  environment for the AutoEnvScaling loop with useful reward spread across solver attempts,
  grounded in actual solver trajectories.
---

You are the **environment proposer** in AutoEnvScaling. **Follow the `create-task` skill
for ALL the mechanics** of building a valid Harbor task — scaffold with `harbor task
init`, design the verifier, write the environment + solution, configure task.toml. This
overlay ONLY adds *what kind* of task to build (the curriculum target) and how to validate
it before submission. Do NOT hand-roll the task directory — run `harbor task init` and
edit the scaffold.

## The objective: create within-solver reward spread
Author ONE deterministic task targeting the solver's actual failure modes. The PRIMARY
objective is spread across its completed attempt rewards (`reward_std`, equivalently
reward variance); mean `avg_reward` is only a SECONDARY difficulty-coverage check. `[0,
1]` and `[0.5, 0.5]` both average 0.5, but only the first has group-relative learning
signal; the second has zero GRPO signal. Calibrate bidirectionally: ease task types whose
attempts are all zero and harden types whose attempts are all one, while preferring
variants whose genuine solver decisions produce different outcomes.

## Ground the task in real signal — read these first (you are adaptive, not random)
1. **`past_trajectories/index.jsonl`** — use shell/tool calls to find the latest
   `provenance:"self"` round and inspect the solver's `reward_std` plus raw `rewards`.
   Then open actual `past_trajectories/<env_id>/<solver>/*.json` attempts and read the
   trajectory plus `outcome.verifier_stdout`; the captured tool trajectory must
   demonstrate these reads.
2. **`past_environments/<env_id>/`** — how each attempted env is BUILT (canonical Harbor
   layout: `instruction.md`, `tests/test_outputs.py` = real verifier design,
   `environment/Dockerfile` + build script = how state is planted, `task.toml` =
   taxonomy), keyed to the trajectories above so you see the task, how it's built, AND how
   the solver did on it. These are your **EXEMPLARS** for structure & verifier design —
   adapt one toward the solver's gap rather than inventing from scratch. Do NOT duplicate:
   cover a *different* weakness, at sometimes-solvable difficulty.
3. **Web search** — ONLY when the runtime preamble explicitly lists a web tool
   (`web_search`/`web_fetch` or `web_search.py`). If no such preamble is present, you have
   no web access; do not attempt searches.

## Verifier template (when the round mandates it)
A managed fraction-scoring verifier lives at `/opt/verifier_template.sh` (reward = pytest
tests passed / total, via CTRF). If your instructions mandate the managed template: `cp
/opt/verifier_template.sh tests/test.sh` VERBATIM (do not edit it) and design
`tests/test_outputs.py` with meaningful pytest checks that diagnose real stages. Gate
milestones so genuine solver decisions can create across-attempt spread; do not make a
low-bar additive checklist that gives every attempt identical partial credit.

## What makes the env USEFUL
- **Hard but solvable**: the oracle (`solution/solve.sh`) passes the verifier
  deterministically; a **no-op must FAIL** (not trivially gameable).
- **Real skill gap**: multi-step terminal/SWE work — exact flags/paths/config, state that
  must be tracked. No trivia; nothing underspecified or unsolvable.
- **Self-contained & deterministic**: no network at solve time (`network_mode =
  "no-network"`); the managed template writes a **fractional reward 0..1**
  (tests-passed/total) to `/logs/verifier/reward.txt`. Fractional checks may diagnose
  failure stages, but do not make an additive checklist that every attempt clears to the
  same partial score. Gate meaningful milestones. Never manufacture spread with
  randomness, flaky timing, underspecification, or nondeterministic tests. Oracle must
  score 1.0; a no-op must remain below full credit.

## Validation
Validate and iterate before submitting:
```bash
# 1. Oracle MUST pass (env is genuinely solvable) — reward=1.0, exit 0:
python3 /opt/tools/validate.py <task> oracle
# 2. No-op MUST fail (env is not gameable):
cp <task>/solution/solve.sh /tmp/solve.bak && printf '#!/bin/sh\nexit 0\n' > <task>/solution/solve.sh
python3 /opt/tools/validate.py <task> oracle    # expect reward<1 (non-zero exit)
cp /tmp/solve.bak <task>/solution/solve.sh          # restore the real oracle
```
If (1) ≠ reward=1.0: solve.sh and the verifier disagree — fix and rerun. If (2) does NOT
fail: the verifier passes without real work — make it check the actual artifact. Submit
only when **oracle-passes ∧ no-op-fails**.

## Where to write
Scaffold with `harbor task init`, then make the finished **canonical task** land at the
env path your instructions give — i.e. `env_NNN/` itself must contain `environment/`,
`solution/`, `tests/`, `instruction.md`, `task.toml` (move the scaffold up if `harbor task
init <name>` created a `<name>/` subdir). Create no persistent artifacts outside it.
\end{lstlisting}

\section{Execution During Environment Generation and Verifier Limitations}
\label{app:execution-validation}

\subsection{Reference Execution Ablation}
\label{sec:abl-selfvalidate}

To examine the role of execution in Section~\ref{sec:env-scaling}, we compare generation with
and without a reference-solution self-check. Unlike the comparison in Section~\ref{sec:env-scaling},
this ablation runs with solver calibration, so its rates differ from those in Table~\ref{tab:threeway}.
Within the ablation, both conditions use the same assignments for each model. Enabling the check improves valid-environment yield for all five models
(Table~\ref{tab:selfvalidate}). The agent can observe failures and repair the environment before
submission, which accounts for much of the benefit of the coding-agent interface.

\begin{table}[htbp]
\centering
\caption{\textbf{Executing the reference solution improves valid-environment rate.} Valid-environment rate
(\%) with the generation self-check disabled or enabled.}
\label{tab:selfvalidate}
\small
\setlength{\tabcolsep}{10pt}
\begin{tabular}{@{}lrr@{}}
\toprule
Proposer & Self-check disabled & Self-check enabled \\
\midrule
Claude Opus 5 & 52.0 & \textbf{94.0} \\
GPT-5.6-sol & 19.5 & \textbf{84.8} \\
GPT-5.6-terra & 45.2 & \textbf{93.5} \\
GPT-5.6-luna & 40.1 & \textbf{85.3} \\
DeepSeek-V4-Flash & 19.4 & \textbf{73.0} \\
\bottomrule
\end{tabular}
\end{table}

\subsection{Decontamination}
\label{app:decontamination}

Following Tmax~\citep{ivison2026tmax}, we measure train--test contamination as $n$-gram overlap
between the task descriptions of generated environments and those of our evaluation benchmarks,
Terminal-Bench~2.1 and Terminal-Bench~4.0. Using a sliding window of $n{=}13$ tokens with stride $1$,
we flag a generated task if any of its windows matches at least one $13$-gram from a benchmark task.
This check measures textual overlap; it does not rule out paraphrased copies or other forms of
contamination.

\section{Training and Calibration Settings}
\label{app:training-settings}

\subsection{Shared-Model Training}

Table~\ref{tab:training-settings} gives the settings for the recursive-training experiment in
Section~\ref{sec:rsi}. The active pool is reviewed every $16$ solver updates, and tasks that
no longer meet the learning criteria are removed selectively.

\begin{table}[htbp]
\centering
\caption{\textbf{Settings for shared-model recursive training.}}
\label{tab:training-settings}
\small
\setlength{\tabcolsep}{12pt}
\begin{tabular}{@{}lr@{}}
\toprule
Setting & Value \\
\midrule
Learning rate & $10^{-6}$ \\
Solver rollout batch & 16 environments $\times$ 16 trajectories \\
Proposer rollout group & 128 proposer trajectories \\
Proposer generation interval ($k$) & 16 solver updates \\
Active environment pool & 256 tasks \\
Pool review interval & 16 solver updates \\
DPPO total-variation threshold & 0.1 \\
KL / entropy coefficients & 0 / 0 \\
\bottomrule
\end{tabular}
\end{table}

\paragraph{Proposer rewards and updates.}
Proposer rewards follow Equation~\ref{eq:proposer-reward}, using the validation and
solver-calibration checks described in Section~\ref{sec:m-gates}.
Each proposer group contains $128$ multi-turn trajectories sampled for the same generation
assignment. We compute the advantage baseline from their mean reward, separately from the
solver's per-environment groups. Rewards are mean-centered without division by the group
standard deviation.
The solver trains at every step. Every $16$ solver updates, the proposer generates new
environments and trains on the resulting proposer trajectories. No additional role-specific
loss weighting is used.
All $128$ trajectories in a proposer group share the same generation assignment. As for the
solver, each trajectory's advantage is applied to the tokens the model generates.

\paragraph{Rollout collection.}
We train with Miles on 4--8 B200 GPUs. Each solver update uses $16$ environment groups with
$16$ trajectories per group. We oversample and discard groups with no reward variation,
collecting additional groups until the batch is filled. Unfinished terminal trajectories are
carried across updates, preserving their interaction history and live sandbox state. They
resume with the updated policy, while stored sampling log-probabilities are retained for
previously generated tokens.

\paragraph{Evaluation ranges.}
Training comparisons match the number of solver updates, not total compute or elapsed time.
The full \system{} run and Tmax are evaluated through $288$ solver updates. The proposer-training
ablation is evaluated through $160$ updates, so the three-way comparison uses that common endpoint:
Terminal-Bench~2.1 scores are $48.9$, $44.9$, and $40.1$ for the full system, Tmax, and the ablation.
The ablation peaks at $44.1$ at update $96$. On Terminal-Bench~4.0, the full system reaches $3.2$
at update $192$ and retains that score through update $288$.
Rollout-collection summaries use updates $0$--$162$, the range logged for both methods, with matched hardware
and rollout concurrency. Reported retention is total accepted groups divided by total collected
groups; collection time is the arithmetic mean of the recorded minutes per update.

\subsection{Proposer Validity During Shared-Policy Training}
\label{app:proposer-validity}

\begin{figure}[htbp]
\centering
\includegraphics[width=0.8\linewidth]{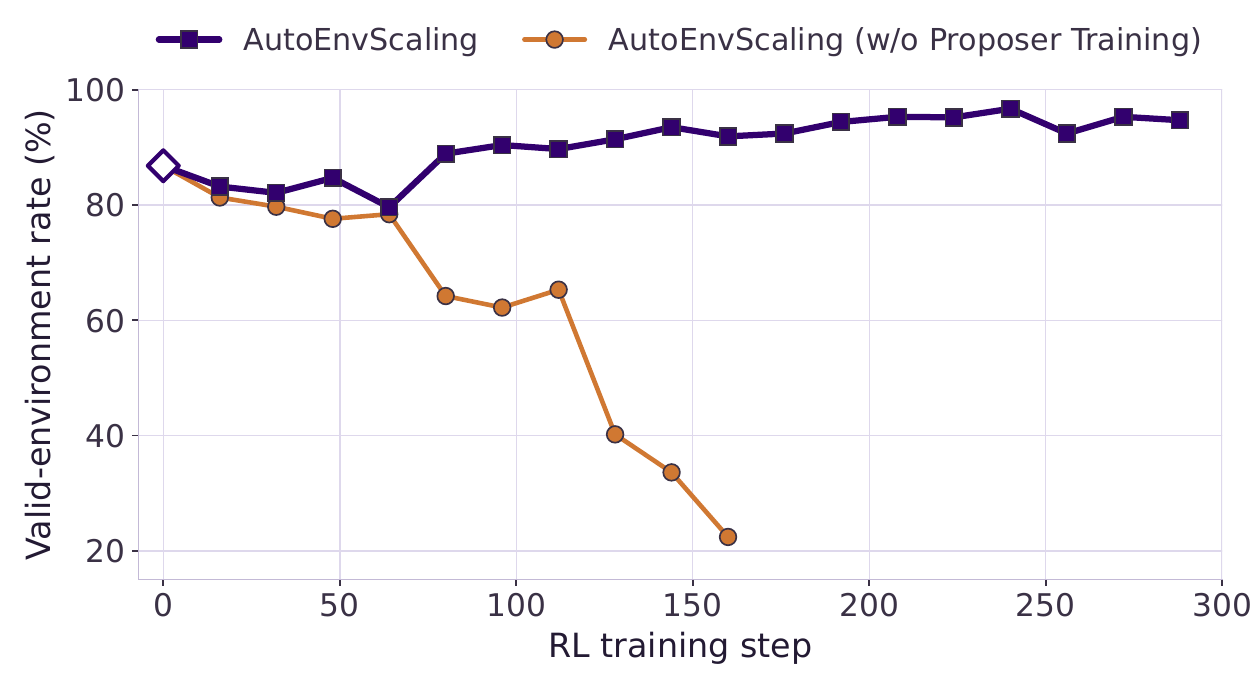}
\caption{\textbf{Proposer training keeps the shared policy's environments valid.} Valid-environment
rate of \propmodel{} acting as proposer during shared-policy training. The open diamond marks the
shared Cold-Start checkpoint. The run without proposer training ends at step 160 because training
collapsed.}
\label{fig:proposer-validity}
\end{figure}

Figure~\ref{fig:proposer-validity} tracks the valid-environment rate of the shared policy acting as
proposer during the run in Section~\ref{sec:rsi}. Both runs start from the Cold-Start checkpoint at
$86.8\%$. With proposer training, the rate dips to $79.6\%$ at step 64, rises above $88\%$ from step
80 onward, and ends at $94.7\%$. Without proposer training, the same weights still change through
solver updates, and the rate falls to $22.4\%$ by step 160, when training collapses.

\subsection{Evaluation Harness and Resources}
\label{app:eval-harness}

All Terminal-Bench results in this paper use the Pi coding agent in sandboxes with 4 CPUs and 8~GiB
of memory. Model cards report higher absolute scores for the same models, for example 40.5 for
Qwen3.5-35B-A3B and 51.5 for Qwen3.6-35B-A3B on Terminal-Bench~2.0. Those numbers use the
Terminus-2 harness with a 3-hour timeout, in sandboxes with 32 CPUs and 48--64~GB of
memory.\footnote{\url{https://huggingface.co/Qwen/Qwen3.5-35B-A3B},
\url{https://huggingface.co/Qwen/Qwen3.6-35B-A3B}} In our tests, the harness accounts for most of
this gap, and the smaller sandboxes for part of the rest. Every comparison in this paper uses the
same harness and resources for all methods. Our Tmax reproduction on Qwen3.5-9B (27.2 and 29.2 on
Terminal-Bench~2.1) is also close to the 28.8 that \citet{ivison2026tmax} report with a different
harness.

\subsection{Solver-Guided Calibration}
\label{app:calibration}

Admission requires reference reward $r=1$ and do-nothing reward $r<0.5$, in addition to valid files
and a successful container build. The task review runs \texttt{harbor check} with the official
Terminal-Bench task-implementation rubric, graded by the proposer's own model. We apply every
criterion except \texttt{binary\_reward}, which requires rewards of exactly $0$ or $1$, because our
tests give partial credit. A task that fails a criterion returns to the proposer for revision. Calibration targets mean reward in $[0.25,0.75]$ with reward standard deviation at least $0.1$.
It begins with four solver rollouts and expands to eight when the mean is within $0.15$ of a
band boundary. Tasks below the band receive one diagnostic rollout with reference-solution hints
to guide revision. Each assignment permits up to two revisions. Revised environments repeat the admission checks before entering the training pool.

\section{Cold-Start Controls}
\label{app:initialization-controls}

Cold-Start fine-tunes \propmodel{} on $600$ high-quality proposer trajectories from Claude Opus~5,
DeepSeek-V4-Flash, and Kimi-K3.

\begin{figure}[htbp]
\centering
\begin{minipage}[t]{0.485\linewidth}
\centering
\includegraphics[width=\linewidth]{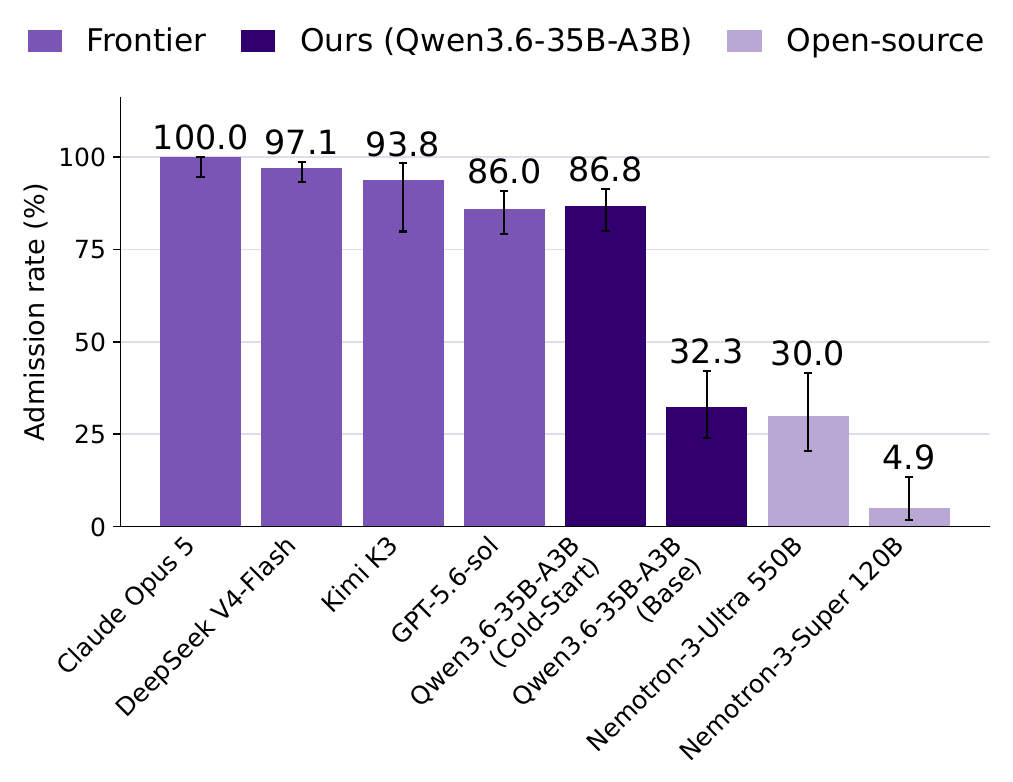}
\caption{\textbf{Cold-Start brings environment generation performance close to frontier models.}
Valid-environment rates across frontier models and Qwen3.6-35B-A3B before and after Cold-Start training.
Error bars show $95\%$ Wilson intervals; assignment suites vary across models.}
\label{fig:proposing}
\end{minipage}\hfill
\begin{minipage}[t]{0.485\linewidth}
\centering
\includegraphics[width=\linewidth]{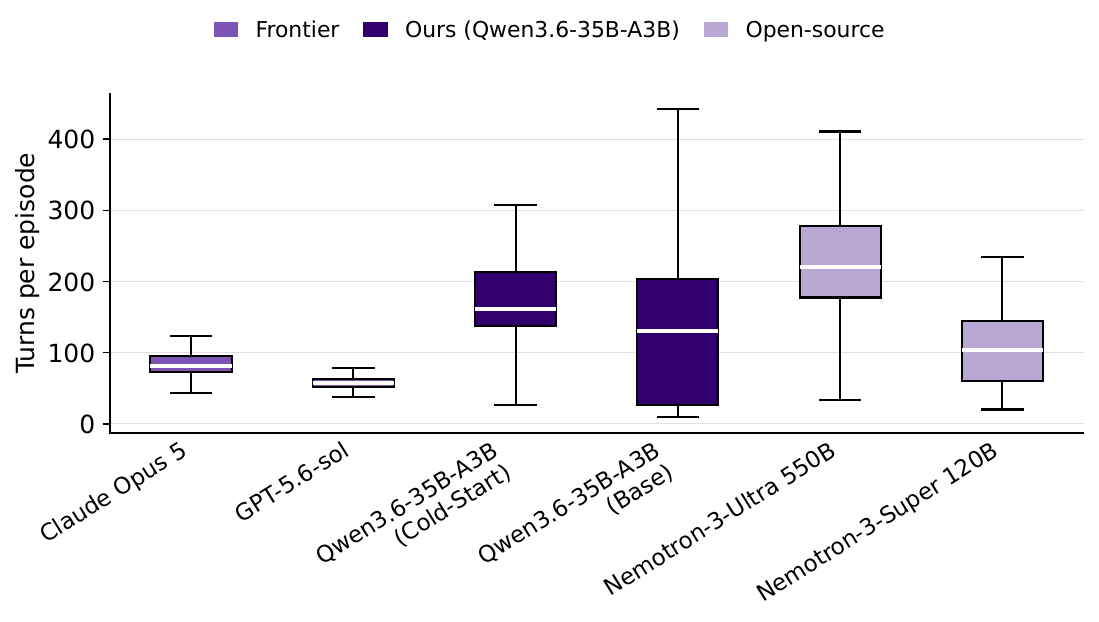}
\caption{\textbf{Proposer episode depth.} Turns per episode for frontier, Cold-Start,
and base models. Boxes show the interquartile range and median; whiskers extend to the
furthest observations within $1.5$ times the interquartile range. Outliers are omitted.
Opus uses an earlier assignment suite; the other models use the shared $136$-assignment suite.}
\label{fig:authoring-depth}
\end{minipage}
\end{figure}

The comparison in Figure~\ref{fig:proposing} gives the base and Cold-Start \propmodel{}
models and GPT-5.6-sol the same $136$ environment-generation assignments. Admission requires
a successful image build, a passing reference solution, and a failing do-nothing solution.

We evaluate \propmodel{} before and after Cold-Start training on environment-generation
trajectories. A paired control uses 60 generation assignments, and a separate evaluation measures
solving performance on Terminal-Bench~2.1 (Table~\ref{tab:initialization-controls}).

\begin{table}[htbp]
\centering
\caption{\textbf{Cold-Start improves environment generation without a detected change in solving performance.}
Environment acceptance uses the same 60 assignments for both models. Solving uses five attempts
on each of 89 Terminal-Bench~2.1 tasks. All values are percentages.}
\label{tab:initialization-controls}
\small
\setlength{\tabcolsep}{8pt}
\begin{tabular}{@{}lrrr@{}}
\toprule
Model & \shortstack{Environment\\acceptance} & \shortstack{TB2.1\\avg@5} & \shortstack{TB2.1\\pass@5} \\
\midrule
Base & 33.3 & 40.0 & 59.6 \\
Cold-Start & \textbf{76.7} & 41.1 & 66.3 \\
\bottomrule
\end{tabular}
\end{table}

Of the 38 disagreements in task acceptance, 32 favor the Cold-Start model and six favor the base.
The mean paired change in solving performance is 1.1 percentage points
(95\% bootstrap interval: $[-4.9,7.4]$; paired sign test $p=0.88$).
The pass@5 difference is also not statistically significant ($p=0.15$).

\paragraph{Training-pool feasibility.}
In a separate pool-generation experiment, the Cold-Start model supplies 16 tasks with
non-degenerate solver rewards (standard deviation $\geq 0.1$), enough for one rollout batch.
The base model supplies four trainable tasks among 13 evaluated admissions, with one admission
unevaluated. Even if that task were trainable, the pool would remain below the 16-task minimum.

\section{Environment Generation on Two Further Seed Cases}
\label{app:authoring-styles}

Figure~\ref{fig:authoring-styles} in Section~\ref{sec:qualitative} shows one seed case. The two
below use the same template and the same three proposers.

\begin{figure}[htbp]
\centering
\includegraphics[width=\linewidth]{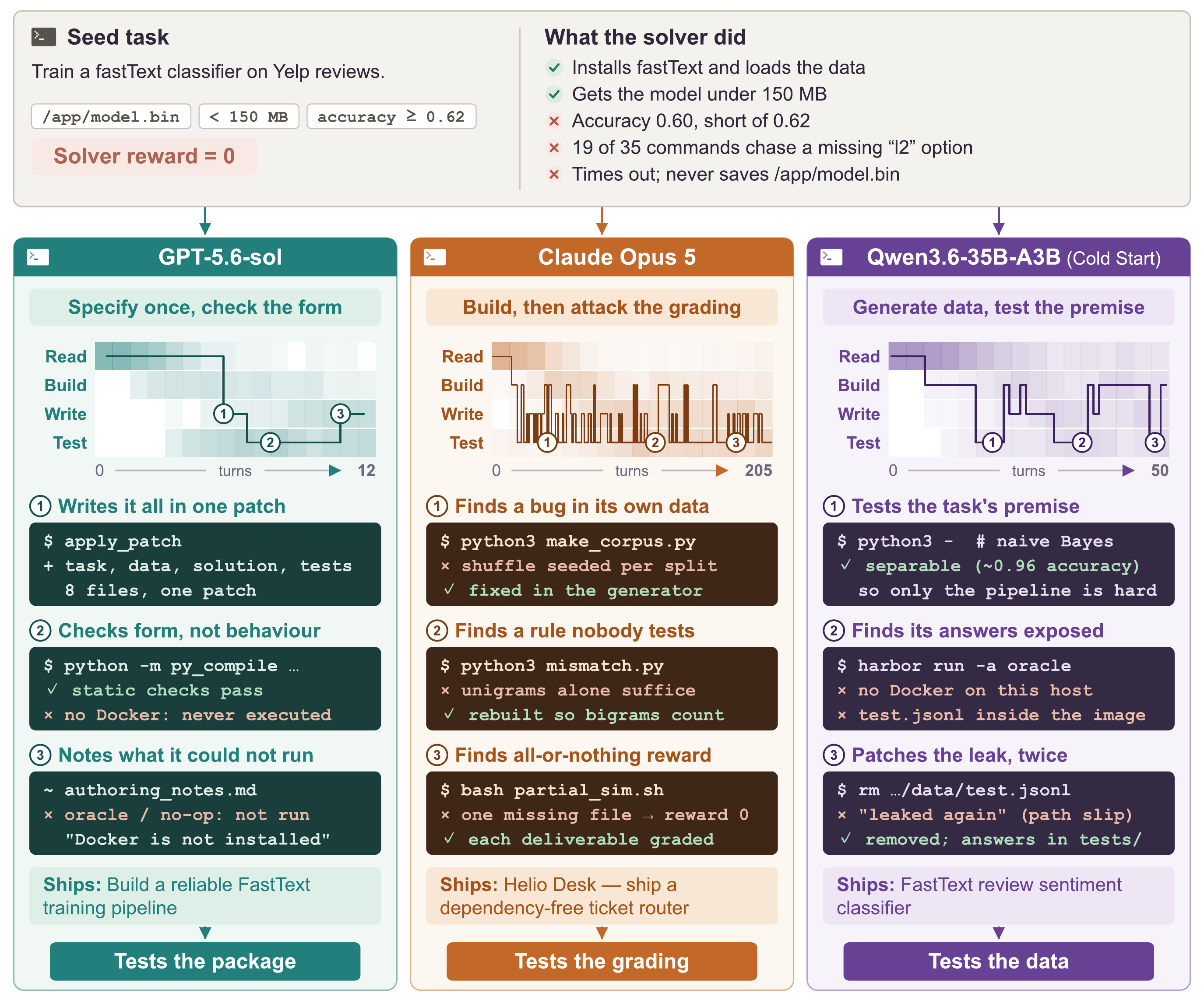}
\caption{\textbf{fastText seed case.} The solver trained a fastText classifier on $650$k Yelp
reviews, measured $0.60$ against a target of $0.62$, spent $19$ of $35$ commands chasing a
non-existent option, and timed out without writing the model file. No Docker daemon was
available to GPT-5.6-sol, so it ran static checks only. Panels follow
Figure~\ref{fig:authoring-styles}.}
\label{fig:authoring-styles-fasttext}
\end{figure}

\begin{figure}[htbp]
\centering
\includegraphics[width=\linewidth]{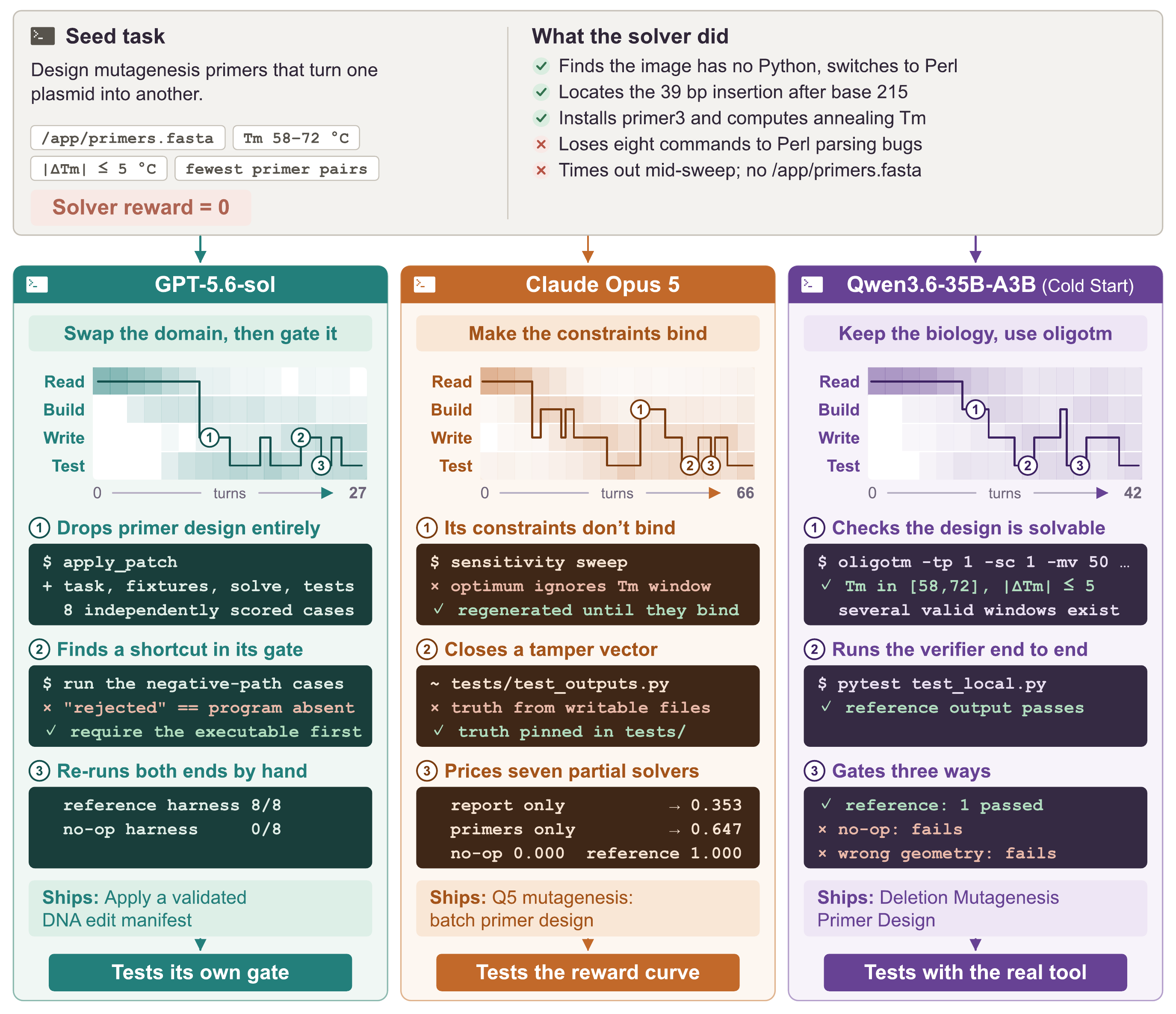}
\caption{\textbf{Primer-design seed case.} The hardest seed task: design Q5 mutagenesis primers
under melting-temperature constraints, graded against primer3's \texttt{oligotm}. The solver
switched to Perl (no Python in the image), located the insertion, computed annealing
temperatures, and timed out. GPT-5.6-sol drops primer design for a generic FASTA pipeline;
Claude Opus~5 and Cold-Start \propmodel{} both keep the biology. Panels follow
Figure~\ref{fig:authoring-styles}.}
\label{fig:authoring-styles-dna}
\end{figure}

\subsection{Labelling protocol}
\label{app:authoring-styles-protocol}

These three seed cases are the only ones in the sample where all three proposers ship a task
without hitting the wall clock, and their seed failures differ: a budget exhausted with nothing
produced, a repair loop that timed out mid-fix, and a domain-hard task the solver almost finished.

A turn is a public text step or tool call. Native thinking is excluded. Cold-Start \propmodel{}'s
traces carry thinking steps and the other two do not, so its public record is thinner than its
actual reasoning. That matters when comparing stated rationale across the three.

All nine drawn paths are labelled by hand, turn by turn, $475$ turns in total. The shaded density
field behind each path comes from a keyword classifier that agrees with the hand labels on $83$ to
$90\%$ of turns per episode. We do not use that classifier for the drawn paths. An earlier version
of it filed Claude Opus~5's difficulty sweeps under \emph{build}, which inverted the reading.

Episodes use the first sample, with two exceptions: Claude Opus~5 on the fastText case, whose first
sample hit the wall clock while waiting on its own reference run, and Cold-Start \propmodel{} on
the primer-design case, whose other samples did not ship. Terminal lines inside the numbered
moments are abridged, and a few label an action instead of quoting a command.

\end{document}